\documentclass[journal=jacsat,manuscript=article]{achemso}

\usepackage[version=3]{mhchem} % Formula subscripts using \ce{}
\usepackage{adjustbox}
\usepackage[pdftex,colorlinks=true,linkcolor=red,citecolor=blue]{hyperref}
\usepackage{graphicx}
\usepackage{hypernat}
\usepackage{csquotes}
\usepackage{xcolor, soul}
\author{L. D. Tamang}
\affiliation[mzu]{Advanced Computation of Functional Materials Research Lab (ACFMRL) Department of Physics, Mizoram University, Aizawl-796004, India}
\alsoaffiliation[puc]{Physical Sciences Research Center (PSRC), Department of Physics, Pachhunga University College,  Aizawl-796001, India}
\author{S. Gurung}
\affiliation{Physical Sciences Research Center (PSRC), Department of Physics, Pachhunga University College,  Aizawl-796001, India}
\author{R. Zosiamliana}
\affiliation{Physical Sciences Research Center (PSRC), Department of Physics, Pachhunga University College,  Aizawl-796001, India}
\author{L. Celestine}
\affiliation{Physical Sciences Research Center (PSRC), Department of Physics, Pachhunga University College,  Aizawl-796001, India}
\author{B. Chettri}
\affiliation{Physical Sciences Research Center (PSRC), Department of Physics, Pachhunga University College,  Aizawl-796001, India}
 \author{Jitendra Pal Singh}
 \affiliation{Department of Sciences, Manav Rachna University, Faridabad, Haryana, 121004, India}
\author{A. Laref}
\affiliation{Department of Physics and Astronomy, College of Science, King Saud University, Riyadh, 11451, Saudi Arabia}
\author{Mukhriddin E. Tursunov}
\author{Avazbek T. Dekhkonov}
\affiliation{National University of Uzbekistan named after Mirzo Ulugbek, Tashkent, Uzbekistan}
\author{D. P. Rai}
\affiliation[mzu]{Advanced Computation of Functional Materials Research Lab (ACFMRL) Department of Physics, Mizoram University, Aizawl-796004, India}
\email{dibyaprakashrai@gmail.com}
\phone{+918132832252}
\fax{XXXXXXXXXX}

\title[An \textsf{achemso} demo]
  {Recent progress on the solid-state materials for photocatalysis\footnote{Recent progress on the solid-state materials for photocatalysis}}

\abbreviations{IR,NMR,UV}

\begin{document}

\begin{tocentry}

\includegraphics[height=6.0cm, width=5.0cm]{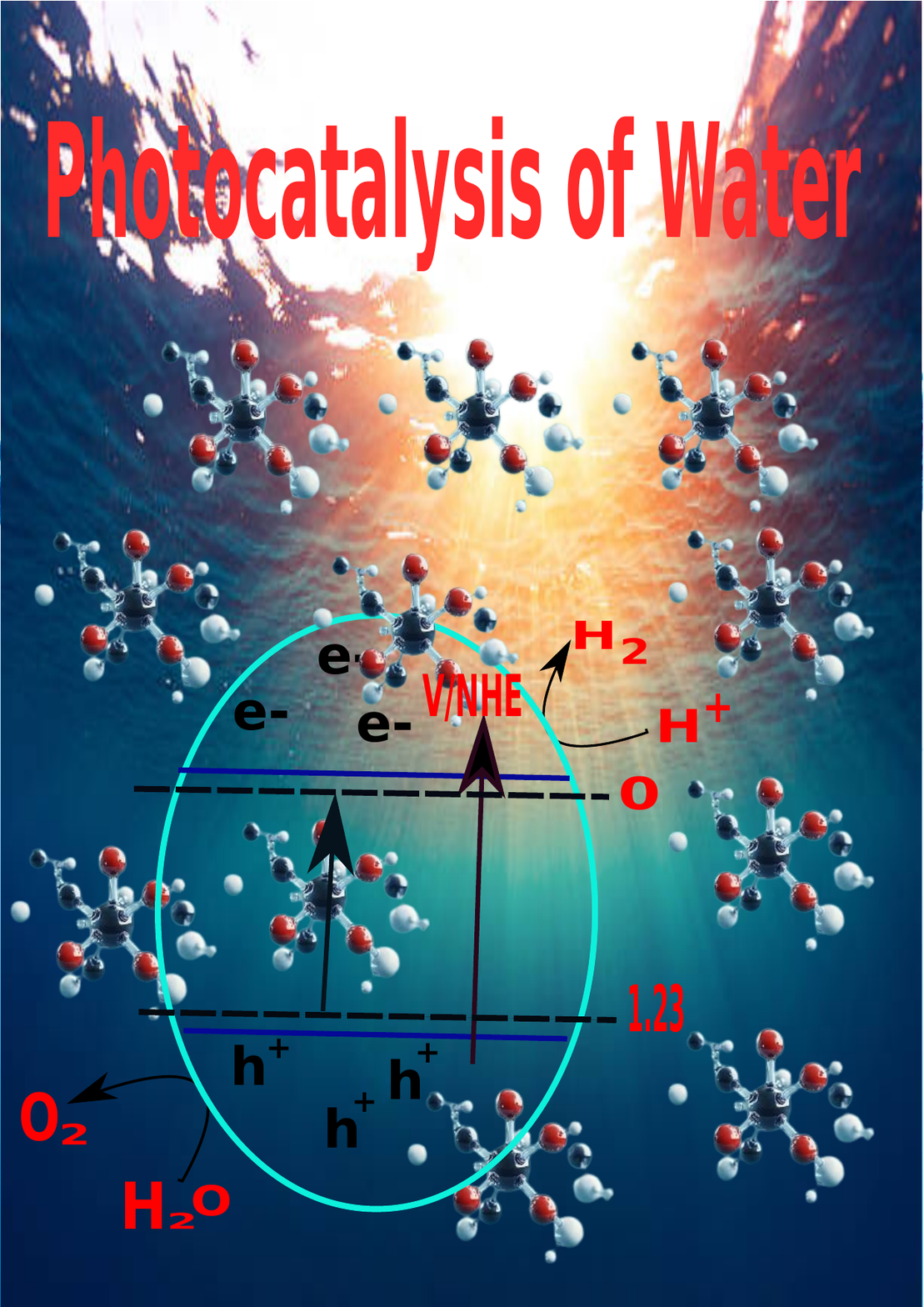}

\end{tocentry}

\begin{abstract}
 Hydrogen is considered an alternative source of energy to fossil fuels for the fulfilment of current energy demands. Photocatalysis initiates the hydrogen evolution reaction which is believed to be the greenest approach to produce hydrogen through clean, safe, and environmentally friendly methods. In this Review, we focus mainly on the comprehensive analysis of the 2D and 3D bulk materials on the basis of their superior photocatalytic activities. However, several literatures have reported the superiority of 2D material over the bulk counterpart in terms of photocatalytic performance owing to their ultrathin layered structures, offer a higher surface-to-volume ratio, flexibility, large active sites for incoming H$_2$O molecules, etc. We have thoroughly analysed the drawbacks of various hydrogen production methods focusing on the photocatalysis mechanism and the processes of evolution of hydrogen. In addition to this, a short overview of the various solid-state materials for photocatalysis that have been developed so far and their mechanisms are discussed. Lastly, we have discussed the recent developments in 2D materials and their composites as promising photocatalysts.
\end{abstract}

\section{Introduction}
The global energy crisis has been foreseen to be inevitable due to the depletion of fossil fuels, due to over consumption, and overpopulation. Fossil fuels are the main contributors of energy to the current energy supply as compared to other forms of energy. For over last 150 years, fossil fuels have been used for different purposes in households to produce heat, in large power plants to generate electricity, to power engines, etc. Almost 80\% of the world's energy is dependent on fossil fuel. However, being a potential energy resource, it has some drawbacks due to over consumption that resulted in the adverse effect on the environment by the formation of carbon dioxide(CO$_2)$ a primary component of greenhouse gases leading to global warming, extreme weather events, rising seas and many more impacts that threaten life sustainability\cite{wang2024natural}. However, an upsurge in population and ongoing industrial development lead to high demands for energy supply. Keeping in mind the impacts of fossil fuels on the environment and the need to balance the demand and supply process of energy supply, the search for a green alternative source is imperative. Hence, a lot of experimental and theoretical research is underway to find a potential energy source that will replace fossil fuels. The new energy source is novel, renewable, reliable and environmentally friendly \cite{keeling1973industrial,suzuki1975extension}.\\
In ancient times, hydrogen was known to them as a shapeless and formless "inflammable air" referred to by alchemists and philosophers who believed it could turn base metals into gold. Around the 1500 s and later in 1671, Swiss physician Paracelsus and Anglo-Irish Robert Boyle both observed this flammable gas by adding sulfuric acid to iron fillings. In 1766, the British scientist Henry Cavendish was credited with the discovery of hydrogen gas, and when he applied a spark to flammable gas, water was produced\cite{west2014henry}. Later in 1783 French chemist Lavoisier did an experiment to form water by reacting oxygen with inflammable gas, he concluded that water is a compound and the decomposition of water gives two individual products. Later, the inflammable by-product was named "hydrogen gas"\cite{lavoisier2001antoine}. From then on a great deal of research has been done to harness the hydrogen gas into a potential fuel\cite{BOCKRIS20132579,weeks1932discovery}.    
Since hydrogen is the simplest element made up of one proton and an electron with a huge amount of energy stored in it and is one of the abundant elements found in the earth's atmosphere in compound form, it is considered a substitute for fossil fuels because of its high energy density, i.e. $(122 kJmol^{-1})$, even greater than the fossil fuels and other sources of energy. One of the most advantageous parts is that hydrogen burning is always water (by-product), making it cleaner, safer, and environmentally friendly\cite{Chalk2006,Omer2008,Furukawa2009,Li2010,Peschka2012}. In nature, there is a large reservoir of water that covers about 71\% of the earth's surface and can be used as a potential source of hydrogen by electrolysis. With the advancement of technology, water can be split into hydrogen and oxygen using various methods and the released hydrogen can be used to meet our energy needs, which includes generation of electricity, power supply to industry, fuel for vehicles, domestic uses, etc.\\
Depending on the materials used for hydrogen production, it can be divided into two categories, conventional and renewable processes\cite{nikolaidis2017comparative}. Basically in conventional processes, fossil fuels are used to obtain hydrogen gas which is environmentally unfavourable, while renewable processes mainly depend on renewable sources of energy, in which the splitting of water is the mostly accepted method for the production of hydrogen. The production of hydrogen via decomposition of water can be done by the following processes:-\\  
\begin{enumerate}
 \item Electrolysis: In this process, electrical energy transforms into chemical energy as electricity is supplied through water. A reaction occurs at the interface of the electrode and electrolyte, facilitated by charge transfer in a component called an electrolyzer\cite{Rostrup-Nielsen2002}. Only 3.9\% of the world's hydrogen demand was able to be provided by this process but on the other hand, it is not considered a green process because normally the electrolyzers are operated by electricity which is mostly generated by burning coal or natural gases, resulting in the emission of CO$_2$ as a result nowadays more and more research are aiming at utilizing renewable harvesting technologies\cite{kwasi2015review}.\\
\item Thermolysis: In this process, water is decomposed using thermal energy from the existing nuclear power plant as waste heat or using concentrated solar power. For this process to happen, a high temperature around $(500-2000^{\circ}C)$ is needed with a series of chemical reactions that eventually give out hydrogen and oxygen as a by-product, but it is also not suitable due to the high energy demand and is not commercialized\cite{Rostrup-Nielsen2002,Chiesa2005}.
\item Photo-biologic: In this process micro-organisms such as green algae and cyanobacteria harness solar energy to dissociate water into oxygen and hydrogen ions which combine directly or indirectly to produce hydrogen gas some microbes break down organic matter using sunlight by producing hydrogen known as photo fermentative hydrogen production this process has a long term potential with environmental friendly way but cannot be utilized at the moment because it is in the initial phase of development and due to low rate of production and low solar to hydrogen efficiency\cite{OfficeofEnergyEfficiency&Renewable}.\\
\item Photolysis: In this process, sunlight is used to break down water into its components without releasing any harmful gases.  
Among the methods mentioned above, photolysis is considered an innovative approach for the production of hydrogen because the breakdown of water occurs when exposed to sunlight which is a renewable source of energy making it a cleaner, greener and safer way to extract hydrogen gas from water molecules and on the other hand most of the processes involved in the production of hydrogen include release of harmful gases or depend on non-renewable sources of energy\cite{Otsuka2003a,Kodama2002,Kang2010}.
\end{enumerate}

To make the photo-catalysis process more vigorous in active decomposition of water in to H$_2$ and O$_2$ an improved approach is opted by adding an appropriate a catalyst in the presence of sunlight. 
Edmond Becquerel discovered this process in 1839, based on natural photosynthesis used by the Autotrophs. Green plants and microorganisms use inorganic compounds like CO$_2$ and H$_2$O to convert them into carbohydrates using solar energy, with the help of green pigments called chlorophyll. By copying this process, artificial photosynthesis was carried out using catalysts(inorganic materials) instead of chlorophyll\cite{bard1995artificial}.
Here, the function of a catalyst is to enhance the reaction rate of hydrogen production by transforming solar energy into a chemical reaction. Any material can be used as a catalyst to harness solar energy such as metals, non-metals, semiconductors, metal-oxides, etc. However the main key factor is choosing a material with the appropriate energy required to decompose a water molecule (i.e., 1.23 eV) in the form of band energy, and the band edges should lie on the appropriate position compared to water reduction and oxidation potentials. Many semiconductors have been used as catalysts because of their intrinsic property which can be modified accordingly for better performance and to increase their efficiency for hydrogen production\cite{Gratzel2001,Wang2009,Zhou2020}. 
The photo-catalysis was first reported in 1955 by Markham and successfully carried out in 1972 by Fujishima and Honda using TiO$_2$ as electrodes\cite{Furukawa2009}. Thus, photo-catalysis of water to produce hydrogen is a hopeful and eco-friendly way for our development  \cite{Kumaravel2019,Teets2011,Gupta2017}. Subsequently, a large number of materials have been studied to evaluate their capability for hydrogen production\cite{gratzel2001photoelectrochemical,wang2009metal,sobczynski1991molybdenum}. The importance of this study is to find the ideal photo-catalyst with proper band alignment, efficient light harnessing properties, stability, non-toxicity, and abundantly available. Traditionally used materials like TiO$_2$, ZnO,CeO$_2$, and WO$_3$ were excellent photo-catalysts however, due to their large band gap and high recombination rate they need to be modified by using various methods or strategies\cite{wold1993photocatalytic,ABRAHAMS1985353,baba1985investigation}. The charge carriers must travel far to get to the active site in a short interval of time is the major drawback faced by the bulk materials thus decreasing the efficiency of hydrogen production\cite{ganguly2018antimicrobial}. The discovery of two-dimensional (2D) materials helps to overcome this problem by shortening the distance as well as lowering the rate of recombination of charge carriers\cite{ida2014recent}. They provide extraordinary electronic, mechanical, chemical, and optoelectronic properties with the ability to harness a wide range of solar energy spectrum\cite{novoselov2005two,zhang2011sulfur}. Graphene is the first 2D material synthesized thus opening a new perspective in the field of nano-engineering\cite{an2011graphene,butler2013progress}. Following the graphene various 2D materials were synthesized like Transition metal dichalcogenides (TMDs) which are generally represented by MX$_2$ where (M: transition metals from group IVB to VIII); X: S, Se and Te (from group VIA) forming a sandwich structure "X-M-X"\cite{manzeli20172d}. Their three distinct phases are 2H(trigonal prismatic), 1T(octahedral), and $1T'$(distorted octahedral) resulting in various electronic properties\cite{chen2020phase,qian2014quantum}, for example in $MoS_2$ its 2H and 1T phases show different characteristics such as semiconducting and metallic respectively. It also has a thickness-dependent electronic band gap which shifts from indirect to direct band gap when it is reduced from bulk to monolayer\cite{doi:10.1021/nl903868w}. Similarly, 2D metal-free graphitic carbon nitrides (g-$C_3N_4$) show excellent photocatalytic properties under visible light\cite{zheng2012graphitic}. There are lots of 2D materials that have been synthesized and are under study due to compatible band gap, a wide range of optical absorption, and large surface area which makes them ideal candidates for hydrogen evolution reaction (HER)\cite{panneri2016copyrolysed,panneri2017role}. Nonetheless, there are drawbacks to these structures i.e., high photo corrosion, unsuitable band gap, structural instability, etc. Different methods are employed such as doping, formation of hetero-junctions, introduction of co-catalyst, and application of strain to increase the efficiency of the photo-catalyst\cite{cai2018preparation,kumar2018sunlight,xia20182d}. Recent developments in the application of 2D materials and their nano-heterostructure composites for photocatalytic hydrogen evolution reactions, as well as their enhancement through the introduction of numerous methodologies and approaches, have been briefly reviewed in this work.  

\section{Basic principles and mechanism of photo-catalysis:}
\begin{figure}[h!]     
	\centering
	\includegraphics[height=8cm,width=9cm]{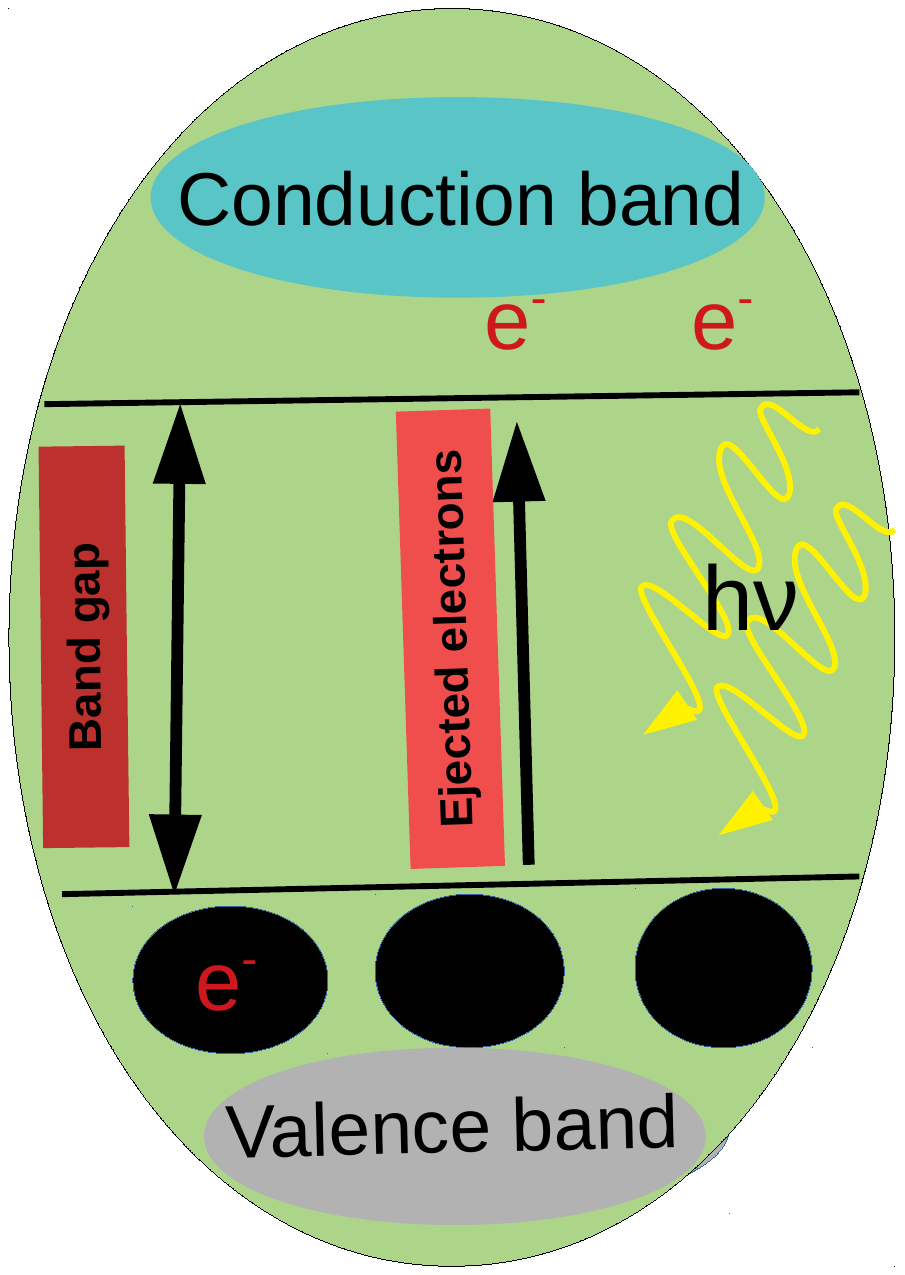}
	\caption{Illustration of the internal photoelectric effect\cite{yang20232d}.}
	\label{fig:effect}
\end{figure}

\begin{figure}[h!]
	\centering
	\includegraphics[height=8cm,width=8.3cm]{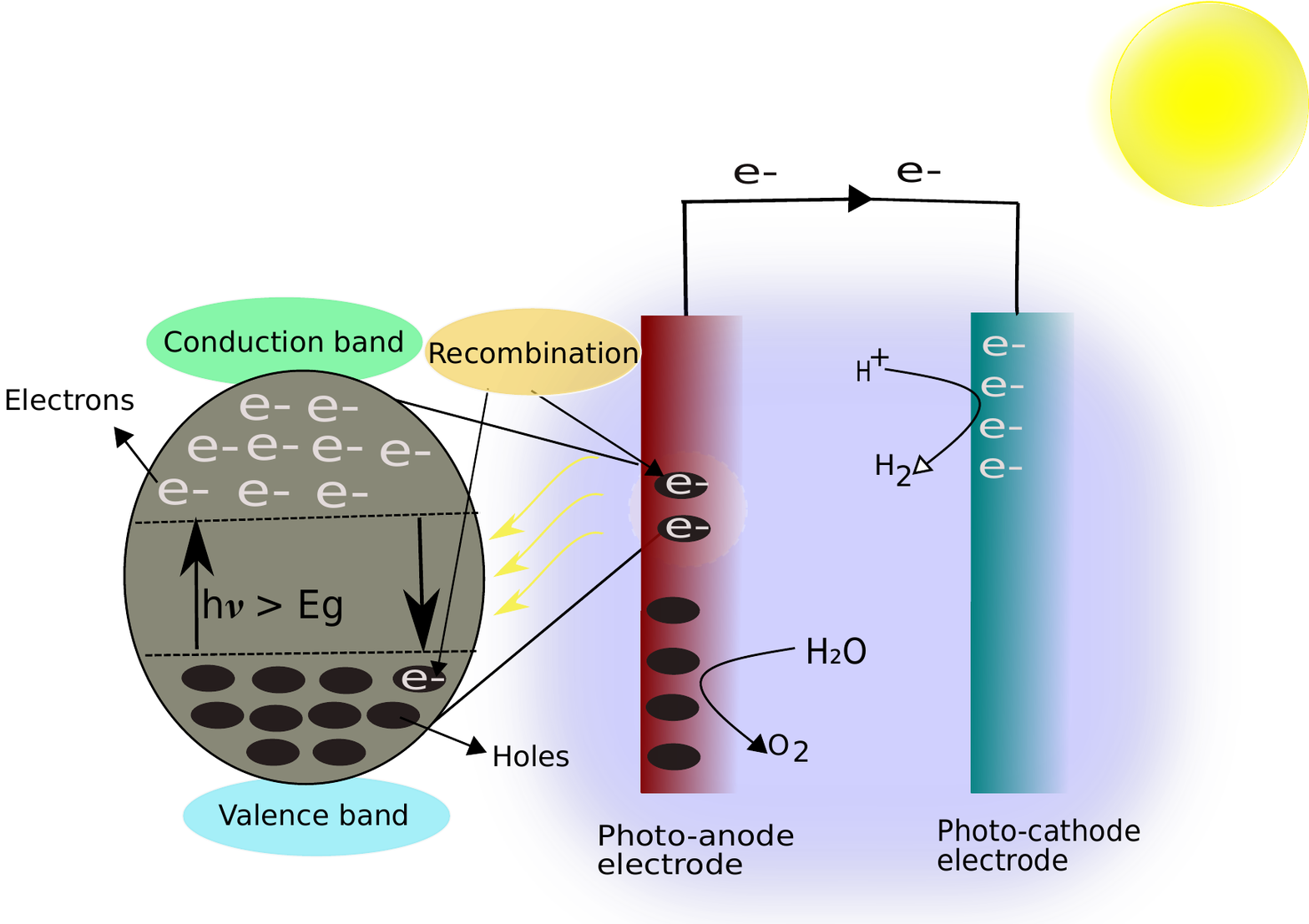}
	\caption{Overall water splitting process}
	\label{fgr:band-region}
\end{figure}

 The main work of a photocatalyst is to absorb light and induce its partners for chemical transformation without changing itself; thus, the reaction is known as a photocatalytic. Numerous uses for photocatalysis exist, including wastewater treatment, self-cleaning air purification glass, tiles, and tents, mainly in environmental fields and energy fields\cite{lacombe2012photocatalysis,zhu2017photocatalysis}. It is the process of using endless renewable sources of energy for the sustainable development of society. Solar energy can be transformed into different forms of energy which then can be utilized for various applications. Photocatalytic reactions can be divided into two categories, i.e., homogeneous and heterogeneous. Heterogeneous photocatalysis provides a platform for the development of an interface between the reactant, the product, and the photocatalysts\cite{ravelli2009photocatalysis}. The energy of incident photons and the catalyst's capacity for absorption are the primary determinants of photocatalytic reactions. 
There are three essential components to water splitting with photocatalysis, and they are:\\
i) Formation of exciton: It is a crucial step in the photocatalytic process. Usually, semi-conductors were used for photocatalysts or materials having the same properties as semiconductors, i.e., the presence of desirable energy band gap; on the other hand, materials like conductors do not have an energy band gap due to overlapping of the conduction-valence band, and in the case of insulators, they have a wide band gap, which means they needed much higher amounts of energy to excite the electrons, which is quite difficult; hence, they are not suitable to use as a photocatalysts When photons of suitable energy strike the surface of the material, some of the valence electrons take in enough energy to enter the conduction band i.e. higher energy levels leave behind the holes in the valence band, thus the separation of holes and electrons will take place creating a pair of energized electrons and holes also known as exciton\cite{liang1970excitons,Li1993}(shown in Figure\ref{fig:effect}) 
below is the summary of the water-splitting reaction :
\begin{equation}
2H_2O \rightarrow  2H_2 + O_2
\label{Eq:Vreax}
\end{equation}
It consists of a reduction reaction at the photo-cathode
\begin{equation}
2H^+ + 2e^- \rightarrow H_2	(0.00 V), 	
\label{Eq:Vreax}
\end{equation}
and the oxidation reaction at photo-anode 
\begin{equation}
H_2O + 4h^* \rightarrow O_2 + 4H^+ (1.229 V).
\end{equation} 
which produces hydrogen and oxygen, respectively.\\
ii) Transfer of exciton: After the formation of charge carriers, they must migrate to the surface of the material by diffusion and drifting phenomena\cite{pil2020charge}. 
a) Diffusion: Here, the charge carrier moves from the surface of higher density to the lower, which is given by Fick's law, for example, diffusion of the electron is given by
\begin{equation}
J_e = qD_e dn/dX
\end{equation}.
where $J_e$ is the current density of electron(current per area),q denotes the elementary charge, $D_e$ is the diffusion coefficient of electrons and $dn/dx$ is the density gradient in direction X. Similarly, diffusion of holes can be expressed as
\begin{equation}
J_h = qD_h dn/dx
\end{equation} 
where $D_h$ is the diffusion coefficient of holes.
b) Drift: The electric field governs the charge carriers' motion. When an electric field is present, holes feel a force directed toward the field, while electrons feel a force directed against the field. The current density induced by the electric field is given by
\begin{equation}
J = pq\mu_hE
\end{equation}
where J is the current density, p represents the density of electrons, q denotes the electronic charge, $\mu_h$ is the mobility constant, and E is the applied electric field. Electrons have greater mobility than holes because they depend on the effective masses of carriers. The electrons have less effective mass than the holes.
Moreover, when the recombination of electrons and holes occurs, they lower the overall reaction rate. It occurs when an electron combines with the holes in the valence band to release energy in the form of photons. However, electrons may become stuck between the forbidden gap or in lattice defects, i.e., the presence of acceptor or donor energy states when doped, which ultimately reduces the number of carriers available for the reaction.
iii) Redox reaction:
Finally, the charge carriers on the surface of the materials are ready to be used for the redox reaction because holes and electrons are the strong oxidizing and reducing agents, respectively. In the case of water splitting, water molecules are oxidized by holes to produce oxygen and electrons reduce the water molecules into hydrogen gases.
The generation of hydrogen gas takes place on the photo-cathode electrode by the following steps:\\
a) The adsorption of hydrogen atom($H^*$) on the surface of materials by an electron is known as the Volmer reaction\cite{enyo1973change}:
\begin{equation}
H_2O + e^- \rightarrow  H^* + OH^-,
\label{Eq:Vreax}
\end{equation}
b) Desorption of hydrogen gas will take place either by Tafel or Heyrovksy reactions.
In the Tafel reaction, two adjacent hydrogen atoms adsorbed on the side of the photocatalysts will combine to produce hydrogen molecules\cite{tilak1977overpotential}.
\begin{equation}
H^* + H^* \rightarrow H_2.
\label{Eq:Treax}
\end{equation}
In the Heyrovsky reaction, the electron will reduce water molecules. Simultaneously, they react with the adsorbed hydrogen atom to give $H_2$ as a product\cite{tilak1976overpotential}.
\begin{equation}
H^* + e^- +  H_2O \rightarrow H_2 + OH^-,
\label{Hreax}
\end{equation}
Thus, the evolution of hydrogen gas is achieved by using appropriate photocatalysts of a suitable band gap to harness solar energy effectively, which increases the reaction rate. The overall photocatalytic process of water splitting is shown in Figure\ref{fgr:band-region}.

\subsection{Criteria for choosing a photo-catalyst materials}
The decomposition of water has a positive Gibbs free energy and is an uphill reaction\cite{doi:10.1021/cr1002326} with $\Delta $G = 237 $kJmol^{-1}$. Hence, the least amount of energy\cite{rahman20162d}($E^{\circ}  = -\Delta G/nF$) required for the general splitting of the water is 1.23 eV, where $E^{\circ}$ is the electrode potential; n denotes the number of electrons per mole and F is the Faraday constant in $Cmol^{-1}$. The proton ($H^+/H_2$) has a reduction potential of (0--0.59 PH, V versus NHE, normal hydrogen electrode), and the water ($O_2/H_2O$) has an oxidation potential of (1.23--0.59 PH, V versus NHE) \cite{ABRAHAMS1985353}. Due to this, 0.41 V is required for hydrogen evolution and 0.82 V is required for oxygen evolution in pure water (PH = 7). Therefore, the semiconductor used as a photocatalyst must have a minimum band gap $E_g =$ of 1.23 eV to proceed with the splitting of water. However, during hydrogen production, an over potential is an unavoidable factor that alters the reaction; therefore, a photocatalysts having a band gap $E_g \ge$ 1.23 eV is preferable. In addition to band-gap requirements, photocatalysts must have a valence band (VB) at a potential more positive than the oxidation potential of the water molecules and a conduction band (CB) at a potential more negative than the reduction potential. 

\subsection{Characteristics of Photo-catalysts:}
\begin{figure}[h]
	\centering
\includegraphics[width=0.44\textwidth]{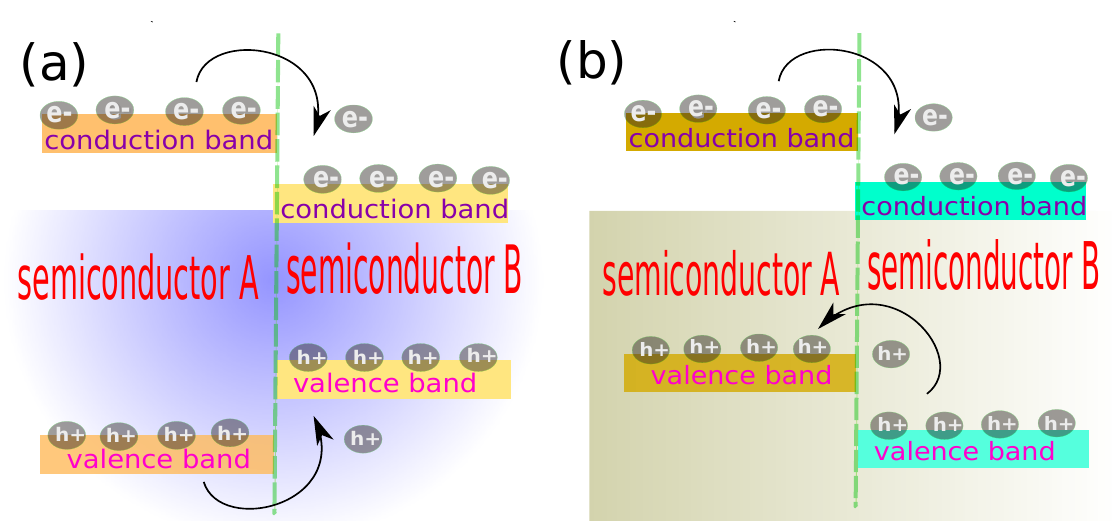}
	\caption{(a) Straddling band gap (Type-I)\cite{li2019photocatalyst} (b) Staggered band gap (Type-II)\cite{li2019photocatalyst}}
	\label{fgr:ht}
\end{figure}

\begin{figure}[h!]
	\centering
	\includegraphics[width=9.6cm, height=10cm]{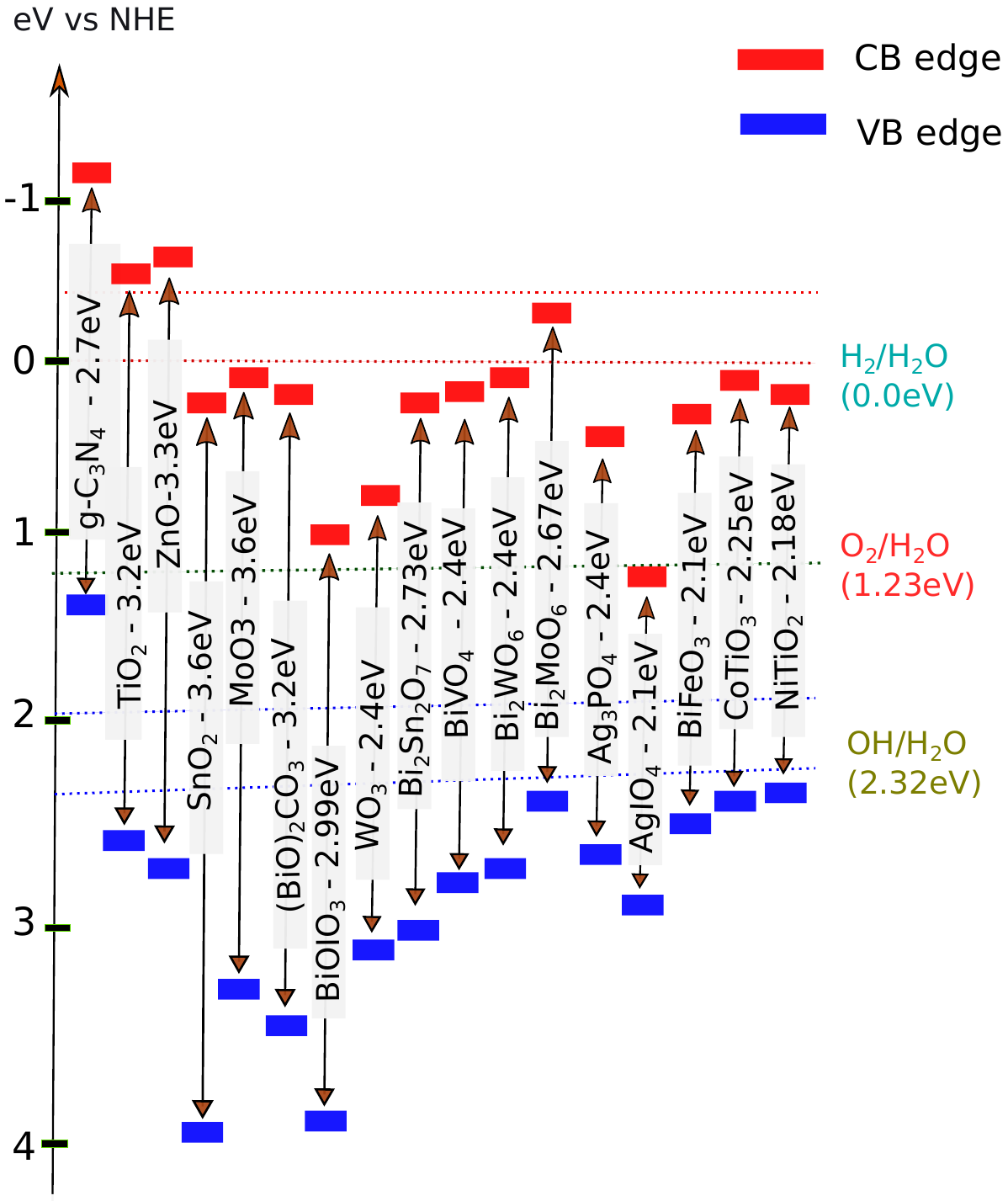}
	\caption{Band alignment of various materials to the hydrogen electrode and water redox energy levels.\cite{sivula2016semiconducting}}
	\label{fgr:bandr}
\end{figure}

During photo-catalysis whole process depends on the photocatalysts from the initialization of the reaction to the final product; therefore, when selecting materials it has to be thoroughly examined and it must satisfy the following requirements, i.e. a variety of light harvesting characteristics, including electrical band structure, lifetimes of excited states, charge transport, charge carrier mobility, chemical inertness, high stability, low cost, and ease of availability. Some of the few properties have been discussed below:
i) Absorption of light: The material needs to be a good harvester of light because it is an intrinsic property of certain materials to absorb light when irradiated by light of certain frequencies. In the case of semiconductors, light absorption is dependent on the material's band gap, which determines their ability to use different regions of the solar spectrum constituting ultraviolet light($<$380 nm)(3\%$-$5\%),
visible light (380- 780 nm)(42\%$-$43\%) and
near infrared light($>$780 nm)(52\%$-$55\%). The materials cultivating visible and near-infrared light have maximum output as compared to materials using ultraviolet light because only 5\% of the solar spectrum is made up of it e.g: materials like TiO$_2$\cite{wang2011hydrogen,pu2013nanostructure}, ZnO\cite{qiu2012secondary}, SnO$_2$\cite{huang2017influence} and WO$_3$\cite{qiu2012hierarchical} were largely used photocatalysts because of their low cost and high chemical stability even though they cannot utilize a wide range of solar spectrum resulting in poor absorption of light. Which can be enhanced by doping with suitable elements like N, O\cite{yan2016nitrogen} or with noble metals or creating oxygen defects\cite{doi:10.1021/am507641k} or forming a hybrid structure\cite{tian2013bi2wo6} or composites\cite{li2016fabrication} leads to better light harvesting ability and ultimately increases their photocatalytic hydrogen production.  
ii) Separation of charges: It is another key mechanism to consider during the photocatalytic process because the rate at which $e^-$ and $h^*$ migrate is $10^{-3 s}$ and $10^{-8 s}$ respectively before the recombination occurs which results in photo corrosion and the shift at the band edges leads to poor photocatalytic activity, so there needs to be better transport of carriers at the site of reactions due to this nanomaterial that is gaining popularity over bulk materials\cite{melman1977excited}. Another merit of using nanomaterials is that photoelectrons will be able to travel thousands of interatomic distances without scattering because of the wide surface area with a large number of carriers per area, thus enhancing the catalytic activity. Another way to prevent recombination is by forming heterojunction structures between different combinations of layers. The mostly formed heterojunctions are:-\\
a)Straddling/Type-I band gap: In which the edge of the conduction band of semiconductor A is higher(more -ve) than that of semiconductor B, whereas the valence band of semiconductor A is (more + ve) than that of semiconductor B; therefore, both the holes and electrons of semiconductor A are diffused in the conduction and valence band of semiconductor B, respectively, as shown in Figure\ref{fgr:ht}(a).
 \\
b)Staggered/Type-II band gap: It is the most used band alignment because it offers the best charge separation due to its relative position of the valence and conduction band, allowing the movement of electrons from the conduction band of semiconductor-A to that of semiconductor-B. However, holes are moved from the valence band of semiconductor-B to that of semiconductor-A as shown in Figure\ref{fgr:ht}(b). However, a special type of band alignment is possible in a staggered band gap, which is commonly called the Z-scheme, in which the holes of semiconductor-A combine with electrons of semiconductor-B at the interface; thus, the holes of semiconductor-B and electrons of semiconductor-A are available for the photocatalytic activity.
In 2019 Z.Zhou, X. Niu, Y.Zhang {\it et al.}\cite{Zhou2019}. Designed a direct Z-scheme hetero junction in a Janus MoSSe/WSeTe compound, showing that it is the best way for charge carrier separation, in which there will be a recombination of the interlayer electron-hole pair between MoSSe and WSeTe monolayers; thus, the carriers on intralayer got separated.
In 2023 T. Wahab, Y. Wang, A. Cammarata {\it et al}\cite{Wahab2023}. formed a hetero junction between $GeC-MX_2$(M= Mo and W; X=S and Se) compound which shows effective charge separation due to the formation of the electric field between the interface of Gec and M$X_2$ mono layers in both the cases only stacking configuration is different in the latter case both electrons and holes moves from higher to lower concentration, etc. Some of the heterostructures with their functional properties are given below in table\ref{table:1}:
\begin{table}[htp]
    \centering
    \caption{Some of the hetero-structures and their functional properties\cite{luo2016recent}.}
    \small %
\begin{tabular}{|p{3.5cm}|p{7.0cm}|p{6.50cm}|}
    \hline
	 \textbf{Heterostructures} & \textbf{Synthesis Methods} & \textbf{Unique properties}   \\
		\hline 
        \hline
MoS$_2$/Mo$_2$C\cite{zhang2020high} & Exfoliation method and a thermal treatment & improved mechanical robustness and good electrical contact \\
		\hline
Pt/MoS$_2$\cite{deng2015triggering} & One-pot chemical method& refine the adsorption behaviour of H atoms \\
		\hline
CoS$_2$/CuS\cite{li2019cus} & Hydrothermal electro-deposition & regulatory effect of heterostructure on electrons and energy bonds \\
		\hline
AU@MoS$_2$\cite{liu2019synthesis} & Colloidal synthesis approach &efficient electron transfer \\
		\hline
MoS$_2$/CNTs\cite{yang2023cnts} & Liquid-phase synthesis approach & high electron mobility \\
		\hline
Co/NCNT/g-C$_3$N$_4$\cite{yang2018co} & Sol-gel process & high surface area and large pore volume \\
		\hline
MoSe$_2$/r-GO/CNT\cite{park2017mose2} & Spray pyrolysis process & high catalytic efficient than binary materials \\
		\hline
CoP/NSGO\cite{lin2016graphene} & Thermal decomposition method & synergistic effect \\
		\hline
Ni$_2$P@CoP\cite{tang2017heterostructured} & Hydrothermal reaction, chemical bath deposition and phosphorization & electronic interaction \\
		\hline
MoS$_2$/Ni-Co LDH\cite{hu2017nanohybridization} & Two-step hydrothermal method &  faster decomposition of water molecules \\
		\hline
MoS$_2$/rGO\cite{ma2014mos} & Solvothermal method & abundant active catalytic edges \\
		\hline
MoP/MoS$_2$\cite{wu2019effective} & Hydrothermal method & $\delta$ $G_H$ becomes near zero \\
		\hline
MoS$_2$/MoSe$_2$\cite{kong2013synthesis} & Electron beam evaporation & high electrical conductivity \\
		\hline
Co-N-Ni$_9$S$_8$/Nb$_2$O$_5$\cite{chandrasekaran2022interface} &Two-step hydrothermal method & regulate local charge distribution and electronic properties \\
		\hline
Cu-N-Ni$_9$S$_8$/Nb$_2$O$_5$\cite{chandrasekaran2022interface} & Two-step hydrothermal method & modulate local charge distribution and electronic properties \\
		\hline
CoP/Co-MOF\cite{li2021building} & electro-deposition and hydrothermal treatment & durability \\
		\hline
Nitrogen-doped MoS$_2$/graphene\cite{tang20183d} & Template method & electro-catalytic performance \\
		\hline
Mo$_2$C/graphene\cite{geng2017direct} & Electrochemical vapor deposition & strong electronic coupling \\ 
		\hline
CuS/graphdiyne\cite{shi2019situ} & Hydrothermal method and in situ polymerization & tight interaction \\
		\hline
C$_3$N$_4$/N-doped graphene\cite{duan2015porous} & vacuum filtration method & porous structure and rich active sites \\
		\hline
C$_3$N$_4$/Bi$_2$O$_2$Se\cite{lin2024improving} & Solution phase self-assembly & Efficient charge separation effect \\
            \hline
Ni-Mo-S/C\cite{wang2018defect} & Hydrothermal method & stability \\
		\hline
CoP/CoMoP\cite{zhang2022heterostructured}& Dissolutioin regrowth and phosphorization & HER over a broad PH range \\
		\hline
MoS$_2$/rGO\cite{choi2021hierarchically} & One-pot hydrothermal method & large surface areas and many active sites \\
		\hline
MoS$_2$/nanoporous gold\cite{tan2014monolayer} & CVD method & low resistance ohmic contact \\
		\hline
MoS$_2$/Mo$_2$C\cite{luo2019morphology} & Hydrothermal method and CVD & spherical morphology is beneficial for the access of reactants and the release of H$_2$ \\
		\hline
Mo-W-P/Carbon cloth\cite{wang2016novel} & Hydrothermal method & strong synergistic effect \\
		\hline
MoS$_2$/Co$_3$O$_4$\cite{qin2020metal} & calcination, hydrothermal method & short diffusion path \\
		\hline
Crystalline/amorphous hollow MoS$_2$\cite{liu2022crystalline} & Solvothermal method & large active sites \\
		\hline
\end{tabular}
 \label{table:1}
\end{table}
iii) Catalysis: The material used as a catalyst must possess a favourable band edge concerning water reduction and oxidation potentials. The excited hole and electron are predetermined by the respective positions of the conduction and valence bands of the semiconductor and the redox levels of a substrate, by these properties of various metal oxides, semiconductor, etc. have been used as photocatalysts in the decomposition of water\cite{Lorente2011,Ameta2018} some of which is shown in the figure\ref{fgr:bandr}.
In general, the presence of transition metal cations with electronic configurations $d^{0}$
$(Ta^{5+}$, $Ti^{4+}$, $Zr^{4+}$, $Nb^{5+}$ and $W^{6+}$ or metal cations with electronic configurations $d^10$ ($(In^{3+}$, $Sn^{4+}$, $Ga^{3+}$, $Ge^{4+})$ and $Mo^{6+})$, etc. is considered as the main materials for efficient photocatalytic processes\cite{qiu2012hierarchical}.  
\section{Nanomaterials as a photocatalysts}  
Nanomaterials due to their large surface area and smaller size(5-100 nm) show excellent chemical, electronic, mechanical and physical properties compared to their bulk counterparts\cite{chiang2015enhanced}. The bright side of nanomaterials is that they can be shaped into various structures such as nanotubes, nanoparticles, nanowires and nanorods, which has shown optimal properties for their applications in photocatalysis\cite{sun2016preparation}, photoresponsiveness \cite{han2022wafer,han2022peel}, and photoluminescence \cite{lin2022direct}. They can be synthesized accordingly into different shapes and sizes based on the properties required using different methods\cite{ming2017removal}. Hence, they can be used in almost every sector such as manufacturing and modification of drugs, to solve environmental issues, industry, electronics, and energy production\cite{khan2019nanoparticles}. The photocatalysis process using nanomaterials has a wide range of applications and can be used for the degradation of environmental pollutants and in energy production. Some nanomaterials show excellent photocatalysis, such as semiconductors\cite{khataee2015photocatalysis,samu2017bandgap}, oxides\cite{andrade2017star,liu2011improvement}, metals\cite{liu2016loading,asapu2017silver}, and graphene\cite{aleksandrzak2017graphitic,zhang2017electrospun} with their enhanced optical properties\cite{li2017optimization,oliveira2015remediation}.  
Some of the recently explored 2D materials for photocatalytic applications are transition metal oxides (eg Zr$O_2$ and W$O_3$)\cite{fu2017two}, transitional metal dichalcogenides\cite{yang20232d}, h-BN\cite{ren20212d}, g-$C_3N_4$\cite{zhang2013enhanced,lin2024improving} and others. 

\subsection{2D materials}
In recent decades, 2D materials have received tremendous attention from researchers in the field of photocatalysis, and they are also considered a promising candidate due to their unique structural and physiochemical properties; moreover, they also provide room for improvement. In 2D materials, its light harnessing property can be tuned by reducing its thickness, and there will be lesser recombination of charge carriers due to shortened diffusion distance and the presence of surface atoms in high proportion available for redox reactions. Depending on their composition, 2D photocatalysts can be differentiated into three types: metal oxides, metal chalcogenides, and metal-free catalysts.
2D photocatalysts can be categorized into two types: layered and non-layered materials.
In layered materials, there exist strong chemical bonds between the atoms in the same layer, but different layers are held together by the weak van der Waals forces\cite{wang2019chemical} eg, transition metals dichalcogenides(TMDs)\cite{zhao2022recent,guo20192d}, hexagonal boron nitride(h-BN)\cite{MORSCHER20063280}, black phosphorus(BP)\cite{li2019black,deng2018progress}, metal-organic layers(MOLs)\cite{bhimanapati2015recent}etc. It can be synthesized from its respective bulk using two approaches, i.e. top-down and bottom-up approaches. Layered materials can be easily reduced into 2D materials by a top-down method using a liquid exfoliation process, in which a particular solvent is used having a surface tension similar to bulk materials. Jonathan N. Coleman {\it et al.\cite{shibata2011titanoniobate,alzakia2021liquid}} successfully exfoliated the bulk crystals of TMDs into 2D nanosheets using the liquid exfoliation method, also showed that using this procedure it is possible to form a hybrid film with enhanced properties.
Similarly, synthesis of 2D $g-C_3N_4$ using water as a solvent from its bulk\cite{yan2017preparation}. The key feature of the direct liquid exfoliation method is to select a desirable solvent, for which Hansen solubility parameters(HSP) theory is used and is given by
\begin{equation}
\begin{split}
R_a=[4(\delta_{D,solv}-\delta_{D,solu})^2+(\delta_{P,solv}-\delta_{P,solu})^2+ \\
(\delta_{H,solv}-\delta_{H,solu})^2]^{0.5}.
\label{Hreax}
\end{split}
\end{equation} 
Where, $\delta_D$ = dispersive of the materials, $\delta_P$ = polar of the materials, $\delta_H$ = hydrogen bonding solubility of the materials and $R_a$ = dispersion of the materials(should be smaller). In addition to the direct liquid exfoliation method several other processes are also used, such as thermal oxidation\cite{acik2013review}, chemical exfoliation\cite{xu2013chemical}, temperature-swing gas\cite{meng2012effect} etc.
On the other hand, non-layered materials like TiO$_2$, WO$_3$, CeO$_2$, Fe$_2$O$_3$ etc and some chalcogenides materials CdS, ZnS, CuS, etc. are formed by strong atomic bonds which crystallize into a 3D structure or bulk materials, due to their anisotropic growth along the 2D directions it has been difficult to exfoliate them. For such materials, bottom-up methods, such as self-assembly, 2D templates, and interface-mediated processes, are used for their exfoliation. Among them, the self-assembly process is a widely used bottom-up method for fabricating 2d materials. Using this process, 2D transition metal oxides such as TiO$_2$\cite{hao2005photocatalytic}, ZnO\cite{zhou2015self} have been synthesized. 

\subsection{Strategies used to improve photocatalytic performance of 2D materials}
Besides their excellent properties still, they are not efficient enough for practical usage or large-scale development. There is still room for improvement in the area of catalytic activity, durability, and selectivity because in 2D materials large surface area will not guarantee the presence of a large number of active sites. Catalytic activity is generally found to originate in the presence of an unsaturated atom on the edges\cite{sun2015atomically}. Therefore, to increase their productivity various methods or techniques have been proposed, for example: To increase their light harvesting properties, some materials are doped with metals or nonmetals\cite{wang2021controllable}, similarly heterostructure\cite{wang20222d} was formed between the layer of materials to prevent recombination, and other methods were used, such as surface engineering\cite{xiong2018surface}, creating vacancies\cite{ma2021vacancy} etc.  
\subsubsection{ Defect engineering:} There are various types of defects in materials induced by manufacturing processes, environmental factors, and intrinsic qualities of the materials. Defects can substantially influence the properties of optical, mechanical, electrical, and, in particular, catalytic processes. Because catalysis is a surface phenomenon, the presence of defects can greatly enhance its activity by introducing active sites for the adsorption of reactants following chemical reactions. So, by engineering the defects of the desired photocatalytic materials with improved efficiency, great stability can be formed by introducing defects into a catalyst or reforming the defect structures\cite{Luo2023}. e.g.: In nanocrystalline Si reforming by annealing at $(250-450)^\circ C$ resulted in better conductivity at the Si interface due to the quantum size effect\cite{Milovzorov2004}. In ZnO, it is seen that the increased photocatalysis resulting from oxygen defects created by solvothermal and thermal treatment processes resulted in the improvement of charge separation\cite{Zheng2007}. Similarly, the types of defects and their application to the few compounds are given in the table\ref{table:II}.

\begin{table}[htp]
	\centering
 \caption{Transition metal dichalcogenides and their properties\cite{marschall2014non}}
 	\begin{tabular}{|p{1.5cm}|p{3.5cm}|p{2.50cm}|p{3.5cm}|p{3.5cm}|}
		\hline
\textbf{TMDs} & \textbf{Defect types} & \textbf{Formation energy} & \textbf{Modulation} & \textbf{Applications} \\
		\hline
MoS$_2$\cite{Peto2018} & O substitution of S & air exposure & reduced Gibbs free energy & improved catalytic performance \\ 
		\hline
PdSe$_2$\cite{Li2022} & O substitution of Se & ozone treatment  & hole doping & improved catalytic performance  \\
		\hline
MoTe$_2$\cite{Cho2015}  & Te vacancies & laser irradiation & phase transition  & improved carrier mobility \\
		\hline
MoS$_2$\cite{Ding2019} & S vacancies & Solution-phase method & band gap reduction   & improved photodynamic efficiency \\ 
		\hline
MoS$_2$\cite{Zhou2020a} & S vacancies Co cluster & synthesis  & reduced Gibbs free energy & improved catalytic performance \\ 
		\hline
MoS$_2$\cite{Zhang2017} & -- & poly(4-styrene sulfonate) treatment & increased electron concentration & lateral homo junction \\ 
		\hline
WSe$_2$\cite{Lu2015}  & O substitution of Se & laser irradiation   & improved conductivity & photo conductivity improved by 150 times \\
		\hline
MoSSe\cite{liu2024quadruple} & Pt doped &- & increase in number of active sites& enhanced photocatalytic activity \\ 
	\end{tabular}
    \label{table:II}
\end{table}

\subsubsection{Doping:}
Since, photocatalysis is a surface phenomenon, its actions are affected by the particle arrangement and its size, surface properties, etc. These factors can be controlled by different preparation methods or by adding impurities. Doping is the best way to change the morphology of photocatalyst materials\cite{Kudo}. Titania has been widely studied from the beginning in the context of photocatalysis because of their versatile nature, low cost, non-toxicity, and outstanding chemical stability; however, due to the wide energy band gap, i.e. 3.0 eV for rutile and 3.2 eV for Anatase, it can only use 5\% of the solar energy. In addition, TiO$_2$ has a high rate of recombination resulting in unfavourable photocatalytic activity, therefore doping with metals or non-metals could be beneficial\cite{Colmenares2006}.
A. Kudo, R. Niishiro, A. Iwase, Akihide Kato, Hideki {\it et al.}. Experimentally showed that the shift in TiO band energy$_2$ to 2.85 eV and 2.00 eV after doping with molybdenum and chromium, respectively, occurs even at low concentration (0.1 at.\%) along with the reduced lifetime of TiO$_2$ $(89.3 \mu s)$ to 30 $\mu s$ when doped by $Cr^{3+}$ and 20 $\mu s$ by Mo$^{5+}$\cite{Wilke1999}. Doping of cations on Zns, TiO$_2$ and SrTiO$_3$ resulted in the shift of light absorption from ultraviolet to visible lights for H$_2$ evolution reaction from an aqueous solution containing sacrificial reagents\cite{Kudo2007}. Titania doped with cobalt ions(Co$^{2+}$) leads to phase transition from Anatase to Rutile and with ferric ions(Fe$^{3+}$) it behaves as doped semiconductor\cite{Bouras2007}.In the case of TiO$_2$ doped with non-metals(N, C, and S) leads to a reduced band gap by the addition of electrons on the valence band or introducing electronic states in between the band gap resulting in better absorption of photons energy and on the other hand effectively separating charges thus improving the photocatalytic process\cite{Yalcin2010}. J. Bloh, R. Dillert, D. Bahemann {\it et al.} prepared ZnO doped with transition metals (Mn, Fe, Co, Ni and Cu) by the solvothermal method containing 3\% of dopant sizes ranging from 200-400 nm. Cu-ZnO has been found to show much better photocatalytic performance than other dopants because it contains the highest photocurrent density$(10.6 \mu A/cm^2)$ which is eight times greater than pure ZnO$(1.2 \mu A/cm^2)$ followed by Mn-ZnO$(9.5 \mu A/cm^2)$, $Cu^{2+}$ ions also act as effective trappers of charge carriers, thus preventing recombination\cite{Bloh2012}. ZnO doped with transition metal oxide like Cu$_2$O, MnO$_2$, and CoO helps harness visible light for photoactivity by reducing the band gap energy\cite{Qi}.    

\subsection{Strain engineering:} It is a new way to tailor the band gap and electronic structures of the photocatalysts. In nanomaterials science, strain is a kind of deformation produced when a crystal is compressed, stretched, or sheared. Based on deformation, strain can be classified as elastic (reversible) or inelastic (irreversible). Strain engineering provides the precise adjustment of the band structure and carrier mobility of photocatalysts\cite{dai2019strain,peng2020strain}. Strain engineering has been used on a large scale to modify the properties of 2D materials because of their adjustable nature. The application of strain in 2D materials is mainly related to the electronic structure of their surfaces that provides the bond strength of the surface adsorbate, as well as various catalytic reactions\cite{zeng2022strain}. The main objective of strain engineering is to improve and optimize the properties of 2D materials through the application of compression or tensile strain.\\
$\gamma$-PN monolayer has a potential for water splitting under application of 10\% tensile deformation with an indirect band gap of 2.85 eV with VBM and CBM located at -6.82 eV and -3.98 eV, respectively, which is suitable for reduction and oxidation reaction under visible light\cite{tan2017novel}. 
 Wu{\it et al} uses first principles calculation along with AIMD simulation which shows the stability of $SnN_3$ monolayer with an indirect band gap of 1.965 eV showing great potential for water splitting. By application of external strain band gap can be changed from indirect to direct along with great carrier mobilities for electrons 376.19 $cm^2V^{-1}s^{-1}$ and holes 769.09 $cm^2V^{-1}s^{-1}$.
Being a good photocatalyst Janus WSSe monolayer with proper band edge position, high mobility of charge carriers, and resistance to photocorossion under visible light. Applying both tensile and compression strain can change the band gap from direct to indirect whereas external tensile strain can enhance the solar to hydrogen efficiency on the other hand uniaxial tensile strain can widen the optical absorption from visible to near-infrared spectrum of solar energy\cite{ju2020janus}. 
GeN$_3$ monolayer exhibits a band gap of 1.962 eV having a high electron and hole carrier mobility with anisotropy behaviour i.e. 1.55 x $10^4 cm^2V^{-1}s^{-1}$ and $1.6 x 10^3 cm^2V^{-1}s^{-1}$ along the armchair and zigzag direction, respectively. It also shows good solar to hydrogen efficiency of 12.63\% under visible regions and its band gap can be changed from indirect to direct by applying external strain and electric field\cite{liu2020gen3}.
Theoretically, it is found that $\beta$ -AuS has an appropriate band gap of 1.79 eV with the appropriate band edge position and high electron mobility of 1.9 x $10^4 cm^2V^{-1}s^{-1}$. Its solar to hydrogen efficiency is 17.21 \% almost equal to the conventional theoretical value 18\%. In addition, its optical and electronic properties can be tuned, applying 8\%. compressive strain can change it to a direct band gap along with an increase in the absorption capacity of visible light. Applying 4\%. compressive strain along different directions, that is, the chair and the zigzag direction resulted in high carrier mobility $2.9 x 10^5 cm^2V^{-1}s^{-1}$ and $1.9 x 10^6 cm^2V^{-1}s^{-1}$ respectively\cite{lv2021strain}.

\subsection{ Surface modification and interface engineering:} In a photo-catalysis, kinetics of photo electron and hole pair plays an important role in the conversion of solar energy to chemical energy from generation to transportation and consumption in the redox reaction\cite{C5CS00064E}. The effectiveness of the above steps determines the overall performance of a photocatalyst. To increase its effectiveness photo-catalytic materials have been upgraded from simple structures to composite  structures\cite{C2CS35355E,C5CC02704G}. The creation of hybrid photocatalysts resulted in the absorption of a wide range of solar energy along with better charge generation, effective transportation, and prevention of photocorrosion. Thus, enhancing the photocatalytic performance, which depends on the interaction between the surface and interface structures of a hybrid photo-catalysts\cite{https://doi.org/10.1002/cnma.201500069,bai2015surface}. The efficiency of both simple and hybrid semiconductors was affected by the surface of the catalyst where the redox reaction takes place. In hybrid semiconductors, another key factor is the interface between two adjacent semiconductors where the transfer of charge carriers takes place. The facets that form the interface and surface structure must be carefully selected because the photocatalytic performance may be affected by the exposed facets on the photocatalyst surface by the following processes:
1) Arrangements of surface atoms that determine the adsorption and activation of reactant molecules\cite{zhou2012catalysis,liu2011crystal}.
2) The electronic state of the surface depends on the surface facets that provide tunable redox abilities for catalytic reaction\cite{JMC}.
3) The properties of light harnessing, charge separation, and transfer of carriers depend on the orientation of a crystal leading to variation in charge density\cite{jiang2012synthesis,10.1246/cl.121148}. As compared to simple semiconductors hybrid systems are complicated due to the involvement of the interface of different components.

\section{Recent Advances in Photocatalytic Hydrogen Evolution Reaction (HER):}

Transition Metal Dichalcogenide: TMDs an interesting layered materials It is seen that in $MOS_2$ only the edge sites were active. So many approaches were introduced to increase the number of active sites. Several fabrications of nanostructures like nanowires, nanorods, and nanosheets were done to increase the surface area of the structures\cite{venkata2016atomically}. In a recent study, $MOS_2$ nanosheets were co-doped with Co and Ni and found to be stable structures that result in an enhancement in light absorption\cite{ma2016ultrathin}. Similarly, Z-scheme photocatalysts $MoSe_2$/graphene/$HfS_2$ and $MoSe_2$/N-doped graphene/$HfS_2$ composites in which graphene acts as a redox mediator, doping graphene with N-type resulted in a widening of the light absorption ability of the compounds\cite{fu2016two}. Peng {\it et al.} studied the 2D InSe, InTe, and GaTe monolayers exhibiting fast migration and efficient exciton separation due to their high electron mobility and small exciton binding energy resulting in better energy conversion efficiencies, ie 1. 20\%, 6. 40\% and 3. 22\% respectively\cite{peng2017computational}. Similarly, the 2D PdPS monolayer was found to be an excellent photocatalyst because it contains a large number of active sites with almost zero Gibbs free energy at the Pd site\cite{jiao2018ab}. In 2D $MoS_2/WS_2$ heterostructure becomes active for HER because of their synergistic effect with significantly lower Gibbs free energy of hydrogen adsorption at 0.20 - 0.24 eV\cite{obligacion20202d}. Guo {\it et al.}introduces a new method to improve the hydrogen production by forming a biphase of wood/CoO and wood/CuS-$MoS_2$ yields 220.74 $\mu$mol$h^{-1}cm^{-2}$ and 3271.49 $\mu$$h^{-1}cm^{-2}$ respectively\cite{guo2021boosting}. \\

MXenes and Novel Composites: In 2011 a new family of 2D NMs was reported that consists of transitional metal carbides, nitrides or carbonitrides represented by the formula$M_(n+1)X_nT_x(n= 1,2,3)$ where M is the transitional metals(M = Mo or Ti), X is (= C/N or CN) and ($T_x$ = -F, -O and -OH) represents functional groups\cite{naguib2023two}.$Ti_3C_2T_x$ is commonly used MXene as a co-catalyst in photocatalytic hydrogen generation\cite{huang2020photocatalytic}. These layered materials can be derived from their parent materials like in Ti$_3$C$_2$ can be synthesized by etching Al layers from different MAX phases(Ti$_2$AlC, V$_2$AlC, NB$_2$AlC), etc. Using aqueous Hydrofluoric Acid at room temperature\cite{naguib201425th}. These materials show good conductivity\cite{kamysbayev2020covalent}, thermodynamic stability\cite{pandey2017two}, and surface termination that facilitates the anchoring of other molecules or nanoparticles\cite{malchik2021mxene}. 
Ruan {\it et al.}\cite{ruan2023enhancing} study by forming the heterogeneous ternary photocatalyst a-$TiO_2/H-TiO_2/Ti_3C_2$ Mxene which is prepared by an electrostatic self-assembly method that produces hydrogen at a rate of 0.387 mmol$h^{-1}$. Using homo-interface engineering, it is seen that light absorption with carrier transportation has been improved.
Ramirez {\it et al\cite{ramirez2023green}.}recently synthesized the $Ti_3C_2$ mxene quantum dots (QD) by using 532 nm laser ablation in liquid suspension resulting in increased optical absorption in the UV region with hydrogen production of $2.02 mmolg^{-1}h^{-1}$.
GU {\it et al.}\cite{gu2023robust} developed a binary heterojunction consisting of CdSe nanorods and $Ti_3C_2$ Mxene by a one-step in situ hydrothermal method. This compound produces hydrogen at a rate of 763.2 $\mu molg^{-1}h^{-1}$, which is 6 times higher than the pristine CdSe which shows a synergistic effect Similarly, a ternary CdS, Mo$S_2$ and $Ti_3C_2$ MXene synthesized by in situ growth method shoes this effect, which reduces photocorossion and produces hydrogen at the rate of 14.88 $mmolh^{-1}g^{-1}$ having a useful life of up to 78 h\cite{wu2023insight}. Another interesting hybrid recently synthesized by Liu {\it et al.} that consists of a series of 2,4-bis[4-(N, N-dibutyl amino)phenyl] squaraine (SQ) derivatives containing different numbers of hydroxyl groups were hybridized with $Ti_3C_2T_X$ MXene nanosheets forming an organic-inorganic hybrid photocatalyst. The best hydrogen production rate is found to be 28.6 $\mu molh^{-1}g^{-1}$. The excellent photocatalytic ability achieved due to the combined effect of the hybrid such as SQ dye is a good light harvester, large active sites of $Ti_3C_2T_X$ and well separation and the mobility of charge carriers due to formation of a heterojunction between SQ and $Ti_3C_2T_X$\cite{wu2023insight}.

Organic Semiconductor: Organic photocatalysts are interesting materials because they provide numerous ways to tune their electronic and structural properties; however, most organic semiconductors work only under ultraviolet light, which is the major drawback. The extensively studied organic photocatalyst $g-C_3N_4$ for the hydrogen evolution reaction was first reported by Antonietti et al.\cite{wanasinghe2024motional} in 2009, its hydrogen production rate was very low, so various methods were used to increase its productivity, such as co-catalysts\cite{kaur2014organic,axelsson2024role}, co-polymerization\cite{moi2023band}, microstructure\cite{hai2023charge} and liquid-assisted approaches\cite{zhang2023optical}. Ultrathin materials $g-C_3N_4$ have been synthesized using thermal polycondensation of the hydrogen bonding network (UDF), resulting in improved light harvesting properties with a large surface area. Ultrathin nanosheets produce hydrogen at a rate of 2.5 X $10^{-4} molh^{-1}$ using visible light\cite{zhang2020ultrathin}.
Modification of $g-C_3N_4$ with Mo$S_2$ layers of 39 $\pm$ 5 nm lateral size with 20\% wt load which generates interfacial contacts with a large number of Mo$S_2$ edge sites and efficient electronic transport phenomena resulted in a high hydrogen yield at a rate of 1497 $\mu molh^{-1}g^{-1}$\cite{koutsouroubi2020interface}. Recently, $CoS-Co(OH)_2$ with polypyrrole (PPY) was designed and embedded in the $g-C_3N_4$ photocatalyst, resulting in improved transport and better carrier separation. The hydrogen production rates of $g-C_3N_4-PPY$, ternary$g-C_3N_4-PPY-CoS$ and Quaternary $g-C_3N_4-PPY-CoS-Co(OH)_2$ are 2.49$\mu molh^{-1}$, 23.27 $\mu molh^{-1}$ and 46.67 $\mu molh^{-1}$, respectively\cite{li2020rational}.
Bai {\it et al.}\cite{bai2020carboxyl} has done the surface modification of $g-C_3N_4$ by adding a fictionalized carboxyl group synthesized by the post-treatment grafting method, resulting in better carrier separation and an improved charge carrier to produce hydrogen 52 times greater than pristine hydrogen and have a large apparent quantum yield of 15.7\% at 420$\pm$ 15 nm.
For the first time, Samanta {\it et al.}\cite{samanta2021solar} synthesized the heterostructure of graphitic carbon nitride (g-CN) / N, S co-doped graphene quantum dots (NSGQDS) by a one-pot pyrolysis process. It shows enhanced charge mobility with low recombination, which gives hydrogen at a rate of 5.24 $mmolh^{-1}g^{-1}$ using sunlight. Another way of forming heterostructures using in situ hydrolysis / calcium with 5\% wt loading of $Nb_2O_5$ resulted in a hydrogen evolution rate of 2.07 $\pm$ 0.03 and 6.77$\pm$0.12 mmol$g^{-1}h^{-1}$ using visible and simulated solar light exposure, respectively, between $g-C_3N_4$ and $Nb_2O_5$ that resulted in the type II heterojunction, which provides a large surface area for the interface between the photocatalyst and water\cite{dong2022situ}. Recently, Chang {\it et al.}\cite{CHANG20236729} designed $g-C_3N_4$ adapted with cyano groups and $K^+$ ions prepared using potassium thioacetate as an electron donor and dehydrogenation agent. The K(0.005)-CN's photocatalyst yields hydrogen at a rate of 1319$\mu molh^{-1}g^{-1}$, due to the presence of electron-withdrawing cyano groups thus enhancing the optical absorption property. Ullah {\it et al.}\cite{ullah2024bimetallic}Studied the rate of hydrogen production by forming a plasmonic heterojunction between $g-C-3N_4$ and bimetallic nitride(NiMoN) gives 75.26 $\mu molh^{-1}$ under visible light (420 nm) irradiation resulted into 5.4 times higher than the pristine $g-C_3N_4$. 

Metal-Organic Framework, Covalent-organic Framework, and their Composites: It is an interesting class of materials that are being studied in the field of photocatalytic and energy storage. Recently, they are gaining a lot of attraction due to their porous structure, large surface area, and their ability to tailor metal clusters or an organic linker\cite{chen2017metal}.
Gong {\it et al.}\cite{gong20222d} reported the S-scheme heterojunction constructed to modify the Ni-based metal-organic framework (Ni-MOF), using different in situ treatments like surface sulfonation, oxidation, and phosphatizing of $NiS_2$, NiO, and NiP respectively. They can act as electron trap centers, thus improving the separation of carriers. The highest hydrogen production of the hybrid is 6.337 mmol$g^{-1}h^{-1}$ with 14.18 times the untreated Ni-MOF. Han {\it et al.}\cite{han2022integrating} studied the metal covalent frameworks (MCOFS) namely RuCOF-ETTA, RuCOF-TPB and RuCOF-ETTBA incorporating $Ru^{II}$ photosensitive tris(2,2'-bipyridine). It is found that each RuCOF has three isostructural covalent organic frameworks interlocking together with $Ru^{II}$ at the center. They show an excellent hydrogen production rate of 20308 $\mu molg^{-1}h^{-1}$. This is because RuCOFs exhibit great stability and better light-harvesting ability. Another method to increase hydrogen production by increasing the excitonic dissociation in COFs Ni-intercalated fluorenone-based COFs (Ni-COF-SCAU-1) at the carbon-nitrogen double bond(C=N) for the first time resulted in enhanced electric field polarisation thus providing mobility of electron and the separation of carriers to take part in the surface hydrogen evolution reaction. Due to this linkage, Ni-COF-SCAU-1 produces hydrogen at the rate of 197.46$mmolg^{-1}h^{-1}$ under visible light irradiation with AQE upto 43.2\% at 420 nm\cite{shen2023efficient}. Recently Li {\it et al.}\cite{li2024hybridization} synthesized the Z-scheme single atom between COF and metal-organic ring through the supramolecular interactions of coral-like COF (S-COF) and photosensitized $Pd_2L_2$ type metal-organic ring (MAC-FA1). The MAC-FA1/S-COF heterojunction resulted in better charge separation, improved light absorption, sluggish recombination, and dispersed Pd active sites, increasing the hydrogen production reaction. The 4\% MAC-FA1/S-COF produces hydrogen at the rate of 100 $mmolg^{-1}h^{-1}$ within 5h. 

Perovskites: It is generally represented by the formula $ABX_3$, where A and B are monovalent and divalent cations, and X is a halide or oxide\cite{porwal2022investigation,amano2022improvement}. It is referred to as organic-inorganic if A cation is an organic molecule and others are referred to as inorganic perovskites\cite{dixit2022theoretical}. Because of their excellent band gaps both inorganic and hybrid materials were intensely investigated for photocatalytic applications. Peng {\it et al.}\cite{peng2017first} enhances the light-harvesting ability of $Sr_2Ta_2O_7$ into the visible region from the ultraviolet region of sunlight by doping with anionic (S) and cationic (V/Nb) resulting in shifting of VBM upward by 1.14 eV and CBM downward by 0.68 eV when doped with (S, Nb) and for (S, Nb) VBM shifted 0.74 eV and CBM by 0.63 eV. Thus, a feasible way to increase its optical absorption ability. Another study done by Xu {\it et al.}\cite{irani2020nature} in which $BiVO_4$  is incorporated by Nitrogen reduces the band gap but also introduces a new energy state ($\ne 0.1 eV$) below CBM. Thus, forming the electron traps or recombination centers ultimately lowers hydrogen production. Recently, a bifunctional double perovskite $Sr_2CoWO_6$ was developed and produces oxygen and hydrogen at the rate of 30 $\mu molh^{1-}g^{-1}$(AQE of 3\% at 420nm) and 30 $\mu molh^{-1}g^{-1}$, respectively, when Pt and Rh were used as cocatalysts.\cite{idris2020novel}. Similarly, $Sr_2CoTaO_6$ is the first reported visible light bifunctional double perovskite. The production rate of hydrogen by $Sr_2CoTaO_6$ can be further enhanced by using Ru$O_2$ Rh as a co-catalyst\cite {idris2020sr2cotao6}.  
It is found that the charge transfer between 2D perovskites and co-catalyst Pt increases with the decreasing length of organic cations in perovskites. The Phenylmethylammonium lead iodide produces the $H_2$ at the rate of 333 $\mu molh^{-1}$ which is greater than the 3D perovskites\cite{wang2021mechanistic}.  
On the other hand, Halide perovskites are excellent photocatalysts for HER but their instability in aqueous solution impacts their applications. Ji {\it et al.}\cite{miodynska2023bi} they used ethanol splitting to avoid the ionization of perovskites, $Cs_3Bi_2I_9$gives the maximum production of hydrogen i.e, 2157.8 $\mu molh^{-1}g^{-1}$ out of Bi-based ($Cs_3Bi_2X_9$ PNs; X = I, Br, Cl) halide perovskites nanosheets. As the number of halogen atoms increases, it helps to reduce the Bi-Bi distance, which helps to transfer and separate excitons. Metal-halide perovskites (MHPs) have excellent optoelectronic properties as a promising candidate for photocatalytic hydrogen evolution reactions. However, their instability and insufficient active site exposure affect their catalytic efficiency. Here, they formed a structure confining $CsPbBr_3$ in the zeolite ZEO-1 (JZO). The prepared $CsPbBr_3@ZEO$-1 increases the hydrogen evolution rate to 1734 $\mu molh^{-1}g^{-1}$ under irradiation of visible light as compared to bulk $CsPbBr_3$ i.e, 11 $\mu molh^{-1}g^{-1}$. Even more hydrogen was produced (4826 $\mu h^{-1}g^{-1}$) by incorporating Pt into nanocrystals $CsPbBr_3$ due to improved charge mobility and separation\cite{gao2024confinement}.

\section{Challenges and Future Directions}
A large number of advances in technologies on how to maximize the light-harnessing property of materials to meet our energy needs. There are various ways to harness solar energy and photocatalysis is one of them. In 2D, nanomaterials and their composites show improved hydrogen production compared to traditional materials. The formation of a heterojunction led to better charge separation, and the use of co-catalyst charge mobility was improved. The visible light absorption was also enhanced by forming a 2D/2D composite heterostructure also resulted in the stability of the material. These novel heterojunction/heterostructures are promising candidates for future use. After the significant progress made in the past decades, however, there still exists a problem to overcome in the field of photocatalytic splitting of water. Besides the synthesis of materials and the challenges, hydrogen storage is another problem facing right now. The idea of combining the hydrogen production unit and the storage into a single entity can serve the realistic goals of using hydrogen energy\cite{liao2012hydrogen}. As reported, Song and his colleagues synthesized the 1D core-shell structure using carbon nitride nanowire and carbon nanotube (CNNW/CNT) as a multifunctional system with production and storage of hydrogen. Here photoelectrons and holes are distributed on the active sites of CNNWs and CNTs respectively. The protons penetrate the CNTs, thus inducing the HER in the CNNWS. Thus produced hydrogen molecules are separated naturally and stored in the nanotube\cite{song2022photocatalytic}. The formation of excitons and their separation are the most critical steps for efficient hydrogen production, that is, protons generated at oxidative sites are needed to travel to the reductive sites of the photocatalyst for hydrogen production\cite{yang2017combining}. As a result, this might lead to the contamination of the hydrogen with oxygen. So future research should also focus on better electron transportation to the oxidative sites. As reported, a multilayered structure of carbon nitride and graphene sheets shows a safer way to generate hydrogen. Hydrogen is generated between graphene sheets, as selective absorption does not allow the passage of other molecules/atoms\cite{zhang2015porous}.
Recently, most of the research has been based on 2D materials like MXenes, COFs, MOFs, perovskites, and graphitic nitride, promising photocatalysts that have the potential even to increase their productivity using various methods like doping, forming heterojunctions, or forming hybrid structures or maybe by implying techniques like surface and interface engineering, intercalation of materials, thus enhancing the rate of hydrogen production to reach the goal of commercialization and using clean fuels for a better future. 

\section{Conclusion}
Over the past few decades, scientific communities have conducted many studies in the field of energy due to the shortage of fossil fuels and the increase in energy demands. Hydrogen has the highest energy density as compared to fossil fuels and is an alternative source of energy. Photo-catalysis of water is considered as one of the cleaner, safer, and environmentally benign methods to produce hydrogen. For hydrogen production, various materials have been synthesized, fabricated, and developed. The main aim was to find or synthesize materials capable of using visible light with the appropriate band gaps suitable for the water-splitting reaction. Recently, the main focus of the study has been shifted toward the NMs and their composites. Different strategies have been introduced for the enhancement of its properties, such as charge separation, light harvesting ability, carrier mobility, preventing recombination, reducing the band gap, and introducing defects or creating vacancies for higher hydrogen production. In this Review, we planned to provide a comprehensive discussion of recent advances made in the photocatalytic hydrogen evolution reaction by 2D NMs. In addition to this, basic knowledge of water splitting is detailed and the need for manipulation of electronic and band gaps is also highlighted. There are still a lot of gaps that need to be filled, and the utilization of this technology on a large scale is one of the major challenges for the future. 

\begin{acknowledgement}

DPR acknowledges Anusandhan National Research Foundation (ANRF), Govt. of India  via Sanction Order No.:CRG/2023/000310, \& dated:10 October, 2024.
%\item The research is partially funded by the Ministry of Science and Higher Education of the Russian Federation as part of World-class Research Center program: Advanced Digital Technologies (contract No. 075-15-2022-312 dated 20.04.2022).
\\ A.  Laref acknowledges support from the "Research Center of  the  Female Scientific and Medical Colleges",  Deanship of Scientific Research, King Saud University.

\end{acknowledgement}

\section*{Author contributions}
All authors contributed equally.
%\textbf{Laku Dorjee Tamang:} Formal analysis, Visualization, Validation, Literature review, Writing-original draft, writing-review \& editing.\\
%\textbf{Zosiamliana Renthlei:} Formal analysis, Visualization, Validation, writing-review \& editing. \\
%\textbf{Shivraj Gurung:} Formal analysis, Visualization, Validation, writing-review \& editing. \\ 
%\textbf{Celestine Lalengmawia:} Formal analysis, Visualization, Validation, writing-review \& editing. \\
%\textbf{Bhanu Chettri:} Formal analysis, Visualization, Validation, writing-review \& editing. \\ 
%\textbf{Jitendra Pal Singh}:Formal analysis, Visualization, Validation, writing-review \& editing. \\		
%\textbf{Amel Laref}:Formal analysis, Visualization, Validation, writing-review \& editing. \\	
%\textbf{Mukhriddin E. Tursunov:}Formal analysis, Visualization, Validation, writing-review \& editing. \\ 
%\textbf{Avazbek T. Dekhkonov:}Formal analysis, Visualization, Validation, writing-review \& editing. \\ 
%\textbf{Dibya Prakash Rai:} Project management, Supervision, Resources, software, Formal analysis, Visualization, Validation, writing-review \& editing. 

\begin{suppinfo}

All the information are included in the main text.

\end{suppinfo}

\bibliography{acs-achemso}

\providecommand{\latin}[1]{#1}
\makeatletter
\providecommand{\doi}
  {\begingroup\let\do\@makeother\dospecials
  \catcode`\{=1 \catcode`\}=2 \doi@aux}
\providecommand{\doi@aux}[1]{\endgroup\texttt{#1}}
\makeatother
\providecommand*\mcitethebibliography{\thebibliography}
\csname @ifundefined\endcsname{endmcitethebibliography}
  {\let\endmcitethebibliography\endthebibliography}{}
\begin{mcitethebibliography}{234}
\providecommand*\natexlab[1]{#1}
\providecommand*\mciteSetBstSublistMode[1]{}
\providecommand*\mciteSetBstMaxWidthForm[2]{}
\providecommand*\mciteBstWouldAddEndPuncttrue
  {\def\EndOfBibitem{\unskip.}}
\providecommand*\mciteBstWouldAddEndPunctfalse
  {\let\EndOfBibitem\relax}
\providecommand*\mciteSetBstMidEndSepPunct[3]{}
\providecommand*\mciteSetBstSublistLabelBeginEnd[3]{}
\providecommand*\EndOfBibitem{}
\mciteSetBstSublistMode{f}
\mciteSetBstMaxWidthForm{subitem}{(\alph{mcitesubitemcount})}
\mciteSetBstSublistLabelBeginEnd
  {\mcitemaxwidthsubitemform\space}
  {\relax}
  {\relax}

\bibitem[Wang and Azam(2024)Wang, and Azam]{wang2024natural}
Wang,~J.; Azam,~W. Natural resource scarcity, fossil fuel energy consumption,
  and total greenhouse gas emissions in top emitting countries.
  \emph{Geoscience Frontiers} \textbf{2024}, \emph{15}, 101757\relax
\mciteBstWouldAddEndPuncttrue
\mciteSetBstMidEndSepPunct{\mcitedefaultmidpunct}
{\mcitedefaultendpunct}{\mcitedefaultseppunct}\relax
\EndOfBibitem
\bibitem[Keeling(1973)]{keeling1973industrial}
Keeling,~C.~D. Industrial production of carbon dioxide from fossil fuels and
  limestone. \emph{Tellus} \textbf{1973}, \emph{25}, 174--198\relax
\mciteBstWouldAddEndPuncttrue
\mciteSetBstMidEndSepPunct{\mcitedefaultmidpunct}
{\mcitedefaultendpunct}{\mcitedefaultseppunct}\relax
\EndOfBibitem
\bibitem[Suzuki(1975)]{suzuki1975extension}
Suzuki,~A. An extension of the H{\"a}fele-Manne model for assessing strategies
  for a transition from fossil fuel to nuclear and solar alternatives.
  \textbf{1975}, \relax
\mciteBstWouldAddEndPunctfalse
\mciteSetBstMidEndSepPunct{\mcitedefaultmidpunct}
{}{\mcitedefaultseppunct}\relax
\EndOfBibitem
\bibitem[West(2014)]{west2014henry}
West,~J.~B. Henry Cavendish (1731--1810): hydrogen, carbon dioxide, water, and
  weighing the world. \emph{American Journal of Physiology-Lung Cellular and
  Molecular Physiology} \textbf{2014}, \emph{307}, L1--L6\relax
\mciteBstWouldAddEndPuncttrue
\mciteSetBstMidEndSepPunct{\mcitedefaultmidpunct}
{\mcitedefaultendpunct}{\mcitedefaultseppunct}\relax
\EndOfBibitem
\bibitem[Lavoisier(2001)]{lavoisier2001antoine}
Lavoisier,~A. Antoine Lavoisier. 2001\relax
\mciteBstWouldAddEndPuncttrue
\mciteSetBstMidEndSepPunct{\mcitedefaultmidpunct}
{\mcitedefaultendpunct}{\mcitedefaultseppunct}\relax
\EndOfBibitem
\bibitem[Bockris(2013)]{BOCKRIS20132579}
Bockris,~J.~O. The hydrogen economy: Its history. \emph{International Journal
  of Hydrogen Energy} \textbf{2013}, \emph{38}, 2579--2588\relax
\mciteBstWouldAddEndPuncttrue
\mciteSetBstMidEndSepPunct{\mcitedefaultmidpunct}
{\mcitedefaultendpunct}{\mcitedefaultseppunct}\relax
\EndOfBibitem
\bibitem[Weeks(1932)]{weeks1932discovery}
Weeks,~M.~E. The discovery of the elements. IV. Three important gases.
  \emph{Journal of Chemical Education} \textbf{1932}, \emph{9}, 215\relax
\mciteBstWouldAddEndPuncttrue
\mciteSetBstMidEndSepPunct{\mcitedefaultmidpunct}
{\mcitedefaultendpunct}{\mcitedefaultseppunct}\relax
\EndOfBibitem
\bibitem[Chalk and Miller(2006)Chalk, and Miller]{Chalk2006}
Chalk,~S.~G.; Miller,~J.~F. {Key challenges and recent progress in batteries,
  fuel cells, and hydrogen storage for clean energy systems}. \emph{J. Power
  Sources} \textbf{2006}, \emph{159}, 73--80\relax
\mciteBstWouldAddEndPuncttrue
\mciteSetBstMidEndSepPunct{\mcitedefaultmidpunct}
{\mcitedefaultendpunct}{\mcitedefaultseppunct}\relax
\EndOfBibitem
\bibitem[Omer(2008)]{Omer2008}
Omer,~A.~M. {Energy, environment and sustainable development}. 2008\relax
\mciteBstWouldAddEndPuncttrue
\mciteSetBstMidEndSepPunct{\mcitedefaultmidpunct}
{\mcitedefaultendpunct}{\mcitedefaultseppunct}\relax
\EndOfBibitem
\bibitem[Furukawa and Yaghi(2009)Furukawa, and Yaghi]{Furukawa2009}
Furukawa,~H.; Yaghi,~O.~M. {Storage of hydrogen, methane, and carbon dioxide in
  highly porous covalent organic frameworks for clean energy applications}.
  \emph{J. Am. Chem. Soc.} \textbf{2009}, \emph{131}, 8875--8883\relax
\mciteBstWouldAddEndPuncttrue
\mciteSetBstMidEndSepPunct{\mcitedefaultmidpunct}
{\mcitedefaultendpunct}{\mcitedefaultseppunct}\relax
\EndOfBibitem
\bibitem[Li and Somorjai(2010)Li, and Somorjai]{Li2010}
Li,~Y.; Somorjai,~G.~A. {Nanoscale advances in catalysis and energy
  applications}. \emph{Nano Lett.} \textbf{2010}, \emph{10}, 2289--2295\relax
\mciteBstWouldAddEndPuncttrue
\mciteSetBstMidEndSepPunct{\mcitedefaultmidpunct}
{\mcitedefaultendpunct}{\mcitedefaultseppunct}\relax
\EndOfBibitem
\bibitem[Peschka(2012)]{Peschka2012}
Peschka,~W. \emph{{Liquid hydrogen: fuel of the future}}; 2012\relax
\mciteBstWouldAddEndPuncttrue
\mciteSetBstMidEndSepPunct{\mcitedefaultmidpunct}
{\mcitedefaultendpunct}{\mcitedefaultseppunct}\relax
\EndOfBibitem
\bibitem[Nikolaidis and Poullikkas(2017)Nikolaidis, and
  Poullikkas]{nikolaidis2017comparative}
Nikolaidis,~P.; Poullikkas,~A. A comparative overview of hydrogen production
  processes. \emph{Renewable and sustainable energy reviews} \textbf{2017},
  \emph{67}, 597--611\relax
\mciteBstWouldAddEndPuncttrue
\mciteSetBstMidEndSepPunct{\mcitedefaultmidpunct}
{\mcitedefaultendpunct}{\mcitedefaultseppunct}\relax
\EndOfBibitem
\bibitem[Rostrup-Nielsen and Rostrup-Nielsen(2002)Rostrup-Nielsen, and
  Rostrup-Nielsen]{Rostrup-Nielsen2002}
Rostrup-Nielsen,~J.~R.; Rostrup-Nielsen,~T. {Large-scale hydrogen production}.
  \emph{CATTECH} \textbf{2002}, \emph{6}, 150--159\relax
\mciteBstWouldAddEndPuncttrue
\mciteSetBstMidEndSepPunct{\mcitedefaultmidpunct}
{\mcitedefaultendpunct}{\mcitedefaultseppunct}\relax
\EndOfBibitem
\bibitem[Kwasi-Effah \latin{et~al.}(2015)Kwasi-Effah, Obanor, and
  Aisien]{kwasi2015review}
Kwasi-Effah,~C.; Obanor,~A.; Aisien,~F. A review on electrolytic method of
  hydrogen production from water. \emph{American Journal of Renewable and
  Sustainable Energy} \textbf{2015}, \emph{1}, 51--57\relax
\mciteBstWouldAddEndPuncttrue
\mciteSetBstMidEndSepPunct{\mcitedefaultmidpunct}
{\mcitedefaultendpunct}{\mcitedefaultseppunct}\relax
\EndOfBibitem
\bibitem[Chiesa \latin{et~al.}(2005)Chiesa, Consonni, Kreutz, and {Robert
  Williams}]{Chiesa2005}
Chiesa,~P.; Consonni,~S.; Kreutz,~T.; {Robert Williams} {Co-production of
  hydrogen, electricity and CO2 from coal with commercially ready technology.
  Part A: Performance and emissions}. \emph{Int. J. Hydrogen Energy}
  \textbf{2005}, \emph{30}, 747--767\relax
\mciteBstWouldAddEndPuncttrue
\mciteSetBstMidEndSepPunct{\mcitedefaultmidpunct}
{\mcitedefaultendpunct}{\mcitedefaultseppunct}\relax
\EndOfBibitem
\bibitem[{Office of Energy Efficiency {\&}
  Renewable}()]{OfficeofEnergyEfficiency&Renewable}
{Office of Energy Efficiency {\&} Renewable} {Hydrogen Production:
  Photobiological | Department of Energy}.
  \url{https://www.energy.gov/eere/fuelcells/hydrogen-production-photobiological}\relax
\mciteBstWouldAddEndPuncttrue
\mciteSetBstMidEndSepPunct{\mcitedefaultmidpunct}
{\mcitedefaultendpunct}{\mcitedefaultseppunct}\relax
\EndOfBibitem
\bibitem[Otsuka \latin{et~al.}(2003)Otsuka, Kaburagi, Yamada, and
  Takenaka]{Otsuka2003a}
Otsuka,~K.; Kaburagi,~T.; Yamada,~C.; Takenaka,~S. {Chemical storage of
  hydrogen by modified iron oxides}. \emph{J. Power Sources} \textbf{2003},
  \emph{122}, 111--121\relax
\mciteBstWouldAddEndPuncttrue
\mciteSetBstMidEndSepPunct{\mcitedefaultmidpunct}
{\mcitedefaultendpunct}{\mcitedefaultseppunct}\relax
\EndOfBibitem
\bibitem[Kodama \latin{et~al.}(2002)Kodama, Shimizu, Satoh, Nakata, and
  Shimizu]{Kodama2002}
Kodama,~T.; Shimizu,~T.; Satoh,~T.; Nakata,~M.; Shimizu,~K.~I. {Stepwise
  production of CO-rich syngas and hydrogen via solar methane reforming by
  using a Ni(II)-ferrite redox system}. \emph{Sol. Energy} \textbf{2002},
  \emph{73}, 363--374\relax
\mciteBstWouldAddEndPuncttrue
\mciteSetBstMidEndSepPunct{\mcitedefaultmidpunct}
{\mcitedefaultendpunct}{\mcitedefaultseppunct}\relax
\EndOfBibitem
\bibitem[Kang \latin{et~al.}(2010)Kang, Kim, Bae, Cho, Kim, Kim, Kim, and
  Park]{Kang2010}
Kang,~K.~S.; Kim,~C.~H.; Bae,~K.~K.; Cho,~W.~C.; Kim,~W.~J.; Kim,~Y.~H.;
  Kim,~S.~H.; Park,~C.~S. {Redox cycling of CuFe2O4 supported on ZrO2 and CeO2
  for two-step methane reforming/water splitting}. \emph{Int. J. Hydrogen
  Energy} \textbf{2010}, \emph{35}, 568--576\relax
\mciteBstWouldAddEndPuncttrue
\mciteSetBstMidEndSepPunct{\mcitedefaultmidpunct}
{\mcitedefaultendpunct}{\mcitedefaultseppunct}\relax
\EndOfBibitem
\bibitem[Bard and Fox(1995)Bard, and Fox]{bard1995artificial}
Bard,~A.~J.; Fox,~M.~A. Artificial photosynthesis: solar splitting of water to
  hydrogen and oxygen. \emph{Accounts of Chemical Research} \textbf{1995},
  \emph{28}, 141--145\relax
\mciteBstWouldAddEndPuncttrue
\mciteSetBstMidEndSepPunct{\mcitedefaultmidpunct}
{\mcitedefaultendpunct}{\mcitedefaultseppunct}\relax
\EndOfBibitem
\bibitem[Gr{\"{a}}tzel(2001)]{Gratzel2001}
Gr{\"{a}}tzel,~M. {Photoelectrochemical cells}. 2001;
  \url{https://www.nature.com/articles/35104607}\relax
\mciteBstWouldAddEndPuncttrue
\mciteSetBstMidEndSepPunct{\mcitedefaultmidpunct}
{\mcitedefaultendpunct}{\mcitedefaultseppunct}\relax
\EndOfBibitem
\bibitem[Wang \latin{et~al.}(2009)Wang, Maeda, Thomas, Takanabe, Xin, Carlsson,
  Domen, and Antonietti]{Wang2009}
Wang,~X.; Maeda,~K.; Thomas,~A.; Takanabe,~K.; Xin,~G.; Carlsson,~J.~M.;
  Domen,~K.; Antonietti,~M. {A metal-free polymeric photocatalyst for hydrogen
  production from water under visible light}. \emph{Nat. Mater.} \textbf{2009},
  \emph{8}, 76--80\relax
\mciteBstWouldAddEndPuncttrue
\mciteSetBstMidEndSepPunct{\mcitedefaultmidpunct}
{\mcitedefaultendpunct}{\mcitedefaultseppunct}\relax
\EndOfBibitem
\bibitem[Zhou \latin{et~al.}(2017)Zhou, Liu, Schmidt, Kahnt, Osvet, Romeis,
  Zolnhofer, Marthala, Guldi, Peukert, Hartmann, Meyer, and Schmuki]{Zhou2020}
Zhou,~X.; Liu,~N.; Schmidt,~J.; Kahnt,~A.; Osvet,~A.; Romeis,~S.;
  Zolnhofer,~E.~M.; Marthala,~V. R.~R.; Guldi,~D.~M.; Peukert,~W.;
  Hartmann,~M.; Meyer,~K.; Schmuki,~P. {Noble-Metal-Free Photocatalytic
  Hydrogen Evolution Activity: The Impact of Ball Milling Anatase Nanopowders
  with TiH2}. \emph{Adv. Mater.} \textbf{2017}, \emph{29}\relax
\mciteBstWouldAddEndPuncttrue
\mciteSetBstMidEndSepPunct{\mcitedefaultmidpunct}
{\mcitedefaultendpunct}{\mcitedefaultseppunct}\relax
\EndOfBibitem
\bibitem[Kumaravel \latin{et~al.}(2019)Kumaravel, Mathew, Bartlett, and
  Pillai]{Kumaravel2019}
Kumaravel,~V.; Mathew,~S.; Bartlett,~J.; Pillai,~S.~C. {Photocatalytic hydrogen
  production using metal doped TiO2: A review of recent advances}. 2019\relax
\mciteBstWouldAddEndPuncttrue
\mciteSetBstMidEndSepPunct{\mcitedefaultmidpunct}
{\mcitedefaultendpunct}{\mcitedefaultseppunct}\relax
\EndOfBibitem
\bibitem[Teets and Nocera(2011)Teets, and Nocera]{Teets2011}
Teets,~T.~S.; Nocera,~D.~G. {Photocatalytic hydrogen production}. \emph{Chem.
  Commun.} \textbf{2011}, \emph{47}, 9268--9274\relax
\mciteBstWouldAddEndPuncttrue
\mciteSetBstMidEndSepPunct{\mcitedefaultmidpunct}
{\mcitedefaultendpunct}{\mcitedefaultseppunct}\relax
\EndOfBibitem
\bibitem[Gupta and Rao(2017)Gupta, and Rao]{Gupta2017}
Gupta,~U.; Rao,~C.~N. {Hydrogen generation by water splitting using MoS2 and
  other transition metal dichalcogenides}. 2017\relax
\mciteBstWouldAddEndPuncttrue
\mciteSetBstMidEndSepPunct{\mcitedefaultmidpunct}
{\mcitedefaultendpunct}{\mcitedefaultseppunct}\relax
\EndOfBibitem
\bibitem[Gr{\"a}tzel(2001)]{gratzel2001photoelectrochemical}
Gr{\"a}tzel,~M. Photoelectrochemical cells. \emph{nature} \textbf{2001},
  \emph{414}, 338--344\relax
\mciteBstWouldAddEndPuncttrue
\mciteSetBstMidEndSepPunct{\mcitedefaultmidpunct}
{\mcitedefaultendpunct}{\mcitedefaultseppunct}\relax
\EndOfBibitem
\bibitem[Wang \latin{et~al.}(2009)Wang, Maeda, Thomas, Takanabe, Xin, Carlsson,
  Domen, and Antonietti]{wang2009metal}
Wang,~X.; Maeda,~K.; Thomas,~A.; Takanabe,~K.; Xin,~G.; Carlsson,~J.~M.;
  Domen,~K.; Antonietti,~M. A metal-free polymeric photocatalyst for hydrogen
  production from water under visible light. \emph{Nature materials}
  \textbf{2009}, \emph{8}, 76--80\relax
\mciteBstWouldAddEndPuncttrue
\mciteSetBstMidEndSepPunct{\mcitedefaultmidpunct}
{\mcitedefaultendpunct}{\mcitedefaultseppunct}\relax
\EndOfBibitem
\bibitem[Sobczynski(1991)]{sobczynski1991molybdenum}
Sobczynski,~A. Molybdenum disulfide as a hydrogen evolution catalyst for water
  photodecomposition on semiconductors. \emph{Journal of Catalysis}
  \textbf{1991}, \emph{131}, 156--166\relax
\mciteBstWouldAddEndPuncttrue
\mciteSetBstMidEndSepPunct{\mcitedefaultmidpunct}
{\mcitedefaultendpunct}{\mcitedefaultseppunct}\relax
\EndOfBibitem
\bibitem[Wold(1993)]{wold1993photocatalytic}
Wold,~A. Photocatalytic properties of titanium dioxide (TiO2). \emph{Chemistry
  of Materials} \textbf{1993}, \emph{5}, 280--283\relax
\mciteBstWouldAddEndPuncttrue
\mciteSetBstMidEndSepPunct{\mcitedefaultmidpunct}
{\mcitedefaultendpunct}{\mcitedefaultseppunct}\relax
\EndOfBibitem
\bibitem[Abrahams \latin{et~al.}(1985)Abrahams, Davidson, and
  Morrison]{ABRAHAMS1985353}
Abrahams,~J.; Davidson,~R.; Morrison,~C.~L. Optimization of the photocatalytic
  properties of titanium dioxide. \emph{Journal of Photochemistry}
  \textbf{1985}, \emph{29}, 353--361\relax
\mciteBstWouldAddEndPuncttrue
\mciteSetBstMidEndSepPunct{\mcitedefaultmidpunct}
{\mcitedefaultendpunct}{\mcitedefaultseppunct}\relax
\EndOfBibitem
\bibitem[Baba \latin{et~al.}(1985)Baba, Nakabayashi, Fujishima, and
  Honda]{baba1985investigation}
Baba,~R.; Nakabayashi,~S.; Fujishima,~A.; Honda,~K. Investigation of the
  mechanism of hydrogen evolution during photocatalytic water decomposition on
  metal-loaded semiconductor powders. \emph{The Journal of Physical Chemistry}
  \textbf{1985}, \emph{89}, 1902--1905\relax
\mciteBstWouldAddEndPuncttrue
\mciteSetBstMidEndSepPunct{\mcitedefaultmidpunct}
{\mcitedefaultendpunct}{\mcitedefaultseppunct}\relax
\EndOfBibitem
\bibitem[Ganguly \latin{et~al.}(2018)Ganguly, Byrne, Breen, and
  Pillai]{ganguly2018antimicrobial}
Ganguly,~P.; Byrne,~C.; Breen,~A.; Pillai,~S.~C. Antimicrobial activity of
  photocatalysts: fundamentals, mechanisms, kinetics and recent advances.
  \emph{Applied Catalysis B: Environmental} \textbf{2018}, \emph{225},
  51--75\relax
\mciteBstWouldAddEndPuncttrue
\mciteSetBstMidEndSepPunct{\mcitedefaultmidpunct}
{\mcitedefaultendpunct}{\mcitedefaultseppunct}\relax
\EndOfBibitem
\bibitem[Ida and Ishihara(2014)Ida, and Ishihara]{ida2014recent}
Ida,~S.; Ishihara,~T. Recent progress in two-dimensional oxide photocatalysts
  for water splitting. \emph{The Journal of Physical Chemistry Letters}
  \textbf{2014}, \emph{5}, 2533--2542\relax
\mciteBstWouldAddEndPuncttrue
\mciteSetBstMidEndSepPunct{\mcitedefaultmidpunct}
{\mcitedefaultendpunct}{\mcitedefaultseppunct}\relax
\EndOfBibitem
\bibitem[Novoselov \latin{et~al.}(2005)Novoselov, Jiang, Schedin, Booth,
  Khotkevich, Morozov, and Geim]{novoselov2005two}
Novoselov,~K.~S.; Jiang,~D.; Schedin,~F.; Booth,~T.; Khotkevich,~V.;
  Morozov,~S.; Geim,~A.~K. Two-dimensional atomic crystals. \emph{Proceedings
  of the National Academy of Sciences} \textbf{2005}, \emph{102},
  10451--10453\relax
\mciteBstWouldAddEndPuncttrue
\mciteSetBstMidEndSepPunct{\mcitedefaultmidpunct}
{\mcitedefaultendpunct}{\mcitedefaultseppunct}\relax
\EndOfBibitem
\bibitem[Zhang \latin{et~al.}(2011)Zhang, Sun, Maeda, Domen, Liu, Antonietti,
  Fu, and Wang]{zhang2011sulfur}
Zhang,~J.; Sun,~J.; Maeda,~K.; Domen,~K.; Liu,~P.; Antonietti,~M.; Fu,~X.;
  Wang,~X. Sulfur-mediated synthesis of carbon nitride: band-gap engineering
  and improved functions for photocatalysis. \emph{Energy \& Environmental
  Science} \textbf{2011}, \emph{4}, 675--678\relax
\mciteBstWouldAddEndPuncttrue
\mciteSetBstMidEndSepPunct{\mcitedefaultmidpunct}
{\mcitedefaultendpunct}{\mcitedefaultseppunct}\relax
\EndOfBibitem
\bibitem[An and Jimmy(2011)An, and Jimmy]{an2011graphene}
An,~X.; Jimmy,~C.~Y. Graphene-based photocatalytic composites. \emph{Rsc
  Advances} \textbf{2011}, \emph{1}, 1426--1434\relax
\mciteBstWouldAddEndPuncttrue
\mciteSetBstMidEndSepPunct{\mcitedefaultmidpunct}
{\mcitedefaultendpunct}{\mcitedefaultseppunct}\relax
\EndOfBibitem
\bibitem[Butler \latin{et~al.}(2013)Butler, Hollen, Cao, Cui, Gupta,
  Guti{\'e}rrez, Heinz, Hong, Huang, Ismach, \latin{et~al.}
  others]{butler2013progress}
Butler,~S.~Z.; Hollen,~S.~M.; Cao,~L.; Cui,~Y.; Gupta,~J.~A.;
  Guti{\'e}rrez,~H.~R.; Heinz,~T.~F.; Hong,~S.~S.; Huang,~J.; Ismach,~A.~F.;
  others Progress, challenges, and opportunities in two-dimensional materials
  beyond graphene. \emph{ACS nano} \textbf{2013}, \emph{7}, 2898--2926\relax
\mciteBstWouldAddEndPuncttrue
\mciteSetBstMidEndSepPunct{\mcitedefaultmidpunct}
{\mcitedefaultendpunct}{\mcitedefaultseppunct}\relax
\EndOfBibitem
\bibitem[Manzeli \latin{et~al.}(2017)Manzeli, Ovchinnikov, Pasquier, Yazyev,
  and Kis]{manzeli20172d}
Manzeli,~S.; Ovchinnikov,~D.; Pasquier,~D.; Yazyev,~O.~V.; Kis,~A. 2D
  transition metal dichalcogenides. \emph{Nature Reviews Materials}
  \textbf{2017}, \emph{2}, 1--15\relax
\mciteBstWouldAddEndPuncttrue
\mciteSetBstMidEndSepPunct{\mcitedefaultmidpunct}
{\mcitedefaultendpunct}{\mcitedefaultseppunct}\relax
\EndOfBibitem
\bibitem[Chen \latin{et~al.}(2020)Chen, Lai, Zhang, Fan, He, Tan, and
  Zhang]{chen2020phase}
Chen,~Y.; Lai,~Z.; Zhang,~X.; Fan,~Z.; He,~Q.; Tan,~C.; Zhang,~H. Phase
  engineering of nanomaterials. \emph{Nature Reviews Chemistry} \textbf{2020},
  \emph{4}, 243--256\relax
\mciteBstWouldAddEndPuncttrue
\mciteSetBstMidEndSepPunct{\mcitedefaultmidpunct}
{\mcitedefaultendpunct}{\mcitedefaultseppunct}\relax
\EndOfBibitem
\bibitem[Qian \latin{et~al.}(2014)Qian, Liu, Fu, and Li]{qian2014quantum}
Qian,~X.; Liu,~J.; Fu,~L.; Li,~J. Quantum spin Hall effect in two-dimensional
  transition metal dichalcogenides. \emph{Science} \textbf{2014}, \emph{346},
  1344--1347\relax
\mciteBstWouldAddEndPuncttrue
\mciteSetBstMidEndSepPunct{\mcitedefaultmidpunct}
{\mcitedefaultendpunct}{\mcitedefaultseppunct}\relax
\EndOfBibitem
\bibitem[Splendiani \latin{et~al.}(2010)Splendiani, Sun, Zhang, Li, Kim, Chim,
  Galli, and Wang]{doi:10.1021/nl903868w}
Splendiani,~A.; Sun,~L.; Zhang,~Y.; Li,~T.; Kim,~J.; Chim,~C.-Y.; Galli,~G.;
  Wang,~F. Emerging Photoluminescence in Monolayer MoS2. \emph{Nano Letters}
  \textbf{2010}, \emph{10}, 1271--1275, PMID: 20229981\relax
\mciteBstWouldAddEndPuncttrue
\mciteSetBstMidEndSepPunct{\mcitedefaultmidpunct}
{\mcitedefaultendpunct}{\mcitedefaultseppunct}\relax
\EndOfBibitem
\bibitem[Zheng \latin{et~al.}(2012)Zheng, Liu, Liang, Jaroniec, and
  Qiao]{zheng2012graphitic}
Zheng,~Y.; Liu,~J.; Liang,~J.; Jaroniec,~M.; Qiao,~S.~Z. Graphitic carbon
  nitride materials: controllable synthesis and applications in fuel cells and
  photocatalysis. \emph{Energy \& Environmental Science} \textbf{2012},
  \emph{5}, 6717--6731\relax
\mciteBstWouldAddEndPuncttrue
\mciteSetBstMidEndSepPunct{\mcitedefaultmidpunct}
{\mcitedefaultendpunct}{\mcitedefaultseppunct}\relax
\EndOfBibitem
\bibitem[Panneri \latin{et~al.}(2016)Panneri, Ganguly, Nair, Mohamed, Warrier,
  and Hareesh]{panneri2016copyrolysed}
Panneri,~S.; Ganguly,~P.; Nair,~B.~N.; Mohamed,~A. A.~P.; Warrier,~K.~G.;
  Hareesh,~U. N.~S. Copyrolysed C3N4-Ag/ZnO ternary heterostructure systems for
  enhanced adsorption and photocatalytic degradation of tetracycline.
  \emph{European Journal of Inorganic Chemistry} \textbf{2016}, \emph{2016},
  5068--5076\relax
\mciteBstWouldAddEndPuncttrue
\mciteSetBstMidEndSepPunct{\mcitedefaultmidpunct}
{\mcitedefaultendpunct}{\mcitedefaultseppunct}\relax
\EndOfBibitem
\bibitem[Panneri \latin{et~al.}(2017)Panneri, Ganguly, Nair, Mohamed, Warrier,
  and Hareesh]{panneri2017role}
Panneri,~S.; Ganguly,~P.; Nair,~B.~N.; Mohamed,~A. A.~P.; Warrier,~K. G.~K.;
  Hareesh,~U. N.~S. Role of precursors on the photophysical properties of
  carbon nitride and its application for antibiotic degradation.
  \emph{Environmental Science and Pollution Research} \textbf{2017}, \emph{24},
  8609--8618\relax
\mciteBstWouldAddEndPuncttrue
\mciteSetBstMidEndSepPunct{\mcitedefaultmidpunct}
{\mcitedefaultendpunct}{\mcitedefaultseppunct}\relax
\EndOfBibitem
\bibitem[Cai \latin{et~al.}(2018)Cai, Luo, Liu, and Cheng]{cai2018preparation}
Cai,~X.; Luo,~Y.; Liu,~B.; Cheng,~H.-M. Preparation of 2D material dispersions
  and their applications. \emph{Chemical Society Reviews} \textbf{2018},
  \emph{47}, 6224--6266\relax
\mciteBstWouldAddEndPuncttrue
\mciteSetBstMidEndSepPunct{\mcitedefaultmidpunct}
{\mcitedefaultendpunct}{\mcitedefaultseppunct}\relax
\EndOfBibitem
\bibitem[Kumar \latin{et~al.}(2018)Kumar, Boukherroub, and
  Shankar]{kumar2018sunlight}
Kumar,~P.; Boukherroub,~R.; Shankar,~K. Sunlight-driven water-splitting using
  two-dimensional carbon based semiconductors. \emph{Journal of Materials
  Chemistry A} \textbf{2018}, \emph{6}, 12876--12931\relax
\mciteBstWouldAddEndPuncttrue
\mciteSetBstMidEndSepPunct{\mcitedefaultmidpunct}
{\mcitedefaultendpunct}{\mcitedefaultseppunct}\relax
\EndOfBibitem
\bibitem[Xia \latin{et~al.}(2018)Xia, Zhu, Cheng, Yu, and Xu]{xia20182d}
Xia,~P.; Zhu,~B.; Cheng,~B.; Yu,~J.; Xu,~J. 2D/2D g-C3N4/MnO2 nanocomposite as
  a direct Z-scheme photocatalyst for enhanced photocatalytic activity.
  \emph{ACS Sustainable Chemistry \& Engineering} \textbf{2018}, \emph{6},
  965--973\relax
\mciteBstWouldAddEndPuncttrue
\mciteSetBstMidEndSepPunct{\mcitedefaultmidpunct}
{\mcitedefaultendpunct}{\mcitedefaultseppunct}\relax
\EndOfBibitem
\bibitem[Lacombe and Keller(2012)Lacombe, and
  Keller]{lacombe2012photocatalysis}
Lacombe,~S.; Keller,~N. Photocatalysis: fundamentals and applications in JEP
  2011. 2012\relax
\mciteBstWouldAddEndPuncttrue
\mciteSetBstMidEndSepPunct{\mcitedefaultmidpunct}
{\mcitedefaultendpunct}{\mcitedefaultseppunct}\relax
\EndOfBibitem
\bibitem[Zhu and Wang(2017)Zhu, and Wang]{zhu2017photocatalysis}
Zhu,~S.; Wang,~D. Photocatalysis: basic principles, diverse forms of
  implementations and emerging scientific opportunities. \emph{Advanced Energy
  Materials} \textbf{2017}, \emph{7}, 1700841\relax
\mciteBstWouldAddEndPuncttrue
\mciteSetBstMidEndSepPunct{\mcitedefaultmidpunct}
{\mcitedefaultendpunct}{\mcitedefaultseppunct}\relax
\EndOfBibitem
\bibitem[Ravelli \latin{et~al.}(2009)Ravelli, Dondi, Fagnoni, and
  Albini]{ravelli2009photocatalysis}
Ravelli,~D.; Dondi,~D.; Fagnoni,~M.; Albini,~A. Photocatalysis. A multi-faceted
  concept for green chemistry. \emph{Chemical Society Reviews} \textbf{2009},
  \emph{38}, 1999--2011\relax
\mciteBstWouldAddEndPuncttrue
\mciteSetBstMidEndSepPunct{\mcitedefaultmidpunct}
{\mcitedefaultendpunct}{\mcitedefaultseppunct}\relax
\EndOfBibitem
\bibitem[Liang(1970)]{liang1970excitons}
Liang,~W. Excitons. \emph{Physics Education} \textbf{1970}, \emph{5}, 226\relax
\mciteBstWouldAddEndPuncttrue
\mciteSetBstMidEndSepPunct{\mcitedefaultmidpunct}
{\mcitedefaultendpunct}{\mcitedefaultseppunct}\relax
\EndOfBibitem
\bibitem[Li(1993)]{Li1993}
Li,~S.~S. \emph{Semiconductor Physical Electronics}; Springer US: Boston, MA,
  1993; pp 213--245\relax
\mciteBstWouldAddEndPuncttrue
\mciteSetBstMidEndSepPunct{\mcitedefaultmidpunct}
{\mcitedefaultendpunct}{\mcitedefaultseppunct}\relax
\EndOfBibitem
\bibitem[Yang \latin{et~al.}(2023)Yang, Fan, Zhang, Mei, Zhu, Qin, Hu, Chen,
  Hau~Ng, Voiry, \latin{et~al.} others]{yang20232d}
Yang,~R.; Fan,~Y.; Zhang,~Y.; Mei,~L.; Zhu,~R.; Qin,~J.; Hu,~J.; Chen,~Z.;
  Hau~Ng,~Y.; Voiry,~D.; others 2D transition metal dichalcogenides for
  photocatalysis. \emph{Angewandte Chemie} \textbf{2023}, \emph{135},
  e202218016\relax
\mciteBstWouldAddEndPuncttrue
\mciteSetBstMidEndSepPunct{\mcitedefaultmidpunct}
{\mcitedefaultendpunct}{\mcitedefaultseppunct}\relax
\EndOfBibitem
\bibitem[Pil’nik \latin{et~al.}(2020)Pil’nik, Chernov, and
  Islamov]{pil2020charge}
Pil’nik,~A.~A.; Chernov,~A.~A.; Islamov,~D.~R. Charge transport mechanism in
  dielectrics: drift and diffusion of trapped charge carriers. \emph{Scientific
  Reports} \textbf{2020}, \emph{10}, 15759\relax
\mciteBstWouldAddEndPuncttrue
\mciteSetBstMidEndSepPunct{\mcitedefaultmidpunct}
{\mcitedefaultendpunct}{\mcitedefaultseppunct}\relax
\EndOfBibitem
\bibitem[Enyo(1973)]{enyo1973change}
Enyo,~M. Change of mechanism of the hydrogen-electrode reaction with
  overpotential—I. distribution of the reaction affinity among constituent
  steps. \emph{Electrochimica Acta} \textbf{1973}, \emph{18}, 155--162\relax
\mciteBstWouldAddEndPuncttrue
\mciteSetBstMidEndSepPunct{\mcitedefaultmidpunct}
{\mcitedefaultendpunct}{\mcitedefaultseppunct}\relax
\EndOfBibitem
\bibitem[Tilak \latin{et~al.}(1977)Tilak, Rader, and
  Conway]{tilak1977overpotential}
Tilak,~B.; Rader,~C.; Conway,~B. Overpotential decay behavior—II. Generalized
  treatment for reaction pathways involving discharge, recombination and
  electrochemical desorption of adsorbed intermediates. \emph{Electrochimica
  Acta} \textbf{1977}, \emph{22}, 1167--1178\relax
\mciteBstWouldAddEndPuncttrue
\mciteSetBstMidEndSepPunct{\mcitedefaultmidpunct}
{\mcitedefaultendpunct}{\mcitedefaultseppunct}\relax
\EndOfBibitem
\bibitem[Tilak and Conway(1976)Tilak, and Conway]{tilak1976overpotential}
Tilak,~B.; Conway,~B. Overpotential decay behavior—I. Complex electrode
  reactions involving adsorption. \emph{Electrochimica Acta} \textbf{1976},
  \emph{21}, 745--752\relax
\mciteBstWouldAddEndPuncttrue
\mciteSetBstMidEndSepPunct{\mcitedefaultmidpunct}
{\mcitedefaultendpunct}{\mcitedefaultseppunct}\relax
\EndOfBibitem
\bibitem[Walter \latin{et~al.}(2010)Walter, Warren, McKone, Boettcher, Mi,
  Santori, and Lewis]{doi:10.1021/cr1002326}
Walter,~M.~G.; Warren,~E.~L.; McKone,~J.~R.; Boettcher,~S.~W.; Mi,~Q.;
  Santori,~E.~A.; Lewis,~N.~S. Solar Water Splitting Cells. \emph{Chemical
  Reviews} \textbf{2010}, \emph{110}, 6446--6473, PMID: 21062097\relax
\mciteBstWouldAddEndPuncttrue
\mciteSetBstMidEndSepPunct{\mcitedefaultmidpunct}
{\mcitedefaultendpunct}{\mcitedefaultseppunct}\relax
\EndOfBibitem
\bibitem[Rahman \latin{et~al.}(2016)Rahman, Kwong, Davey, and
  Qiao]{rahman20162d}
Rahman,~M.~Z.; Kwong,~C.~W.; Davey,~K.; Qiao,~S.~Z. 2D phosphorene as a water
  splitting photocatalyst: fundamentals to applications. \emph{Energy \&
  Environmental Science} \textbf{2016}, \emph{9}, 709--728\relax
\mciteBstWouldAddEndPuncttrue
\mciteSetBstMidEndSepPunct{\mcitedefaultmidpunct}
{\mcitedefaultendpunct}{\mcitedefaultseppunct}\relax
\EndOfBibitem
\bibitem[Wang \latin{et~al.}(2011)Wang, Wang, Ling, Tang, Yang, Fitzmorris,
  Wang, Zhang, and Li]{wang2011hydrogen}
Wang,~G.; Wang,~H.; Ling,~Y.; Tang,~Y.; Yang,~X.; Fitzmorris,~R.~C.; Wang,~C.;
  Zhang,~J.~Z.; Li,~Y. Hydrogen-treated TiO2 nanowire arrays for
  photoelectrochemical water splitting. \emph{Nano letters} \textbf{2011},
  \emph{11}, 3026--3033\relax
\mciteBstWouldAddEndPuncttrue
\mciteSetBstMidEndSepPunct{\mcitedefaultmidpunct}
{\mcitedefaultendpunct}{\mcitedefaultseppunct}\relax
\EndOfBibitem
\bibitem[Pu \latin{et~al.}(2013)Pu, Wang, Chang, Ling, Lin, Fitzmorris, Liu,
  Lu, Tong, Zhang, \latin{et~al.} others]{pu2013nanostructure}
Pu,~Y.-C.; Wang,~G.; Chang,~K.-D.; Ling,~Y.; Lin,~Y.-K.; Fitzmorris,~B.~C.;
  Liu,~C.-M.; Lu,~X.; Tong,~Y.; Zhang,~J.~Z.; others Au nanostructure-decorated
  TiO2 nanowires exhibiting photoactivity across entire UV-visible region for
  photoelectrochemical water splitting. \emph{Nano letters} \textbf{2013},
  \emph{13}, 3817--3823\relax
\mciteBstWouldAddEndPuncttrue
\mciteSetBstMidEndSepPunct{\mcitedefaultmidpunct}
{\mcitedefaultendpunct}{\mcitedefaultseppunct}\relax
\EndOfBibitem
\bibitem[Qiu \latin{et~al.}(2012)Qiu, Yan, Deng, and Yang]{qiu2012secondary}
Qiu,~Y.; Yan,~K.; Deng,~H.; Yang,~S. Secondary branching and nitrogen doping of
  ZnO nanotetrapods: building a highly active network for photoelectrochemical
  water splitting. \emph{Nano letters} \textbf{2012}, \emph{12}, 407--413\relax
\mciteBstWouldAddEndPuncttrue
\mciteSetBstMidEndSepPunct{\mcitedefaultmidpunct}
{\mcitedefaultendpunct}{\mcitedefaultseppunct}\relax
\EndOfBibitem
\bibitem[Huang \latin{et~al.}(2017)Huang, Liu, Wu, and Li]{huang2017influence}
Huang,~J.; Liu,~Y.; Wu,~Y.; Li,~X. Influence of Mn doping on the sensing
  properties of SnO2 nanobelt to ethanol. \emph{American Journal of Analytical
  Chemistry} \textbf{2017}, \emph{8}, 60--71\relax
\mciteBstWouldAddEndPuncttrue
\mciteSetBstMidEndSepPunct{\mcitedefaultmidpunct}
{\mcitedefaultendpunct}{\mcitedefaultseppunct}\relax
\EndOfBibitem
\bibitem[Qiu \latin{et~al.}(2012)Qiu, Xu, Kuang, Sun, and
  Yang]{qiu2012hierarchical}
Qiu,~Y.; Xu,~G.-L.; Kuang,~Q.; Sun,~S.-G.; Yang,~S. Hierarchical WO 3 flowers
  comprising porous single-crystalline nanoplates show enhanced lithium storage
  and photocatalysis. \emph{Nano research} \textbf{2012}, \emph{5},
  826--832\relax
\mciteBstWouldAddEndPuncttrue
\mciteSetBstMidEndSepPunct{\mcitedefaultmidpunct}
{\mcitedefaultendpunct}{\mcitedefaultseppunct}\relax
\EndOfBibitem
\bibitem[Yan \latin{et~al.}(2016)Yan, Zhou, Li, Chen, Zhang, Dong, Xi, and
  Liu]{yan2016nitrogen}
Yan,~J.; Zhou,~C.; Li,~P.; Chen,~B.; Zhang,~S.; Dong,~X.; Xi,~F.; Liu,~J.
  Nitrogen-rich graphitic carbon nitride: controllable nanosheet-like
  morphology, enhanced visible light absorption and superior photocatalytic
  performance. \emph{Colloids and Surfaces A: Physicochemical and Engineering
  Aspects} \textbf{2016}, \emph{508}, 257--264\relax
\mciteBstWouldAddEndPuncttrue
\mciteSetBstMidEndSepPunct{\mcitedefaultmidpunct}
{\mcitedefaultendpunct}{\mcitedefaultseppunct}\relax
\EndOfBibitem
\bibitem[Huang \latin{et~al.}(2014)Huang, Li, Balogun, Liu, Tong, Lu, and
  Ji]{doi:10.1021/am507641k}
Huang,~Y.; Li,~H.; Balogun,~M.-S.; Liu,~W.; Tong,~Y.; Lu,~X.; Ji,~H. Oxygen
  Vacancy Induced Bismuth Oxyiodide with Remarkably Increased Visible-Light
  Absorption and Superior Photocatalytic Performance. \emph{ACS Applied
  Materials \& Interfaces} \textbf{2014}, \emph{6}, 22920--22927, PMID:
  25437430\relax
\mciteBstWouldAddEndPuncttrue
\mciteSetBstMidEndSepPunct{\mcitedefaultmidpunct}
{\mcitedefaultendpunct}{\mcitedefaultseppunct}\relax
\EndOfBibitem
\bibitem[Tian \latin{et~al.}(2013)Tian, Sang, Yu, Jiang, Mu, and
  Liu]{tian2013bi2wo6}
Tian,~J.; Sang,~Y.; Yu,~G.; Jiang,~H.; Mu,~X.; Liu,~H. A Bi2WO6-based hybrid
  photocatalyst with broad spectrum photocatalytic properties under UV,
  visible, and near-infrared irradiation. \emph{Advanced Materials}
  \textbf{2013}, \emph{25}, 5075--5080\relax
\mciteBstWouldAddEndPuncttrue
\mciteSetBstMidEndSepPunct{\mcitedefaultmidpunct}
{\mcitedefaultendpunct}{\mcitedefaultseppunct}\relax
\EndOfBibitem
\bibitem[Li \latin{et~al.}(2016)Li, Liu, Li, and Chen]{li2016fabrication}
Li,~J.; Liu,~Y.; Li,~H.; Chen,~C. Fabrication of g-C3N4/TiO2 composite
  photocatalyst with extended absorption wavelength range and enhanced
  photocatalytic performance. \emph{Journal of Photochemistry and Photobiology
  A: Chemistry} \textbf{2016}, \emph{317}, 151--160\relax
\mciteBstWouldAddEndPuncttrue
\mciteSetBstMidEndSepPunct{\mcitedefaultmidpunct}
{\mcitedefaultendpunct}{\mcitedefaultseppunct}\relax
\EndOfBibitem
\bibitem[Melman \latin{et~al.}(1977)Melman, O'Steen, Prock, and
  Chance]{melman1977excited}
Melman,~P.; O'Steen,~D.; Prock,~A.; Chance,~R. Excited state migration and
  charge transfer in a semiconductor/aromatic solution system. \emph{Chemical
  Physics} \textbf{1977}, \emph{22}, 71--77\relax
\mciteBstWouldAddEndPuncttrue
\mciteSetBstMidEndSepPunct{\mcitedefaultmidpunct}
{\mcitedefaultendpunct}{\mcitedefaultseppunct}\relax
\EndOfBibitem
\bibitem[Li \latin{et~al.}(2019)Li, Gao, Long, and Xiong]{li2019photocatalyst}
Li,~Y.; Gao,~C.; Long,~R.; Xiong,~Y. Photocatalyst design based on
  two-dimensional materials. \emph{Materials today chemistry} \textbf{2019},
  \emph{11}, 197--216\relax
\mciteBstWouldAddEndPuncttrue
\mciteSetBstMidEndSepPunct{\mcitedefaultmidpunct}
{\mcitedefaultendpunct}{\mcitedefaultseppunct}\relax
\EndOfBibitem
\bibitem[Zhou \latin{et~al.}(2019)Zhou, Niu, Zhang, and Wang]{Zhou2019}
Zhou,~Z.; Niu,~X.; Zhang,~Y.; Wang,~J. {Janus MoSSe/WSeTe heterostructures: A
  direct Z-scheme photocatalyst for hydrogen evolution}. \emph{J. Mater. Chem.
  A} \textbf{2019}, \emph{7}, 21835--21842\relax
\mciteBstWouldAddEndPuncttrue
\mciteSetBstMidEndSepPunct{\mcitedefaultmidpunct}
{\mcitedefaultendpunct}{\mcitedefaultseppunct}\relax
\EndOfBibitem
\bibitem[Wahab \latin{et~al.}(2023)Wahab, Wang, and Cammarata]{Wahab2023}
Wahab,~T.; Wang,~Y.; Cammarata,~A. {A first principles study of structural and
  optoelectronic properties and photocatalytic performance of GeC-MX2 (M = Mo
  and W; X = S and Se) van der Waals heterostructures}. \emph{Phys. Chem. Chem.
  Phys.} \textbf{2023}, \emph{25}, 11169--11175\relax
\mciteBstWouldAddEndPuncttrue
\mciteSetBstMidEndSepPunct{\mcitedefaultmidpunct}
{\mcitedefaultendpunct}{\mcitedefaultseppunct}\relax
\EndOfBibitem
\bibitem[Luo \latin{et~al.}(2016)Luo, Liu, and Wang]{luo2016recent}
Luo,~B.; Liu,~G.; Wang,~L. Recent advances in 2D materials for photocatalysis.
  \emph{Nanoscale} \textbf{2016}, \emph{8}, 6904--6920\relax
\mciteBstWouldAddEndPuncttrue
\mciteSetBstMidEndSepPunct{\mcitedefaultmidpunct}
{\mcitedefaultendpunct}{\mcitedefaultseppunct}\relax
\EndOfBibitem
\bibitem[Zhang \latin{et~al.}(2020)Zhang, Luo, Tan, Yu, Yang, Zhang, Yang,
  Cheng, and Liu]{zhang2020high}
Zhang,~C.; Luo,~Y.; Tan,~J.; Yu,~Q.; Yang,~F.; Zhang,~Z.; Yang,~L.;
  Cheng,~H.-M.; Liu,~B. High-throughput production of cheap mineral-based
  two-dimensional electrocatalysts for high-current-density hydrogen evolution.
  \emph{Nature communications} \textbf{2020}, \emph{11}, 3724\relax
\mciteBstWouldAddEndPuncttrue
\mciteSetBstMidEndSepPunct{\mcitedefaultmidpunct}
{\mcitedefaultendpunct}{\mcitedefaultseppunct}\relax
\EndOfBibitem
\bibitem[Deng \latin{et~al.}(2015)Deng, Li, Xiao, Tu, Deng, Yang, Tian, Li,
  Ren, and Bao]{deng2015triggering}
Deng,~J.; Li,~H.; Xiao,~J.; Tu,~Y.; Deng,~D.; Yang,~H.; Tian,~H.; Li,~J.;
  Ren,~P.; Bao,~X. Triggering the electrocatalytic hydrogen evolution activity
  of the inert two-dimensional MoS 2 surface via single-atom metal doping.
  \emph{Energy \& environmental science} \textbf{2015}, \emph{8},
  1594--1601\relax
\mciteBstWouldAddEndPuncttrue
\mciteSetBstMidEndSepPunct{\mcitedefaultmidpunct}
{\mcitedefaultendpunct}{\mcitedefaultseppunct}\relax
\EndOfBibitem
\bibitem[Li \latin{et~al.}(2019)Li, Qian, Du, Wu, Zhang, Li, Li, Wang, and
  Kang]{li2019cus}
Li,~M.; Qian,~Y.; Du,~J.; Wu,~H.; Zhang,~L.; Li,~G.; Li,~K.; Wang,~W.;
  Kang,~D.~J. CuS nanosheets decorated with CoS2 nanoparticles as an efficient
  electrocatalyst for enhanced hydrogen evolution at all pH values. \emph{ACS
  Sustainable Chemistry \& Engineering} \textbf{2019}, \emph{7},
  14016--14022\relax
\mciteBstWouldAddEndPuncttrue
\mciteSetBstMidEndSepPunct{\mcitedefaultmidpunct}
{\mcitedefaultendpunct}{\mcitedefaultseppunct}\relax
\EndOfBibitem
\bibitem[Liu \latin{et~al.}(2019)Liu, Zhang, Gong, Lu, Zhang, Cheng, Ma, Chen,
  Zhao, Chen, \latin{et~al.} others]{liu2019synthesis}
Liu,~Z.; Zhang,~X.; Gong,~Y.; Lu,~Q.; Zhang,~Z.; Cheng,~H.; Ma,~Q.; Chen,~J.;
  Zhao,~M.; Chen,~B.; others Synthesis of MoX 2 (X= Se or S) monolayers with
  high-concentration 1T' phase on 4H/fcc-Au nanorods for hydrogen evolution.
  \emph{Nano Research} \textbf{2019}, \emph{12}, 1301--1305\relax
\mciteBstWouldAddEndPuncttrue
\mciteSetBstMidEndSepPunct{\mcitedefaultmidpunct}
{\mcitedefaultendpunct}{\mcitedefaultseppunct}\relax
\EndOfBibitem
\bibitem[Yang \latin{et~al.}(2023)Yang, Hu, Yang, Chen, Yin, Hao, Sun, Gao,
  Sun, Wang, \latin{et~al.} others]{yang2023cnts}
Yang,~F.; Hu,~P.; Yang,~F.~F.; Chen,~B.; Yin,~F.; Hao,~K.; Sun,~R.; Gao,~L.;
  Sun,~Z.; Wang,~K.; others CNTs bridged basal-plane-active 2H-MoS2 nanosheets
  for efficient robust electrocatalysis. \emph{Small} \textbf{2023}, \emph{19},
  2301468\relax
\mciteBstWouldAddEndPuncttrue
\mciteSetBstMidEndSepPunct{\mcitedefaultmidpunct}
{\mcitedefaultendpunct}{\mcitedefaultseppunct}\relax
\EndOfBibitem
\bibitem[Yang \latin{et~al.}(2018)Yang, Lv, and Cao]{yang2018co}
Yang,~L.; Lv,~Y.; Cao,~D. Co, N-codoped nanotube/graphene 1D/2D heterostructure
  for efficient oxygen reduction and hydrogen evolution reactions.
  \emph{Journal of Materials Chemistry A} \textbf{2018}, \emph{6},
  3926--3932\relax
\mciteBstWouldAddEndPuncttrue
\mciteSetBstMidEndSepPunct{\mcitedefaultmidpunct}
{\mcitedefaultendpunct}{\mcitedefaultseppunct}\relax
\EndOfBibitem
\bibitem[Park \latin{et~al.}(2017)Park, Kim, Park, and Kang]{park2017mose2}
Park,~G.~D.; Kim,~J.~H.; Park,~S.-K.; Kang,~Y.~C. MoSe2 embedded CNT-reduced
  graphene oxide composite microsphere with superior sodium ion storage and
  electrocatalytic hydrogen evolution performances. \emph{ACS Applied Materials
  \& Interfaces} \textbf{2017}, \emph{9}, 10673--10683\relax
\mciteBstWouldAddEndPuncttrue
\mciteSetBstMidEndSepPunct{\mcitedefaultmidpunct}
{\mcitedefaultendpunct}{\mcitedefaultseppunct}\relax
\EndOfBibitem
\bibitem[Lin \latin{et~al.}(2016)Lin, Pan, Zhang, Chen, Sun, Liu, and
  Liu]{lin2016graphene}
Lin,~Y.; Pan,~Y.; Zhang,~J.; Chen,~Y.; Sun,~K.; Liu,~Y.; Liu,~C. Graphene oxide
  co-doped with nitrogen and sulfur and decorated with cobalt phosphide
  nanorods: An efficient hybrid catalyst for electrochemical hydrogen
  evolution. \emph{Electrochimica Acta} \textbf{2016}, \emph{222},
  246--256\relax
\mciteBstWouldAddEndPuncttrue
\mciteSetBstMidEndSepPunct{\mcitedefaultmidpunct}
{\mcitedefaultendpunct}{\mcitedefaultseppunct}\relax
\EndOfBibitem
\bibitem[Tang \latin{et~al.}(2017)Tang, Wang, Guo, Teng, Meyer, and
  Chen]{tang2017heterostructured}
Tang,~W.; Wang,~J.; Guo,~L.; Teng,~X.; Meyer,~T.~J.; Chen,~Z. Heterostructured
  arrays of Ni x P/S/Se nanosheets on Co x P/S/Se nanowires for efficient
  hydrogen evolution. \emph{ACS applied materials \& interfaces} \textbf{2017},
  \emph{9}, 41347--41353\relax
\mciteBstWouldAddEndPuncttrue
\mciteSetBstMidEndSepPunct{\mcitedefaultmidpunct}
{\mcitedefaultendpunct}{\mcitedefaultseppunct}\relax
\EndOfBibitem
\bibitem[Hu \latin{et~al.}(2017)Hu, Zhang, Jiang, Lin, An, Zhou, Leung, and
  Yang]{hu2017nanohybridization}
Hu,~J.; Zhang,~C.; Jiang,~L.; Lin,~H.; An,~Y.; Zhou,~D.; Leung,~M.~K.; Yang,~S.
  Nanohybridization of MoS2 with layered double hydroxides efficiently
  synergizes the hydrogen evolution in alkaline media. \emph{Joule}
  \textbf{2017}, \emph{1}, 383--393\relax
\mciteBstWouldAddEndPuncttrue
\mciteSetBstMidEndSepPunct{\mcitedefaultmidpunct}
{\mcitedefaultendpunct}{\mcitedefaultseppunct}\relax
\EndOfBibitem
\bibitem[Ma \latin{et~al.}(2014)Ma, Qi, Chen, Bao, Yin, Wu, Luo, Wei, Zhang,
  and Zhang]{ma2014mos}
Ma,~C.-B.; Qi,~X.; Chen,~B.; Bao,~S.; Yin,~Z.; Wu,~X.-J.; Luo,~Z.; Wei,~J.;
  Zhang,~H.-L.; Zhang,~H. MoS 2 nanoflower-decorated reduced graphene oxide
  paper for high-performance hydrogen evolution reaction. \emph{Nanoscale}
  \textbf{2014}, \emph{6}, 5624--5629\relax
\mciteBstWouldAddEndPuncttrue
\mciteSetBstMidEndSepPunct{\mcitedefaultmidpunct}
{\mcitedefaultendpunct}{\mcitedefaultseppunct}\relax
\EndOfBibitem
\bibitem[Wu \latin{et~al.}(2019)Wu, Gu, Xie, Tian, Yan, Wang, Zhang, Cai, and
  Fu]{wu2019effective}
Wu,~A.; Gu,~Y.; Xie,~Y.; Tian,~C.; Yan,~H.; Wang,~D.; Zhang,~X.; Cai,~Z.;
  Fu,~H. Effective electrocatalytic hydrogen evolution in neutral medium based
  on 2D MoP/MoS2 heterostructure nanosheets. \emph{ACS applied materials \&
  interfaces} \textbf{2019}, \emph{11}, 25986--25995\relax
\mciteBstWouldAddEndPuncttrue
\mciteSetBstMidEndSepPunct{\mcitedefaultmidpunct}
{\mcitedefaultendpunct}{\mcitedefaultseppunct}\relax
\EndOfBibitem
\bibitem[Kong \latin{et~al.}(2013)Kong, Wang, Cha, Pasta, Koski, Yao, and
  Cui]{kong2013synthesis}
Kong,~D.; Wang,~H.; Cha,~J.~J.; Pasta,~M.; Koski,~K.~J.; Yao,~J.; Cui,~Y.
  Synthesis of MoS2 and MoSe2 films with vertically aligned layers. \emph{Nano
  letters} \textbf{2013}, \emph{13}, 1341--1347\relax
\mciteBstWouldAddEndPuncttrue
\mciteSetBstMidEndSepPunct{\mcitedefaultmidpunct}
{\mcitedefaultendpunct}{\mcitedefaultseppunct}\relax
\EndOfBibitem
\bibitem[Chandrasekaran \latin{et~al.}(2022)Chandrasekaran, Li, Zhuang, Sui,
  Xiao, Fan, Aravindan, Bowen, Lu, and Liu]{chandrasekaran2022interface}
Chandrasekaran,~S.; Li,~N.; Zhuang,~Y.; Sui,~L.; Xiao,~Z.; Fan,~D.;
  Aravindan,~V.; Bowen,~C.; Lu,~H.; Liu,~Y. Interface charge density modulation
  of a lamellar-like spatially separated Ni9S8 nanosheet/Nb2O5 nanobelt
  heterostructure catalyst coupled with nitrogen and metal (M= Co, Fe, or Cu)
  atoms to accelerate acidic and alkaline hydrogen evolution reactions.
  \emph{Chemical Engineering Journal} \textbf{2022}, \emph{431}, 134073\relax
\mciteBstWouldAddEndPuncttrue
\mciteSetBstMidEndSepPunct{\mcitedefaultmidpunct}
{\mcitedefaultendpunct}{\mcitedefaultseppunct}\relax
\EndOfBibitem
\bibitem[Li \latin{et~al.}(2021)Li, Zhou, Chen, Liu, Huo, and
  Yi]{li2021building}
Li,~Q.; Zhou,~Y.; Chen,~C.; Liu,~Q.; Huo,~J.; Yi,~H. Building CoP/Co-MOF
  heterostructure in 2D nanosheets for improving electrocatalytic hydrogen
  evolution over a wide pH range. \emph{Journal of Electroanalytical Chemistry}
  \textbf{2021}, \emph{895}, 115514\relax
\mciteBstWouldAddEndPuncttrue
\mciteSetBstMidEndSepPunct{\mcitedefaultmidpunct}
{\mcitedefaultendpunct}{\mcitedefaultseppunct}\relax
\EndOfBibitem
\bibitem[Tang \latin{et~al.}(2018)Tang, Zhong, Zhang, Wang, and
  Zhang]{tang20183d}
Tang,~C.; Zhong,~L.; Zhang,~B.; Wang,~H.-F.; Zhang,~Q. 3D mesoporous van der
  Waals heterostructures for trifunctional energy electrocatalysis.
  \emph{Advanced materials} \textbf{2018}, \emph{30}, 1705110\relax
\mciteBstWouldAddEndPuncttrue
\mciteSetBstMidEndSepPunct{\mcitedefaultmidpunct}
{\mcitedefaultendpunct}{\mcitedefaultseppunct}\relax
\EndOfBibitem
\bibitem[Geng \latin{et~al.}(2017)Geng, Zhao, Chen, Sun, Fu, Chen, Liu, Zhou,
  and Loh]{geng2017direct}
Geng,~D.; Zhao,~X.; Chen,~Z.; Sun,~W.; Fu,~W.; Chen,~J.; Liu,~W.; Zhou,~W.;
  Loh,~K.~P. Direct synthesis of large-area 2D Mo2C on in situ grown graphene.
  \emph{Advanced Materials} \textbf{2017}, \emph{29}, 1700072\relax
\mciteBstWouldAddEndPuncttrue
\mciteSetBstMidEndSepPunct{\mcitedefaultmidpunct}
{\mcitedefaultendpunct}{\mcitedefaultseppunct}\relax
\EndOfBibitem
\bibitem[Shi \latin{et~al.}(2019)Shi, Fan, Du, Fu, Dong, Xie, Zhao, Wang, and
  Yuan]{shi2019situ}
Shi,~G.; Fan,~Z.; Du,~L.; Fu,~X.; Dong,~C.; Xie,~W.; Zhao,~D.; Wang,~M.;
  Yuan,~M. In situ construction of graphdiyne/CuS heterostructures for
  efficient hydrogen evolution reaction. \emph{Materials Chemistry Frontiers}
  \textbf{2019}, \emph{3}, 821--828\relax
\mciteBstWouldAddEndPuncttrue
\mciteSetBstMidEndSepPunct{\mcitedefaultmidpunct}
{\mcitedefaultendpunct}{\mcitedefaultseppunct}\relax
\EndOfBibitem
\bibitem[Duan \latin{et~al.}(2015)Duan, Chen, Jaroniec, and
  Qiao]{duan2015porous}
Duan,~J.; Chen,~S.; Jaroniec,~M.; Qiao,~S.~Z. Porous C3N4 nanolayers@
  N-graphene films as catalyst electrodes for highly efficient hydrogen
  evolution. \emph{ACS nano} \textbf{2015}, \emph{9}, 931--940\relax
\mciteBstWouldAddEndPuncttrue
\mciteSetBstMidEndSepPunct{\mcitedefaultmidpunct}
{\mcitedefaultendpunct}{\mcitedefaultseppunct}\relax
\EndOfBibitem
\bibitem[Lin \latin{et~al.}(2024)Lin, Zhao, Xiao, Sattar, Tang, Nairan, Guo,
  Xia, Canali, Khan, \latin{et~al.} others]{lin2024improving}
Lin,~C.; Zhao,~X.; Xiao,~Y.; Sattar,~S.; Tang,~L.; Nairan,~A.; Guo,~Y.;
  Xia,~M.; Canali,~C.~M.; Khan,~U.; others Improving photocatalytic hydrogen
  generation of g-C3N4 via efficient charge separation imposed by Bi2O2Se
  nanosheets. \emph{Carbon} \textbf{2024}, \emph{218}, 118721\relax
\mciteBstWouldAddEndPuncttrue
\mciteSetBstMidEndSepPunct{\mcitedefaultmidpunct}
{\mcitedefaultendpunct}{\mcitedefaultseppunct}\relax
\EndOfBibitem
\bibitem[Wang \latin{et~al.}(2018)Wang, Tang, Wang, Li, Cui, and
  Zhang]{wang2018defect}
Wang,~H.-F.; Tang,~C.; Wang,~B.; Li,~B.-Q.; Cui,~X.; Zhang,~Q. Defect-rich
  carbon fiber electrocatalysts with porous graphene skin for flexible
  solid-state zinc--air batteries. \emph{Energy Storage Materials}
  \textbf{2018}, \emph{15}, 124--130\relax
\mciteBstWouldAddEndPuncttrue
\mciteSetBstMidEndSepPunct{\mcitedefaultmidpunct}
{\mcitedefaultendpunct}{\mcitedefaultseppunct}\relax
\EndOfBibitem
\bibitem[Zhang \latin{et~al.}(2022)Zhang, Wang, Yuan, Fang, and
  Wang]{zhang2022heterostructured}
Zhang,~T.; Wang,~Y.; Yuan,~J.; Fang,~K.; Wang,~A.-j. Heterostructured
  CoP{\textperiodcentered} CoMoP nanocages as advanced electrocatalysts for
  efficient hydrogen evolution over a wide pH range. \emph{Journal of colloid
  and interface science} \textbf{2022}, \emph{615}, 465--474\relax
\mciteBstWouldAddEndPuncttrue
\mciteSetBstMidEndSepPunct{\mcitedefaultmidpunct}
{\mcitedefaultendpunct}{\mcitedefaultseppunct}\relax
\EndOfBibitem
\bibitem[Choi \latin{et~al.}(2021)Choi, Lee, Kim, Park, Jang, Ahn, Sohn, Park,
  Hong, and Lee]{choi2021hierarchically}
Choi,~H.; Lee,~S.; Kim,~M.-C.; Park,~Y.; Jang,~A.-R.; Ahn,~W.; Sohn,~J.~I.;
  Park,~J.~B.; Hong,~J.; Lee,~Y.-W. Hierarchically ordinated two-dimensional
  MoS2 nanosheets on three-dimensional reduced graphene oxide aerogels as
  highly active and stable catalysts for hydrogen evolution reaction.
  \emph{Catalysts} \textbf{2021}, \emph{11}, 182\relax
\mciteBstWouldAddEndPuncttrue
\mciteSetBstMidEndSepPunct{\mcitedefaultmidpunct}
{\mcitedefaultendpunct}{\mcitedefaultseppunct}\relax
\EndOfBibitem
\bibitem[Tan \latin{et~al.}(2014)Tan, Liu, Chen, Cong, Ito, Han, Guo, Tang,
  Fujita, Hirata, \latin{et~al.} others]{tan2014monolayer}
Tan,~Y.; Liu,~P.; Chen,~L.; Cong,~W.; Ito,~Y.; Han,~J.; Guo,~X.; Tang,~Z.;
  Fujita,~T.; Hirata,~A.; others Monolayer MoS2 films supported by 3D
  nanoporous metals for high-efficiency electrocatalytic hydrogen production.
  \emph{Advanced Materials (Deerfield Beach, Fla.)} \textbf{2014}, \emph{26},
  8023--8028\relax
\mciteBstWouldAddEndPuncttrue
\mciteSetBstMidEndSepPunct{\mcitedefaultmidpunct}
{\mcitedefaultendpunct}{\mcitedefaultseppunct}\relax
\EndOfBibitem
\bibitem[Luo \latin{et~al.}(2019)Luo, Tang, Khan, Yu, Cheng, Zou, and
  Liu]{luo2019morphology}
Luo,~Y.; Tang,~L.; Khan,~U.; Yu,~Q.; Cheng,~H.-M.; Zou,~X.; Liu,~B. Morphology
  and surface chemistry engineering toward pH-universal catalysts for hydrogen
  evolution at high current density. \emph{Nature communications}
  \textbf{2019}, \emph{10}, 269\relax
\mciteBstWouldAddEndPuncttrue
\mciteSetBstMidEndSepPunct{\mcitedefaultmidpunct}
{\mcitedefaultendpunct}{\mcitedefaultseppunct}\relax
\EndOfBibitem
\bibitem[Wang \latin{et~al.}(2016)Wang, Xu, Rao, Xu, Chen, Zhang, Kuang, and
  Su]{wang2016novel}
Wang,~X.-D.; Xu,~Y.-F.; Rao,~H.-S.; Xu,~W.-J.; Chen,~H.-Y.; Zhang,~W.-X.;
  Kuang,~D.-B.; Su,~C.-Y. Novel porous molybdenum tungsten phosphide hybrid
  nanosheets on carbon cloth for efficient hydrogen evolution. \emph{Energy \&
  Environmental Science} \textbf{2016}, \emph{9}, 1468--1475\relax
\mciteBstWouldAddEndPuncttrue
\mciteSetBstMidEndSepPunct{\mcitedefaultmidpunct}
{\mcitedefaultendpunct}{\mcitedefaultseppunct}\relax
\EndOfBibitem
\bibitem[Qin \latin{et~al.}(2020)Qin, Wang, Wu, Han, Wu, Zhang, Tian, Shen, Li,
  and Wang]{qin2020metal}
Qin,~C.; Wang,~B.; Wu,~N.; Han,~C.; Wu,~C.; Zhang,~X.; Tian,~Q.; Shen,~S.;
  Li,~P.; Wang,~Y. Metal-organic frameworks derived porous Co3O4 dodecahedeons
  with abundant active Co3+ for ppb-level CO gas sensing. \emph{Applied Surface
  Science} \textbf{2020}, \emph{506}, 144900\relax
\mciteBstWouldAddEndPuncttrue
\mciteSetBstMidEndSepPunct{\mcitedefaultmidpunct}
{\mcitedefaultendpunct}{\mcitedefaultseppunct}\relax
\EndOfBibitem
\bibitem[Liu \latin{et~al.}(2022)Liu, Nie, Yuan, Li, Liu, Chong, Du, and
  Huang]{liu2022crystalline}
Liu,~Z.; Nie,~K.; Yuan,~Y.; Li,~B.; Liu,~P.; Chong,~S.; Du,~Y.; Huang,~W.
  Crystalline/Amorphous Heterophase with Self-Assembled Hollow Structure for
  Highly Efficient Electrochemical Hydrogen Production. \emph{CCS Chemistry}
  \textbf{2022}, \emph{4}, 3391--3401\relax
\mciteBstWouldAddEndPuncttrue
\mciteSetBstMidEndSepPunct{\mcitedefaultmidpunct}
{\mcitedefaultendpunct}{\mcitedefaultseppunct}\relax
\EndOfBibitem
\bibitem[Lorente \latin{et~al.}(2011)Lorente, Pe{\~{n}}a, and
  Herguido]{Lorente2011}
Lorente,~E.; Pe{\~{n}}a,~J.~A.; Herguido,~J. {Cycle behaviour of iron ores in
  the steam-iron process}. \emph{Int. J. Hydrogen Energy} \textbf{2011},
  \emph{36}, 7043--7050\relax
\mciteBstWouldAddEndPuncttrue
\mciteSetBstMidEndSepPunct{\mcitedefaultmidpunct}
{\mcitedefaultendpunct}{\mcitedefaultseppunct}\relax
\EndOfBibitem
\bibitem[Ameta \latin{et~al.}(2018)Ameta, Solanki, Benjamin, and
  Ameta]{Ameta2018}
Ameta,~R.; Solanki,~M.~S.; Benjamin,~S.; Ameta,~S.~C. \emph{Adv. Oxid. Process.
  Wastewater Treat. Emerg. Green Chem. Technol.}; Elsevier Inc., 2018; pp
  135--175\relax
\mciteBstWouldAddEndPuncttrue
\mciteSetBstMidEndSepPunct{\mcitedefaultmidpunct}
{\mcitedefaultendpunct}{\mcitedefaultseppunct}\relax
\EndOfBibitem
\bibitem[Sivula and Van De~Krol(2016)Sivula, and Van
  De~Krol]{sivula2016semiconducting}
Sivula,~K.; Van De~Krol,~R. Semiconducting materials for photoelectrochemical
  energy conversion. \emph{Nature Reviews Materials} \textbf{2016}, \emph{1},
  1--16\relax
\mciteBstWouldAddEndPuncttrue
\mciteSetBstMidEndSepPunct{\mcitedefaultmidpunct}
{\mcitedefaultendpunct}{\mcitedefaultseppunct}\relax
\EndOfBibitem
\bibitem[Chiang and Lin(2015)Chiang, and Lin]{chiang2015enhanced}
Chiang,~M.-Y.; Lin,~H.-N. Enhanced photocatalysis of ZnO nanowires co-modified
  with cuprous oxide and silver nanoparticles. \emph{Materials Letters}
  \textbf{2015}, \emph{160}, 440--443\relax
\mciteBstWouldAddEndPuncttrue
\mciteSetBstMidEndSepPunct{\mcitedefaultmidpunct}
{\mcitedefaultendpunct}{\mcitedefaultseppunct}\relax
\EndOfBibitem
\bibitem[Sun \latin{et~al.}(2016)Sun, Zhu, Zheng, Jiang, Yin, Wang, Qiu, Yuan,
  Wu, Wu, \latin{et~al.} others]{sun2016preparation}
Sun,~G.; Zhu,~C.; Zheng,~J.; Jiang,~B.; Yin,~H.; Wang,~H.; Qiu,~S.; Yuan,~J.;
  Wu,~M.; Wu,~W.; others Preparation of spherical and dendritic CdS@ TiO2
  hollow double-shelled nanoparticles for photocatalysis. \emph{Materials
  Letters} \textbf{2016}, \emph{166}, 113--115\relax
\mciteBstWouldAddEndPuncttrue
\mciteSetBstMidEndSepPunct{\mcitedefaultmidpunct}
{\mcitedefaultendpunct}{\mcitedefaultseppunct}\relax
\EndOfBibitem
\bibitem[Han \latin{et~al.}(2022)Han, Shawkat, Lee, Park, Li, Chung, Yoo,
  Mofid, Sattar, Choudhary, \latin{et~al.} others]{han2022wafer}
Han,~S.~S.; Shawkat,~M.~S.; Lee,~Y.~H.; Park,~G.; Li,~H.; Chung,~H.-S.;
  Yoo,~C.; Mofid,~S.~A.; Sattar,~S.; Choudhary,~N.; others Wafer-Scale Anion
  Exchange Conversion of Nonlayered PtS Films to van der Waals Two-Dimensional
  PtTe2 Layers with Negative Photoresponsiveness. \emph{Chemistry of Materials}
  \textbf{2022}, \emph{34}, 6996--7005\relax
\mciteBstWouldAddEndPuncttrue
\mciteSetBstMidEndSepPunct{\mcitedefaultmidpunct}
{\mcitedefaultendpunct}{\mcitedefaultseppunct}\relax
\EndOfBibitem
\bibitem[Han \latin{et~al.}(2022)Han, Ko, Shawkat, Shum, Bae, Chung, Ma,
  Sattar, Hafiz, Mahfuz, \latin{et~al.} others]{han2022peel}
Han,~S.~S.; Ko,~T.-J.; Shawkat,~M.~S.; Shum,~A.~K.; Bae,~T.-S.; Chung,~H.-S.;
  Ma,~J.; Sattar,~S.; Hafiz,~S.~B.; Mahfuz,~M. M.~A.; others Peel-and-Stick
  Integration of Atomically Thin Nonlayered PtS Semiconductors for
  Multidimensionally Stretchable Electronic Devices. \emph{ACS Applied
  Materials \& Interfaces} \textbf{2022}, \emph{14}, 20268--20279\relax
\mciteBstWouldAddEndPuncttrue
\mciteSetBstMidEndSepPunct{\mcitedefaultmidpunct}
{\mcitedefaultendpunct}{\mcitedefaultseppunct}\relax
\EndOfBibitem
\bibitem[Lin \latin{et~al.}(2022)Lin, Cai, Fu, Sattar, Wang, Wan, Tseng, Yang,
  Aljarb, Jiang, \latin{et~al.} others]{lin2022direct}
Lin,~C.; Cai,~L.; Fu,~J.-H.; Sattar,~S.; Wang,~Q.; Wan,~Y.; Tseng,~C.-C.;
  Yang,~C.-W.; Aljarb,~A.; Jiang,~K.; others Direct band gap in multilayer
  transition metal dichalcogenide nanoscrolls with enhanced photoluminescence.
  \emph{ACS Materials Letters} \textbf{2022}, \emph{4}, 1547--1555\relax
\mciteBstWouldAddEndPuncttrue
\mciteSetBstMidEndSepPunct{\mcitedefaultmidpunct}
{\mcitedefaultendpunct}{\mcitedefaultseppunct}\relax
\EndOfBibitem
\bibitem[Ming \latin{et~al.}(2017)Ming, Davies, Liu, Caillol, \latin{et~al.}
  others]{ming2017removal}
Ming,~T.; Davies,~P.; Liu,~W.; Caillol,~S.; others Removal of non-CO2
  greenhouse gases by large-scale atmospheric solar photocatalysis.
  \emph{Progress in Energy and Combustion Science} \textbf{2017}, \emph{60},
  68--96\relax
\mciteBstWouldAddEndPuncttrue
\mciteSetBstMidEndSepPunct{\mcitedefaultmidpunct}
{\mcitedefaultendpunct}{\mcitedefaultseppunct}\relax
\EndOfBibitem
\bibitem[Khan \latin{et~al.}(2019)Khan, Saeed, and Khan]{khan2019nanoparticles}
Khan,~I.; Saeed,~K.; Khan,~I. Nanoparticles: Properties, applications and
  toxicities. \emph{Arabian journal of chemistry} \textbf{2019}, \emph{12},
  908--931\relax
\mciteBstWouldAddEndPuncttrue
\mciteSetBstMidEndSepPunct{\mcitedefaultmidpunct}
{\mcitedefaultendpunct}{\mcitedefaultseppunct}\relax
\EndOfBibitem
\bibitem[Khataee \latin{et~al.}(2015)Khataee, Arefi-Oskoui, Fathinia, Fazli,
  Hojaghan, Hanifehpour, and Joo]{khataee2015photocatalysis}
Khataee,~A.; Arefi-Oskoui,~S.; Fathinia,~M.; Fazli,~A.; Hojaghan,~A.~S.;
  Hanifehpour,~Y.; Joo,~S.~W. Photocatalysis of sulfasalazine using Gd-doped
  PbSe nanoparticles under visible light irradiation: kinetics, intermediate
  identification and phyto-toxicological studies. \emph{Journal of Industrial
  and Engineering Chemistry} \textbf{2015}, \emph{30}, 134--146\relax
\mciteBstWouldAddEndPuncttrue
\mciteSetBstMidEndSepPunct{\mcitedefaultmidpunct}
{\mcitedefaultendpunct}{\mcitedefaultseppunct}\relax
\EndOfBibitem
\bibitem[Samu \latin{et~al.}(2017)Samu, Veres, Endr{\H{o}}di, Varga, Rajeshwar,
  and Jan{\'a}ky]{samu2017bandgap}
Samu,~G.~F.; Veres,~{\'A}.; Endr{\H{o}}di,~B.; Varga,~E.; Rajeshwar,~K.;
  Jan{\'a}ky,~C. Bandgap-engineered quaternary MxBi2- xTi2O7 (M: Fe, Mn)
  semiconductor nanoparticles: solution combustion synthesis, characterization,
  and photocatalysis. \emph{Applied Catalysis B: Environmental} \textbf{2017},
  \emph{208}, 148--160\relax
\mciteBstWouldAddEndPuncttrue
\mciteSetBstMidEndSepPunct{\mcitedefaultmidpunct}
{\mcitedefaultendpunct}{\mcitedefaultseppunct}\relax
\EndOfBibitem
\bibitem[Andrade \latin{et~al.}(2017)Andrade, Nascimento, Lima, Teixeira-Neto,
  Costa, and Gimenez]{andrade2017star}
Andrade,~G.~R.; Nascimento,~C.~C.; Lima,~Z.~M.; Teixeira-Neto,~E.;
  Costa,~L.~P.; Gimenez,~I.~F. Star-shaped ZnO/Ag hybrid nanostructures for
  enhanced photocatalysis and antibacterial activity. \emph{Applied Surface
  Science} \textbf{2017}, \emph{399}, 573--582\relax
\mciteBstWouldAddEndPuncttrue
\mciteSetBstMidEndSepPunct{\mcitedefaultmidpunct}
{\mcitedefaultendpunct}{\mcitedefaultseppunct}\relax
\EndOfBibitem
\bibitem[Liu \latin{et~al.}(2011)Liu, Xie, Li, Chen, Zou, and
  Zeng]{liu2011improvement}
Liu,~Y.; Xie,~C.; Li,~H.; Chen,~H.; Zou,~T.; Zeng,~D. Improvement of gaseous
  pollutant photocatalysis with WO3/TiO2 heterojunctional-electrical layered
  system. \emph{Journal of Hazardous Materials} \textbf{2011}, \emph{196},
  52--58\relax
\mciteBstWouldAddEndPuncttrue
\mciteSetBstMidEndSepPunct{\mcitedefaultmidpunct}
{\mcitedefaultendpunct}{\mcitedefaultseppunct}\relax
\EndOfBibitem
\bibitem[Liu \latin{et~al.}(2016)Liu, Zhang, Chen, and Wen]{liu2016loading}
Liu,~M.; Zhang,~D.-X.; Chen,~S.; Wen,~T. Loading Ag nanoparticles on Cd (II)
  boron imidazolate framework for photocatalysis. \emph{Journal of Solid State
  Chemistry} \textbf{2016}, \emph{237}, 32--35\relax
\mciteBstWouldAddEndPuncttrue
\mciteSetBstMidEndSepPunct{\mcitedefaultmidpunct}
{\mcitedefaultendpunct}{\mcitedefaultseppunct}\relax
\EndOfBibitem
\bibitem[Asapu \latin{et~al.}(2017)Asapu, Claes, Bals, Denys, Detavernier,
  Lenaerts, and Verbruggen]{asapu2017silver}
Asapu,~R.; Claes,~N.; Bals,~S.; Denys,~S.; Detavernier,~C.; Lenaerts,~S.;
  Verbruggen,~S.~W. Silver-polymer core-shell nanoparticles for ultrastable
  plasmon-enhanced photocatalysis. \emph{Applied Catalysis B: Environmental}
  \textbf{2017}, \emph{200}, 31--38\relax
\mciteBstWouldAddEndPuncttrue
\mciteSetBstMidEndSepPunct{\mcitedefaultmidpunct}
{\mcitedefaultendpunct}{\mcitedefaultseppunct}\relax
\EndOfBibitem
\bibitem[Aleksandrzak \latin{et~al.}(2017)Aleksandrzak, Kukulka, and
  Mijowska]{aleksandrzak2017graphitic}
Aleksandrzak,~M.; Kukulka,~W.; Mijowska,~E. Graphitic carbon nitride/graphene
  oxide/reduced graphene oxide nanocomposites for photoluminescence and
  photocatalysis. \emph{Applied Surface Science} \textbf{2017}, \emph{398},
  56--62\relax
\mciteBstWouldAddEndPuncttrue
\mciteSetBstMidEndSepPunct{\mcitedefaultmidpunct}
{\mcitedefaultendpunct}{\mcitedefaultseppunct}\relax
\EndOfBibitem
\bibitem[Zhang \latin{et~al.}(2017)Zhang, Zhang, Xie, Guo, Lyu, Li, Sun, Wang,
  and Guo]{zhang2017electrospun}
Zhang,~L.; Zhang,~Q.; Xie,~H.; Guo,~J.; Lyu,~H.; Li,~Y.; Sun,~Z.; Wang,~H.;
  Guo,~Z. Electrospun titania nanofibers segregated by graphene oxide for
  improved visible light photocatalysis. \emph{Applied Catalysis B:
  Environmental} \textbf{2017}, \emph{201}, 470--478\relax
\mciteBstWouldAddEndPuncttrue
\mciteSetBstMidEndSepPunct{\mcitedefaultmidpunct}
{\mcitedefaultendpunct}{\mcitedefaultseppunct}\relax
\EndOfBibitem
\bibitem[Li \latin{et~al.}(2017)Li, Zhang, Nie, Shao, and
  Hu]{li2017optimization}
Li,~B.; Zhang,~B.; Nie,~S.; Shao,~L.; Hu,~L. Optimization of plasmon-induced
  photocatalysis in electrospun Au/CeO2 hybrid nanofibers for selective
  oxidation of benzyl alcohol. \emph{Journal of Catalysis} \textbf{2017},
  \emph{348}, 256--264\relax
\mciteBstWouldAddEndPuncttrue
\mciteSetBstMidEndSepPunct{\mcitedefaultmidpunct}
{\mcitedefaultendpunct}{\mcitedefaultseppunct}\relax
\EndOfBibitem
\bibitem[Oliveira \latin{et~al.}(2015)Oliveira, Ferreira, Bertazzoli, and
  Longo]{oliveira2015remediation}
Oliveira,~H.~G.; Ferreira,~L.~H.; Bertazzoli,~R.; Longo,~C. Remediation of
  17-$\alpha$-ethinylestradiol aqueous solution by photocatalysis and
  electrochemically-assisted photocatalysis using TiO2 and TiO2/WO3 electrodes
  irradiated by a solar simulator. \emph{Water research} \textbf{2015},
  \emph{72}, 305--314\relax
\mciteBstWouldAddEndPuncttrue
\mciteSetBstMidEndSepPunct{\mcitedefaultmidpunct}
{\mcitedefaultendpunct}{\mcitedefaultseppunct}\relax
\EndOfBibitem
\bibitem[Fu \latin{et~al.}(2017)Fu, Li, Luo, and Yang]{fu2017two}
Fu,~C.-F.; Li,~X.; Luo,~Q.; Yang,~J. Two-dimensional multilayer M 2 CO 2 (M=
  Sc, Zr, Hf) as photocatalysts for hydrogen production from water splitting: A
  first principles study. \emph{Journal of Materials Chemistry A}
  \textbf{2017}, \emph{5}, 24972--24980\relax
\mciteBstWouldAddEndPuncttrue
\mciteSetBstMidEndSepPunct{\mcitedefaultmidpunct}
{\mcitedefaultendpunct}{\mcitedefaultseppunct}\relax
\EndOfBibitem
\bibitem[Ren and Innocenzi(2021)Ren, and Innocenzi]{ren20212d}
Ren,~J.; Innocenzi,~P. 2D boron nitride heterostructures: recent advances and
  future challenges. \emph{Small Structures} \textbf{2021}, \emph{2},
  2100068\relax
\mciteBstWouldAddEndPuncttrue
\mciteSetBstMidEndSepPunct{\mcitedefaultmidpunct}
{\mcitedefaultendpunct}{\mcitedefaultseppunct}\relax
\EndOfBibitem
\bibitem[Zhang \latin{et~al.}(2013)Zhang, Xie, Wang, Zhang, Pan, and
  Xie]{zhang2013enhanced}
Zhang,~X.; Xie,~X.; Wang,~H.; Zhang,~J.; Pan,~B.; Xie,~Y. Enhanced
  photoresponsive ultrathin graphitic-phase C3N4 nanosheets for bioimaging.
  \emph{Journal of the American Chemical Society} \textbf{2013}, \emph{135},
  18--21\relax
\mciteBstWouldAddEndPuncttrue
\mciteSetBstMidEndSepPunct{\mcitedefaultmidpunct}
{\mcitedefaultendpunct}{\mcitedefaultseppunct}\relax
\EndOfBibitem
\bibitem[Wang \latin{et~al.}(2019)Wang, Sun, and Liu]{wang2019chemical}
Wang,~X.; Sun,~Y.; Liu,~K. Chemical and structural stability of 2D layered
  materials. \emph{2D Materials} \textbf{2019}, \emph{6}, 042001\relax
\mciteBstWouldAddEndPuncttrue
\mciteSetBstMidEndSepPunct{\mcitedefaultmidpunct}
{\mcitedefaultendpunct}{\mcitedefaultseppunct}\relax
\EndOfBibitem
\bibitem[Zhao \latin{et~al.}(2022)Zhao, Yan, and Lee]{zhao2022recent}
Zhao,~Y.; Yan,~Y.; Lee,~J.-M. Recent progress on transition metal diselenides
  from formation and modification to applications. \emph{Nanoscale}
  \textbf{2022}, \emph{14}, 1075--1095\relax
\mciteBstWouldAddEndPuncttrue
\mciteSetBstMidEndSepPunct{\mcitedefaultmidpunct}
{\mcitedefaultendpunct}{\mcitedefaultseppunct}\relax
\EndOfBibitem
\bibitem[Guo \latin{et~al.}(2019)Guo, Xiao, Wang, and Zhang]{guo20192d}
Guo,~B.; Xiao,~Q.-l.; Wang,~S.-h.; Zhang,~H. 2D layered materials: synthesis,
  nonlinear optical properties, and device applications. \emph{Laser \&
  Photonics Reviews} \textbf{2019}, \emph{13}, 1800327\relax
\mciteBstWouldAddEndPuncttrue
\mciteSetBstMidEndSepPunct{\mcitedefaultmidpunct}
{\mcitedefaultendpunct}{\mcitedefaultseppunct}\relax
\EndOfBibitem
\bibitem[Morscher \latin{et~al.}(2006)Morscher, Corso, Greber, and
  Osterwalder]{MORSCHER20063280}
Morscher,~M.; Corso,~M.; Greber,~T.; Osterwalder,~J. Formation of single layer
  h-BN on Pd(111). \emph{Surface Science} \textbf{2006}, \emph{600},
  3280--3284\relax
\mciteBstWouldAddEndPuncttrue
\mciteSetBstMidEndSepPunct{\mcitedefaultmidpunct}
{\mcitedefaultendpunct}{\mcitedefaultseppunct}\relax
\EndOfBibitem
\bibitem[Li \latin{et~al.}(2019)Li, Lai, Zeng, Huang, Qin, Zhang, Cheng, Liu,
  Yi, Zhou, \latin{et~al.} others]{li2019black}
Li,~B.; Lai,~C.; Zeng,~G.; Huang,~D.; Qin,~L.; Zhang,~M.; Cheng,~M.; Liu,~X.;
  Yi,~H.; Zhou,~C.; others Black phosphorus, a rising star 2D nanomaterial in
  the post-graphene era: synthesis, properties, modifications, and
  photocatalysis applications. \emph{Small} \textbf{2019}, \emph{15},
  1804565\relax
\mciteBstWouldAddEndPuncttrue
\mciteSetBstMidEndSepPunct{\mcitedefaultmidpunct}
{\mcitedefaultendpunct}{\mcitedefaultseppunct}\relax
\EndOfBibitem
\bibitem[Deng \latin{et~al.}(2018)Deng, Frisenda, Li, Chen, Castellanos-Gomez,
  and Xia]{deng2018progress}
Deng,~B.; Frisenda,~R.; Li,~C.; Chen,~X.; Castellanos-Gomez,~A.; Xia,~F.
  Progress on black phosphorus photonics. \emph{Advanced Optical Materials}
  \textbf{2018}, \emph{6}, 1800365\relax
\mciteBstWouldAddEndPuncttrue
\mciteSetBstMidEndSepPunct{\mcitedefaultmidpunct}
{\mcitedefaultendpunct}{\mcitedefaultseppunct}\relax
\EndOfBibitem
\bibitem[Bhimanapati \latin{et~al.}(2015)Bhimanapati, Lin, Meunier, Jung, Cha,
  Das, Xiao, Son, Strano, Cooper, \latin{et~al.} others]{bhimanapati2015recent}
Bhimanapati,~G.~R.; Lin,~Z.; Meunier,~V.; Jung,~Y.; Cha,~J.; Das,~S.; Xiao,~D.;
  Son,~Y.; Strano,~M.~S.; Cooper,~V.~R.; others Recent advances in
  two-dimensional materials beyond graphene. \emph{ACS nano} \textbf{2015},
  \emph{9}, 11509--11539\relax
\mciteBstWouldAddEndPuncttrue
\mciteSetBstMidEndSepPunct{\mcitedefaultmidpunct}
{\mcitedefaultendpunct}{\mcitedefaultseppunct}\relax
\EndOfBibitem
\bibitem[Shibata \latin{et~al.}(2011)Shibata, Takanashi, Nakamura, Fukuda,
  Ebina, and Sasaki]{shibata2011titanoniobate}
Shibata,~T.; Takanashi,~G.; Nakamura,~T.; Fukuda,~K.; Ebina,~Y.; Sasaki,~T.
  Titanoniobate and niobate nanosheet photocatalysts: superior photoinduced
  hydrophilicity and enhanced thermal stability of unilamellar Nb 3 O 8
  nanosheet. \emph{Energy \& Environmental Science} \textbf{2011}, \emph{4},
  535--542\relax
\mciteBstWouldAddEndPuncttrue
\mciteSetBstMidEndSepPunct{\mcitedefaultmidpunct}
{\mcitedefaultendpunct}{\mcitedefaultseppunct}\relax
\EndOfBibitem
\bibitem[Alzakia and Tan(2021)Alzakia, and Tan]{alzakia2021liquid}
Alzakia,~F.~I.; Tan,~S.~C. Liquid-exfoliated 2D materials for optoelectronic
  applications. \emph{Advanced Science} \textbf{2021}, \emph{8}, 2003864\relax
\mciteBstWouldAddEndPuncttrue
\mciteSetBstMidEndSepPunct{\mcitedefaultmidpunct}
{\mcitedefaultendpunct}{\mcitedefaultseppunct}\relax
\EndOfBibitem
\bibitem[Yan \latin{et~al.}(2017)Yan, Han, Qian, Liu, Dong, and
  Xi]{yan2017preparation}
Yan,~J.; Han,~X.; Qian,~J.; Liu,~J.; Dong,~X.; Xi,~F. Preparation of 2D
  graphitic carbon nitride nanosheets by a green exfoliation approach and the
  enhanced photocatalytic performance. \emph{Journal of Materials Science}
  \textbf{2017}, \emph{52}, 13091--13102\relax
\mciteBstWouldAddEndPuncttrue
\mciteSetBstMidEndSepPunct{\mcitedefaultmidpunct}
{\mcitedefaultendpunct}{\mcitedefaultseppunct}\relax
\EndOfBibitem
\bibitem[Acik and Chabal(2013)Acik, and Chabal]{acik2013review}
Acik,~M.; Chabal,~Y.~J. A review on thermal exfoliation of graphene oxide.
  \emph{Journal of Materials Science Research} \textbf{2013}, \emph{2},
  101\relax
\mciteBstWouldAddEndPuncttrue
\mciteSetBstMidEndSepPunct{\mcitedefaultmidpunct}
{\mcitedefaultendpunct}{\mcitedefaultseppunct}\relax
\EndOfBibitem
\bibitem[Xu \latin{et~al.}(2013)Xu, Zhang, Shi, and Zhu]{xu2013chemical}
Xu,~J.; Zhang,~L.; Shi,~R.; Zhu,~Y. Chemical exfoliation of graphitic carbon
  nitride for efficient heterogeneous photocatalysis. \emph{Journal of
  Materials Chemistry A} \textbf{2013}, \emph{1}, 14766--14772\relax
\mciteBstWouldAddEndPuncttrue
\mciteSetBstMidEndSepPunct{\mcitedefaultmidpunct}
{\mcitedefaultendpunct}{\mcitedefaultseppunct}\relax
\EndOfBibitem
\bibitem[Meng and Park(2012)Meng, and Park]{meng2012effect}
Meng,~L.-Y.; Park,~S.-J. Effect of exfoliation temperature on carbon dioxide
  capture of graphene nanoplates. \emph{Journal of colloid and interface
  science} \textbf{2012}, \emph{386}, 285--290\relax
\mciteBstWouldAddEndPuncttrue
\mciteSetBstMidEndSepPunct{\mcitedefaultmidpunct}
{\mcitedefaultendpunct}{\mcitedefaultseppunct}\relax
\EndOfBibitem
\bibitem[Hao \latin{et~al.}(2005)Hao, Pan, and Wang]{hao2005photocatalytic}
Hao,~W.; Pan,~F.; Wang,~T. Photocatalytic activity TiO2 granular films prepared
  by layer-by-layer self-assembly method. \emph{Journal of materials science}
  \textbf{2005}, \emph{40}, 1251--1253\relax
\mciteBstWouldAddEndPuncttrue
\mciteSetBstMidEndSepPunct{\mcitedefaultmidpunct}
{\mcitedefaultendpunct}{\mcitedefaultseppunct}\relax
\EndOfBibitem
\bibitem[Zhou \latin{et~al.}(2015)Zhou, Zhang, Wang, Miao, Gu, and
  Jiao]{zhou2015self}
Zhou,~H.; Zhang,~H.; Wang,~Y.; Miao,~Y.; Gu,~L.; Jiao,~Z. Self-assembly and
  template-free synthesis of ZnO hierarchical nanostructures and their
  photocatalytic properties. \emph{Journal of colloid and interface science}
  \textbf{2015}, \emph{448}, 367--373\relax
\mciteBstWouldAddEndPuncttrue
\mciteSetBstMidEndSepPunct{\mcitedefaultmidpunct}
{\mcitedefaultendpunct}{\mcitedefaultseppunct}\relax
\EndOfBibitem
\bibitem[Sun \latin{et~al.}(2015)Sun, Gao, Lei, and Xie]{sun2015atomically}
Sun,~Y.; Gao,~S.; Lei,~F.; Xie,~Y. Atomically-thin two-dimensional sheets for
  understanding active sites in catalysis. \emph{Chemical Society Reviews}
  \textbf{2015}, \emph{44}, 623--636\relax
\mciteBstWouldAddEndPuncttrue
\mciteSetBstMidEndSepPunct{\mcitedefaultmidpunct}
{\mcitedefaultendpunct}{\mcitedefaultseppunct}\relax
\EndOfBibitem
\bibitem[Wang \latin{et~al.}(2021)Wang, Xia, Wang, Zhou, Liu, Zhang, Wang,
  Huang, Chen, Wu, \latin{et~al.} others]{wang2021controllable}
Wang,~Z.; Xia,~H.; Wang,~P.; Zhou,~X.; Liu,~C.; Zhang,~Q.; Wang,~F.; Huang,~M.;
  Chen,~S.; Wu,~P.; others Controllable doping in 2D layered materials.
  \emph{Advanced Materials} \textbf{2021}, \emph{33}, 2104942\relax
\mciteBstWouldAddEndPuncttrue
\mciteSetBstMidEndSepPunct{\mcitedefaultmidpunct}
{\mcitedefaultendpunct}{\mcitedefaultseppunct}\relax
\EndOfBibitem
\bibitem[Wang \latin{et~al.}(2022)Wang, Ding, Arif, Jiang, and
  Zeng]{wang20222d}
Wang,~Y.; Ding,~Z.; Arif,~N.; Jiang,~W.-C.; Zeng,~Y.-J. 2D material based
  heterostructures for solar light driven photocatalytic H 2 production.
  \emph{Materials Advances} \textbf{2022}, \emph{3}, 3389--3417\relax
\mciteBstWouldAddEndPuncttrue
\mciteSetBstMidEndSepPunct{\mcitedefaultmidpunct}
{\mcitedefaultendpunct}{\mcitedefaultseppunct}\relax
\EndOfBibitem
\bibitem[Xiong \latin{et~al.}(2018)Xiong, Di, Xia, Zhu, and
  Li]{xiong2018surface}
Xiong,~J.; Di,~J.; Xia,~J.; Zhu,~W.; Li,~H. Surface defect engineering in 2D
  nanomaterials for photocatalysis. \emph{Advanced Functional Materials}
  \textbf{2018}, \emph{28}, 1801983\relax
\mciteBstWouldAddEndPuncttrue
\mciteSetBstMidEndSepPunct{\mcitedefaultmidpunct}
{\mcitedefaultendpunct}{\mcitedefaultseppunct}\relax
\EndOfBibitem
\bibitem[Ma \latin{et~al.}(2021)Ma, Qiu, Zhang, and Xing]{ma2021vacancy}
Ma,~Y.; Qiu,~B.; Zhang,~J.; Xing,~M. Vacancy engineering of ultrathin 2D
  materials for photocatalytic CO2 reduction. \emph{ChemNanoMat} \textbf{2021},
  \emph{7}, 368--379\relax
\mciteBstWouldAddEndPuncttrue
\mciteSetBstMidEndSepPunct{\mcitedefaultmidpunct}
{\mcitedefaultendpunct}{\mcitedefaultseppunct}\relax
\EndOfBibitem
\bibitem[Luo and Wu(2023)Luo, and Wu]{Luo2023}
Luo,~Y.; Wu,~Y. {Defect Engineering of Nanomaterials for Catalysis}. 2023;
  \url{https://www.mdpi.com/2079-4991/13/6/1116/htm
  https://www.mdpi.com/2079-4991/13/6/1116}\relax
\mciteBstWouldAddEndPuncttrue
\mciteSetBstMidEndSepPunct{\mcitedefaultmidpunct}
{\mcitedefaultendpunct}{\mcitedefaultseppunct}\relax
\EndOfBibitem
\bibitem[Milovzorov(2004)]{Milovzorov2004}
Milovzorov,~D. {Defect engineering and control in nanocrystalline silicon}.
  Proc. - Electrochem. Soc. 2004; pp 226--233\relax
\mciteBstWouldAddEndPuncttrue
\mciteSetBstMidEndSepPunct{\mcitedefaultmidpunct}
{\mcitedefaultendpunct}{\mcitedefaultseppunct}\relax
\EndOfBibitem
\bibitem[Zheng \latin{et~al.}(2007)Zheng, Chen, Zhan, Lin, Zheng, Wei, Zhu, and
  Zhu]{Zheng2007}
Zheng,~Y.; Chen,~C.; Zhan,~Y.; Lin,~X.; Zheng,~Q.; Wei,~K.; Zhu,~J.; Zhu,~Y.
  {Luminescence and photocatalytic activity of ZnO nanocrystals: Correlation
  between structure and property}. \emph{Inorg. Chem.} \textbf{2007},
  \emph{46}, 6675--6682\relax
\mciteBstWouldAddEndPuncttrue
\mciteSetBstMidEndSepPunct{\mcitedefaultmidpunct}
{\mcitedefaultendpunct}{\mcitedefaultseppunct}\relax
\EndOfBibitem
\bibitem[Marschall and Wang(2014)Marschall, and Wang]{marschall2014non}
Marschall,~R.; Wang,~L. Non-metal doping of transition metal oxides for
  visible-light photocatalysis. \emph{Catalysis Today} \textbf{2014},
  \emph{225}, 111--135\relax
\mciteBstWouldAddEndPuncttrue
\mciteSetBstMidEndSepPunct{\mcitedefaultmidpunct}
{\mcitedefaultendpunct}{\mcitedefaultseppunct}\relax
\EndOfBibitem
\bibitem[Pető \latin{et~al.}(2018)Pető, Oll{\'{a}}r, Vancs{\'{o}}, Popov,
  Magda, Dobrik, Hwang, Sorokin, and Tapaszt{\'{o}}]{Peto2018}
Pető,~J.; Oll{\'{a}}r,~T.; Vancs{\'{o}},~P.; Popov,~Z.~I.; Magda,~G.~Z.;
  Dobrik,~G.; Hwang,~C.; Sorokin,~P.~B.; Tapaszt{\'{o}},~L. {Spontaneous doping
  of the basal plane of MoS 2 single layers through oxygen substitution under
  ambient conditions}. \emph{Nat. Chem.} \textbf{2018}, \emph{10},
  1246--1251\relax
\mciteBstWouldAddEndPuncttrue
\mciteSetBstMidEndSepPunct{\mcitedefaultmidpunct}
{\mcitedefaultendpunct}{\mcitedefaultseppunct}\relax
\EndOfBibitem
\bibitem[Li \latin{et~al.}(2022)Li, Chen, Yu, Li, Xiong, Pam, Zhang, and
  Ang]{Li2022}
Li,~Y.; Chen,~S.; Yu,~Z.; Li,~S.; Xiong,~Y.; Pam,~M.~E.; Zhang,~Y.~W.;
  Ang,~K.~W. {In-Memory Computing using Memristor Arrays with Ultrathin 2D
  PdSeOx/PdSe2 Heterostructure}. \emph{Adv. Mater.} \textbf{2022}, \emph{34},
  2201488\relax
\mciteBstWouldAddEndPuncttrue
\mciteSetBstMidEndSepPunct{\mcitedefaultmidpunct}
{\mcitedefaultendpunct}{\mcitedefaultseppunct}\relax
\EndOfBibitem
\bibitem[Cho \latin{et~al.}(2015)Cho, Kim, Kim, Zhao, Seok, Keum, Baik, Choe,
  Chang, Suenaga, Kim, Lee, and Yang]{Cho2015}
Cho,~S.; Kim,~S.; Kim,~J.~H.; Zhao,~J.; Seok,~J.; Keum,~D.~H.; Baik,~J.;
  Choe,~D.~H.; Chang,~K.~J.; Suenaga,~K.; Kim,~S.~W.; Lee,~Y.~H.; Yang,~H.
  {Phase patterning for ohmic homojunction contact in MoTe2}. \emph{Science
  (80-. ).} \textbf{2015}, \emph{349}, 625--628\relax
\mciteBstWouldAddEndPuncttrue
\mciteSetBstMidEndSepPunct{\mcitedefaultmidpunct}
{\mcitedefaultendpunct}{\mcitedefaultseppunct}\relax
\EndOfBibitem
\bibitem[Ding \latin{et~al.}(2019)Ding, Peng, Zhou, Gong, Slaven, Loh, Lim, and
  Leong]{Ding2019}
Ding,~X.; Peng,~F.; Zhou,~J.; Gong,~W.; Slaven,~G.; Loh,~K.~P.; Lim,~C.~T.;
  Leong,~D.~T. {Defect engineered bioactive transition metals dichalcogenides
  quantum dots}. \emph{Nat. Commun.} \textbf{2019}, \emph{10}, 1--13\relax
\mciteBstWouldAddEndPuncttrue
\mciteSetBstMidEndSepPunct{\mcitedefaultmidpunct}
{\mcitedefaultendpunct}{\mcitedefaultseppunct}\relax
\EndOfBibitem
\bibitem[Zhou \latin{et~al.}(2020)Zhou, Zhang, Song, Lin, Zhou, Suenaga, Zhou,
  Liu, Liu, Lou, and Fan]{Zhou2020a}
Zhou,~Y.; Zhang,~J.; Song,~E.; Lin,~J.; Zhou,~J.; Suenaga,~K.; Zhou,~W.;
  Liu,~Z.; Liu,~J.; Lou,~J.; Fan,~H.~J. {Enhanced performance of in-plane
  transition metal dichalcogenides monolayers by configuring local atomic
  structures}. \emph{Nat. Commun.} \textbf{2020}, \emph{11}, 1--8\relax
\mciteBstWouldAddEndPuncttrue
\mciteSetBstMidEndSepPunct{\mcitedefaultmidpunct}
{\mcitedefaultendpunct}{\mcitedefaultseppunct}\relax
\EndOfBibitem
\bibitem[Zhang \latin{et~al.}(2017)Zhang, Liao, Liu, Kang, Zhang, Du, Li,
  Zhang, Xiao, Liu, Ou, Liu, Gu, and Zhang]{Zhang2017}
Zhang,~X.; Liao,~Q.; Liu,~S.; Kang,~Z.; Zhang,~Z.; Du,~J.; Li,~F.; Zhang,~S.;
  Xiao,~J.; Liu,~B.; Ou,~Y.; Liu,~X.; Gu,~L.; Zhang,~Y.
  {Poly(4-styrenesulfonate)-induced sulfur vacancy self-healing strategy for
  monolayer MoS 2 homojunction photodiode}. \emph{Nat. Commun.} \textbf{2017},
  \emph{8}, 1--8\relax
\mciteBstWouldAddEndPuncttrue
\mciteSetBstMidEndSepPunct{\mcitedefaultmidpunct}
{\mcitedefaultendpunct}{\mcitedefaultseppunct}\relax
\EndOfBibitem
\bibitem[Lu \latin{et~al.}(2015)Lu, Carvalho, Chan, Liu, Liu, Tok, Loh, {Castro
  Neto}, and Sow]{Lu2015}
Lu,~J.; Carvalho,~A.; Chan,~X.~K.; Liu,~H.; Liu,~B.; Tok,~E.~S.; Loh,~K.~P.;
  {Castro Neto},~A.~H.; Sow,~C.~H. {Atomic healing of defects in transition
  metal dichalcogenides}. \emph{Nano Lett.} \textbf{2015}, \emph{15},
  3524--3532\relax
\mciteBstWouldAddEndPuncttrue
\mciteSetBstMidEndSepPunct{\mcitedefaultmidpunct}
{\mcitedefaultendpunct}{\mcitedefaultseppunct}\relax
\EndOfBibitem
\bibitem[Liu \latin{et~al.}(2024)Liu, Chen, Zhao, Long, Chen, Lu, and
  Chen]{liu2024quadruple}
Liu,~P.; Chen,~H.; Zhao,~C.; Long,~D.; Chen,~W.; Lu,~M.; Chen,~X. A quadruple
  transition metal dichalcogenide for variously synergetic electron behaviors
  during photocatalytic carbon dioxide reduction. \emph{Applied Surface
  Science} \textbf{2024}, \emph{659}, 159887\relax
\mciteBstWouldAddEndPuncttrue
\mciteSetBstMidEndSepPunct{\mcitedefaultmidpunct}
{\mcitedefaultendpunct}{\mcitedefaultseppunct}\relax
\EndOfBibitem
\bibitem[Kudo \latin{et~al.}(2007)Kudo, Niishiro, Iwase, and Kato]{Kudo}
Kudo,~A.; Niishiro,~R.; Iwase,~A.; Kato,~H. {Effects of doping of metal cations
  on morphology, activity, and visible light response of photocatalysts}.
  \emph{Chem. Phys.} \textbf{2007}, \emph{339}, 104--110\relax
\mciteBstWouldAddEndPuncttrue
\mciteSetBstMidEndSepPunct{\mcitedefaultmidpunct}
{\mcitedefaultendpunct}{\mcitedefaultseppunct}\relax
\EndOfBibitem
\bibitem[Colmenares \latin{et~al.}(2006)Colmenares, Aramend{\'{i}}a, Marinas,
  Marinas, and Urbano]{Colmenares2006}
Colmenares,~J.~C.; Aramend{\'{i}}a,~M.~A.; Marinas,~A.; Marinas,~J.~M.;
  Urbano,~F.~J. {Synthesis, characterization and photocatalytic activity of
  different metal-doped titania systems}. \emph{Appl. Catal. A Gen.}
  \textbf{2006}, \emph{306}, 120--127\relax
\mciteBstWouldAddEndPuncttrue
\mciteSetBstMidEndSepPunct{\mcitedefaultmidpunct}
{\mcitedefaultendpunct}{\mcitedefaultseppunct}\relax
\EndOfBibitem
\bibitem[Wilke and Breuer(1999)Wilke, and Breuer]{Wilke1999}
Wilke,~K.; Breuer,~H.~D. {The influence of transition metal doping on the
  physical and photocatalytic properties of titania}. \emph{J. Photochem.
  Photobiol. A Chem.} \textbf{1999}, \emph{121}, 49--53\relax
\mciteBstWouldAddEndPuncttrue
\mciteSetBstMidEndSepPunct{\mcitedefaultmidpunct}
{\mcitedefaultendpunct}{\mcitedefaultseppunct}\relax
\EndOfBibitem
\bibitem[Kudo \latin{et~al.}(2007)Kudo, Niishiro, Iwase, and Kato]{Kudo2007}
Kudo,~A.; Niishiro,~R.; Iwase,~A.; Kato,~H. {Effects of doping of metal cations
  on morphology, activity, and visible light response of photocatalysts}.
  \emph{Chem. Phys.} \textbf{2007}, \emph{339}, 104--110\relax
\mciteBstWouldAddEndPuncttrue
\mciteSetBstMidEndSepPunct{\mcitedefaultmidpunct}
{\mcitedefaultendpunct}{\mcitedefaultseppunct}\relax
\EndOfBibitem
\bibitem[Bouras \latin{et~al.}(2007)Bouras, Stathatos, and Lianos]{Bouras2007}
Bouras,~P.; Stathatos,~E.; Lianos,~P. {Pure versus metal-ion-doped
  nanocrystalline titania for photocatalysis}. \emph{Appl. Catal. B Environ.}
  \textbf{2007}, \emph{73}, 51--59\relax
\mciteBstWouldAddEndPuncttrue
\mciteSetBstMidEndSepPunct{\mcitedefaultmidpunct}
{\mcitedefaultendpunct}{\mcitedefaultseppunct}\relax
\EndOfBibitem
\bibitem[Yal{\c{c}}in \latin{et~al.}(2010)Yal{\c{c}}in, Kili{\c{c}}, and
  {\c{C}}inar]{Yalcin2010}
Yal{\c{c}}in,~Y.; Kili{\c{c}},~M.; {\c{C}}inar,~Z. {The role of non-metal
  doping in TiO2 photocatalysis}. J. Adv. Oxid. Technol. 2010; pp
  281--296\relax
\mciteBstWouldAddEndPuncttrue
\mciteSetBstMidEndSepPunct{\mcitedefaultmidpunct}
{\mcitedefaultendpunct}{\mcitedefaultseppunct}\relax
\EndOfBibitem
\bibitem[Bloh \latin{et~al.}(2012)Bloh, Dillert, and Bahnemann]{Bloh2012}
Bloh,~J.~Z.; Dillert,~R.; Bahnemann,~D.~W. {Designing optimal metal-doped
  photocatalysts: Correlation between photocatalytic activity, doping ratio,
  and particle size}. \emph{J. Phys. Chem. C} \textbf{2012}, \emph{116},
  25558--25562\relax
\mciteBstWouldAddEndPuncttrue
\mciteSetBstMidEndSepPunct{\mcitedefaultmidpunct}
{\mcitedefaultendpunct}{\mcitedefaultseppunct}\relax
\EndOfBibitem
\bibitem[Qi \latin{et~al.}(2020)Qi, Xing, Zada, Li, Wang, yuan Liu, Lin, and
  Wang]{Qi}
Qi,~K.; Xing,~X.; Zada,~A.; Li,~M.; Wang,~Q.; yuan Liu,~S.; Lin,~H.; Wang,~G.
  {Transition metal doped ZnO nanoparticles with enhanced photocatalytic and
  antibacterial performances: Experimental and DFT studies}. \emph{Ceram. Int.}
  \textbf{2020}, \emph{46}, 1494--1502\relax
\mciteBstWouldAddEndPuncttrue
\mciteSetBstMidEndSepPunct{\mcitedefaultmidpunct}
{\mcitedefaultendpunct}{\mcitedefaultseppunct}\relax
\EndOfBibitem
\bibitem[Dai \latin{et~al.}(2019)Dai, Liu, and Zhang]{dai2019strain}
Dai,~Z.; Liu,~L.; Zhang,~Z. Strain engineering of 2D materials: issues and
  opportunities at the interface. \emph{Advanced Materials} \textbf{2019},
  \emph{31}, 1805417\relax
\mciteBstWouldAddEndPuncttrue
\mciteSetBstMidEndSepPunct{\mcitedefaultmidpunct}
{\mcitedefaultendpunct}{\mcitedefaultseppunct}\relax
\EndOfBibitem
\bibitem[Peng \latin{et~al.}(2020)Peng, Chen, Fan, Srolovitz, and
  Lei]{peng2020strain}
Peng,~Z.; Chen,~X.; Fan,~Y.; Srolovitz,~D.~J.; Lei,~D. Strain engineering of 2D
  semiconductors and graphene: from strain fields to band-structure tuning and
  photonic applications. \emph{Light: Science \& Applications} \textbf{2020},
  \emph{9}, 190\relax
\mciteBstWouldAddEndPuncttrue
\mciteSetBstMidEndSepPunct{\mcitedefaultmidpunct}
{\mcitedefaultendpunct}{\mcitedefaultseppunct}\relax
\EndOfBibitem
\bibitem[Zeng \latin{et~al.}(2022)Zeng, Adit~Maark, and
  Peterson]{zeng2022strain}
Zeng,~C.; Adit~Maark,~T.; Peterson,~A.~A. Strain in catalysis: Rationalizing
  material, adsorbate, and site susceptibilities to biaxial lattice strain.
  \emph{The Journal of Physical Chemistry C} \textbf{2022}, \emph{126},
  20892--20902\relax
\mciteBstWouldAddEndPuncttrue
\mciteSetBstMidEndSepPunct{\mcitedefaultmidpunct}
{\mcitedefaultendpunct}{\mcitedefaultseppunct}\relax
\EndOfBibitem
\bibitem[Tan \latin{et~al.}(2017)Tan, Ji, Dong, Liu, Hou, and Li]{tan2017novel}
Tan,~X.; Ji,~Y.; Dong,~H.; Liu,~M.; Hou,~T.; Li,~Y. A novel metal-free
  two-dimensional material for photocatalytic water splitting--phosphorus
  nitride ($\gamma$-PN). \emph{RSC advances} \textbf{2017}, \emph{7},
  50239--50245\relax
\mciteBstWouldAddEndPuncttrue
\mciteSetBstMidEndSepPunct{\mcitedefaultmidpunct}
{\mcitedefaultendpunct}{\mcitedefaultseppunct}\relax
\EndOfBibitem
\bibitem[Ju \latin{et~al.}(2020)Ju, Bie, Tang, Shang, and Kou]{ju2020janus}
Ju,~L.; Bie,~M.; Tang,~X.; Shang,~J.; Kou,~L. Janus WSSe monolayer: an
  excellent photocatalyst for overall water splitting. \emph{ACS applied
  materials \& interfaces} \textbf{2020}, \emph{12}, 29335--29343\relax
\mciteBstWouldAddEndPuncttrue
\mciteSetBstMidEndSepPunct{\mcitedefaultmidpunct}
{\mcitedefaultendpunct}{\mcitedefaultseppunct}\relax
\EndOfBibitem
\bibitem[Liu \latin{et~al.}(2020)Liu, Shen, Gao, Lv, Ma, Wu, Wang, and
  Zhou]{liu2020gen3}
Liu,~J.; Shen,~Y.; Gao,~X.; Lv,~L.; Ma,~Y.; Wu,~S.; Wang,~X.; Zhou,~Z. GeN3
  monolayer: A promising 2D high-efficiency photo-hydrolytic catalyst with High
  carrier mobility transport anisotropy. \emph{Applied Catalysis B:
  Environmental} \textbf{2020}, \emph{279}, 119368\relax
\mciteBstWouldAddEndPuncttrue
\mciteSetBstMidEndSepPunct{\mcitedefaultmidpunct}
{\mcitedefaultendpunct}{\mcitedefaultseppunct}\relax
\EndOfBibitem
\bibitem[Lv \latin{et~al.}(2021)Lv, Shen, Gao, Liu, Wu, Ma, Wang, Gong, and
  Zhou]{lv2021strain}
Lv,~L.; Shen,~Y.; Gao,~X.; Liu,~J.; Wu,~S.; Ma,~Y.; Wang,~X.; Gong,~D.;
  Zhou,~Z. Strain engineering on the electrical properties and photocatalytic
  activity in gold sulfide monolayer. \emph{Applied Surface Science}
  \textbf{2021}, \emph{546}, 149066\relax
\mciteBstWouldAddEndPuncttrue
\mciteSetBstMidEndSepPunct{\mcitedefaultmidpunct}
{\mcitedefaultendpunct}{\mcitedefaultseppunct}\relax
\EndOfBibitem
\bibitem[Bai \latin{et~al.}(2015)Bai, Jiang, Zhang, and Xiong]{C5CS00064E}
Bai,~S.; Jiang,~J.; Zhang,~Q.; Xiong,~Y. Steering charge kinetics in
  photocatalysis: intersection of materials syntheses{,} characterization
  techniques and theoretical simulations. \emph{Chem. Soc. Rev.} \textbf{2015},
  \emph{44}, 2893--2939\relax
\mciteBstWouldAddEndPuncttrue
\mciteSetBstMidEndSepPunct{\mcitedefaultmidpunct}
{\mcitedefaultendpunct}{\mcitedefaultseppunct}\relax
\EndOfBibitem
\bibitem[Qu and Duan(2013)Qu, and Duan]{C2CS35355E}
Qu,~Y.; Duan,~X. Progress{,} challenge and perspective of heterogeneous
  photocatalysts. \emph{Chem. Soc. Rev.} \textbf{2013}, \emph{42},
  2568--2580\relax
\mciteBstWouldAddEndPuncttrue
\mciteSetBstMidEndSepPunct{\mcitedefaultmidpunct}
{\mcitedefaultendpunct}{\mcitedefaultseppunct}\relax
\EndOfBibitem
\bibitem[Bai and Xiong(2015)Bai, and Xiong]{C5CC02704G}
Bai,~S.; Xiong,~Y. Some recent developments in surface and interface design for
  photocatalytic and electrocatalytic hybrid structures. \emph{Chem. Commun.}
  \textbf{2015}, \emph{51}, 10261--10271\relax
\mciteBstWouldAddEndPuncttrue
\mciteSetBstMidEndSepPunct{\mcitedefaultmidpunct}
{\mcitedefaultendpunct}{\mcitedefaultseppunct}\relax
\EndOfBibitem
\bibitem[Bai \latin{et~al.}(2015)Bai, Jiang, Li, and
  Xiong]{https://doi.org/10.1002/cnma.201500069}
Bai,~S.; Jiang,~W.; Li,~Z.; Xiong,~Y. Surface and Interface Engineering in
  Photocatalysis. \emph{ChemNanoMat} \textbf{2015}, \emph{1}, 223--239\relax
\mciteBstWouldAddEndPuncttrue
\mciteSetBstMidEndSepPunct{\mcitedefaultmidpunct}
{\mcitedefaultendpunct}{\mcitedefaultseppunct}\relax
\EndOfBibitem
\bibitem[Bai \latin{et~al.}(2015)Bai, Jiang, Li, and Xiong]{bai2015surface}
Bai,~S.; Jiang,~W.; Li,~Z.; Xiong,~Y. Surface and interface engineering in
  photocatalysis. \emph{ChemNanoMat} \textbf{2015}, \emph{1}, 223--239\relax
\mciteBstWouldAddEndPuncttrue
\mciteSetBstMidEndSepPunct{\mcitedefaultmidpunct}
{\mcitedefaultendpunct}{\mcitedefaultseppunct}\relax
\EndOfBibitem
\bibitem[Zhou and Li(2012)Zhou, and Li]{zhou2012catalysis}
Zhou,~K.; Li,~Y. Catalysis based on nanocrystals with well-defined facets.
  \emph{Angewandte Chemie International Edition} \textbf{2012}, \emph{51},
  602--613\relax
\mciteBstWouldAddEndPuncttrue
\mciteSetBstMidEndSepPunct{\mcitedefaultmidpunct}
{\mcitedefaultendpunct}{\mcitedefaultseppunct}\relax
\EndOfBibitem
\bibitem[Liu \latin{et~al.}(2011)Liu, Jimmy, Lu, and Cheng]{liu2011crystal}
Liu,~G.; Jimmy,~C.~Y.; Lu,~G. Q.~M.; Cheng,~H.-M. Crystal facet engineering of
  semiconductor photocatalysts: motivations, advances and unique properties.
  \emph{Chemical Communications} \textbf{2011}, \emph{47}, 6763--6783\relax
\mciteBstWouldAddEndPuncttrue
\mciteSetBstMidEndSepPunct{\mcitedefaultmidpunct}
{\mcitedefaultendpunct}{\mcitedefaultseppunct}\relax
\EndOfBibitem
\bibitem[Liu \latin{et~al.}(2017)Liu, Lai, Li, Kang, Liu, Cao, and Yao]{JMC}
Liu,~L.; Lai,~Y.; Li,~H.; Kang,~L.; Liu,~J.; Cao,~Z.; Yao,~J. The role of
  dissolution in the synthesis of high-activity organic nanocatalysts in a wet
  chemical reaction. \emph{Journal of Materials Chemistry A} \textbf{2017},
  \emph{5}, 8029--8036\relax
\mciteBstWouldAddEndPuncttrue
\mciteSetBstMidEndSepPunct{\mcitedefaultmidpunct}
{\mcitedefaultendpunct}{\mcitedefaultseppunct}\relax
\EndOfBibitem
\bibitem[Jiang \latin{et~al.}(2012)Jiang, Zhao, Xiao, and
  Zhang]{jiang2012synthesis}
Jiang,~J.; Zhao,~K.; Xiao,~X.; Zhang,~L. Synthesis and facet-dependent
  photoreactivity of BiOCl single-crystalline nanosheets. \emph{Journal of the
  American Chemical Society} \textbf{2012}, \emph{134}, 4473--4476\relax
\mciteBstWouldAddEndPuncttrue
\mciteSetBstMidEndSepPunct{\mcitedefaultmidpunct}
{\mcitedefaultendpunct}{\mcitedefaultseppunct}\relax
\EndOfBibitem
\bibitem[Zhao and Cao(2013)Zhao, and Cao]{10.1246/cl.121148}
Zhao,~Q.; Cao,~T. {Large-area Synthesis of Single-crystal PbTiO3 Nanobelts and
  Nanoflakes}. \emph{Chemistry Letters} \textbf{2013}, \emph{42},
  338--340\relax
\mciteBstWouldAddEndPuncttrue
\mciteSetBstMidEndSepPunct{\mcitedefaultmidpunct}
{\mcitedefaultendpunct}{\mcitedefaultseppunct}\relax
\EndOfBibitem
\bibitem[Venkata~Subbaiah \latin{et~al.}(2016)Venkata~Subbaiah, Saji, and
  Tiwari]{venkata2016atomically}
Venkata~Subbaiah,~Y.; Saji,~K.; Tiwari,~A. Atomically thin MoS2: a versatile
  nongraphene 2D material. \emph{Advanced Functional Materials} \textbf{2016},
  \emph{26}, 2046--2069\relax
\mciteBstWouldAddEndPuncttrue
\mciteSetBstMidEndSepPunct{\mcitedefaultmidpunct}
{\mcitedefaultendpunct}{\mcitedefaultseppunct}\relax
\EndOfBibitem
\bibitem[Ma \latin{et~al.}(2016)Ma, Li, An, Feng, Chi, Liu, Zhang, and
  Sun]{ma2016ultrathin}
Ma,~X.; Li,~J.; An,~C.; Feng,~J.; Chi,~Y.; Liu,~J.; Zhang,~J.; Sun,~Y.
  Ultrathin Co (Ni)-doped MoS 2 nanosheets as catalytic promoters enabling
  efficient solar hydrogen production. \emph{Nano Research} \textbf{2016},
  \emph{9}, 2284--2293\relax
\mciteBstWouldAddEndPuncttrue
\mciteSetBstMidEndSepPunct{\mcitedefaultmidpunct}
{\mcitedefaultendpunct}{\mcitedefaultseppunct}\relax
\EndOfBibitem
\bibitem[Fu \latin{et~al.}(2016)Fu, Luo, Li, and Yang]{fu2016two}
Fu,~C.-F.; Luo,~Q.; Li,~X.; Yang,~J. Two-dimensional van der Waals
  nanocomposites as Z-scheme type photocatalysts for hydrogen production from
  overall water splitting. \emph{Journal of Materials Chemistry A}
  \textbf{2016}, \emph{4}, 18892--18898\relax
\mciteBstWouldAddEndPuncttrue
\mciteSetBstMidEndSepPunct{\mcitedefaultmidpunct}
{\mcitedefaultendpunct}{\mcitedefaultseppunct}\relax
\EndOfBibitem
\bibitem[Peng \latin{et~al.}(2017)Peng, Xiong, Sa, Zhou, Wen, Wu, Anpo, and
  Sun]{peng2017computational}
Peng,~Q.; Xiong,~R.; Sa,~B.; Zhou,~J.; Wen,~C.; Wu,~B.; Anpo,~M.; Sun,~Z.
  Computational mining of photocatalysts for water splitting hydrogen
  production: two-dimensional InSe-family monolayers. \emph{Catalysis Science
  \& Technology} \textbf{2017}, \emph{7}, 2744--2752\relax
\mciteBstWouldAddEndPuncttrue
\mciteSetBstMidEndSepPunct{\mcitedefaultmidpunct}
{\mcitedefaultendpunct}{\mcitedefaultseppunct}\relax
\EndOfBibitem
\bibitem[Jiao \latin{et~al.}(2018)Jiao, Ma, Zhou, Ng, Bell, Tretiak, and
  Du]{jiao2018ab}
Jiao,~Y.; Ma,~F.; Zhou,~L.; Ng,~Y.~H.; Bell,~J.; Tretiak,~S.; Du,~A. Ab initio
  study of two-dimensional PdPS as an ideal light harvester and promising
  catalyst for hydrogen evolution reaction. \emph{Materials today energy}
  \textbf{2018}, \emph{7}, 136--140\relax
\mciteBstWouldAddEndPuncttrue
\mciteSetBstMidEndSepPunct{\mcitedefaultmidpunct}
{\mcitedefaultendpunct}{\mcitedefaultseppunct}\relax
\EndOfBibitem
\bibitem[Obligacion and Putungan(2020)Obligacion, and
  Putungan]{obligacion20202d}
Obligacion,~J. K.~A.; Putungan,~D.~B. 2D 1T'-MoS2-WS2 van der Waals
  heterostructure for hydrogen evolution reaction: dispersion-corrected density
  functional theory calculations. \emph{Materials Research Express}
  \textbf{2020}, \emph{7}, 075506\relax
\mciteBstWouldAddEndPuncttrue
\mciteSetBstMidEndSepPunct{\mcitedefaultmidpunct}
{\mcitedefaultendpunct}{\mcitedefaultseppunct}\relax
\EndOfBibitem
\bibitem[Guo \latin{et~al.}(2021)Guo, Li, Li, and Wei]{guo2021boosting}
Guo,~S.; Li,~X.; Li,~J.; Wei,~B. Boosting photocatalytic hydrogen production
  from water by photothermally induced biphase systems. \emph{Nature
  communications} \textbf{2021}, \emph{12}, 1343\relax
\mciteBstWouldAddEndPuncttrue
\mciteSetBstMidEndSepPunct{\mcitedefaultmidpunct}
{\mcitedefaultendpunct}{\mcitedefaultseppunct}\relax
\EndOfBibitem
\bibitem[Naguib \latin{et~al.}(2023)Naguib, Kurtoglu, Presser, Lu, Niu, Heon,
  Hultman, Gogotsi, and Barsoum]{naguib2023two}
Naguib,~M.; Kurtoglu,~M.; Presser,~V.; Lu,~J.; Niu,~J.; Heon,~M.; Hultman,~L.;
  Gogotsi,~Y.; Barsoum,~M.~W. \emph{MXenes}; Jenny Stanford Publishing, 2023;
  pp 15--29\relax
\mciteBstWouldAddEndPuncttrue
\mciteSetBstMidEndSepPunct{\mcitedefaultmidpunct}
{\mcitedefaultendpunct}{\mcitedefaultseppunct}\relax
\EndOfBibitem
\bibitem[Huang \latin{et~al.}(2020)Huang, Li, Li, Ren, Wang, Wang, and
  Meng]{huang2020photocatalytic}
Huang,~K.; Li,~C.; Li,~H.; Ren,~G.; Wang,~L.; Wang,~W.; Meng,~X. Photocatalytic
  applications of two-dimensional Ti3C2 MXenes: a review. \emph{ACS Applied
  Nano Materials} \textbf{2020}, \emph{3}, 9581--9603\relax
\mciteBstWouldAddEndPuncttrue
\mciteSetBstMidEndSepPunct{\mcitedefaultmidpunct}
{\mcitedefaultendpunct}{\mcitedefaultseppunct}\relax
\EndOfBibitem
\bibitem[Naguib \latin{et~al.}(2014)Naguib, Mochalin, Barsoum, and
  Gogotsi]{naguib201425th}
Naguib,~M.; Mochalin,~V.~N.; Barsoum,~M.~W.; Gogotsi,~Y. 25th anniversary
  article: MXenes: a new family of two-dimensional materials. \emph{Advanced
  materials} \textbf{2014}, \emph{26}, 992--1005\relax
\mciteBstWouldAddEndPuncttrue
\mciteSetBstMidEndSepPunct{\mcitedefaultmidpunct}
{\mcitedefaultendpunct}{\mcitedefaultseppunct}\relax
\EndOfBibitem
\bibitem[Kamysbayev \latin{et~al.}(2020)Kamysbayev, Filatov, Hu, Rui, Lagunas,
  Wang, Klie, and Talapin]{kamysbayev2020covalent}
Kamysbayev,~V.; Filatov,~A.~S.; Hu,~H.; Rui,~X.; Lagunas,~F.; Wang,~D.;
  Klie,~R.~F.; Talapin,~D.~V. Covalent surface modifications and
  superconductivity of two-dimensional metal carbide MXenes. \emph{Science}
  \textbf{2020}, \emph{369}, 979--983\relax
\mciteBstWouldAddEndPuncttrue
\mciteSetBstMidEndSepPunct{\mcitedefaultmidpunct}
{\mcitedefaultendpunct}{\mcitedefaultseppunct}\relax
\EndOfBibitem
\bibitem[Pandey and Thygesen(2017)Pandey, and Thygesen]{pandey2017two}
Pandey,~M.; Thygesen,~K.~S. Two-dimensional MXenes as catalysts for
  electrochemical hydrogen evolution: A computational screening study.
  \emph{The Journal of Physical Chemistry C} \textbf{2017}, \emph{121},
  13593--13598\relax
\mciteBstWouldAddEndPuncttrue
\mciteSetBstMidEndSepPunct{\mcitedefaultmidpunct}
{\mcitedefaultendpunct}{\mcitedefaultseppunct}\relax
\EndOfBibitem
\bibitem[Malchik \latin{et~al.}(2021)Malchik, Shpigel, Levi, Penki, Gavriel,
  Bergman, Turgeman, Aurbach, and Gogotsi]{malchik2021mxene}
Malchik,~F.; Shpigel,~N.; Levi,~M.~D.; Penki,~T.~R.; Gavriel,~B.; Bergman,~G.;
  Turgeman,~M.; Aurbach,~D.; Gogotsi,~Y. MXene conductive binder for improving
  performance of sodium-ion anodes in water-in-salt electrolyte. \emph{Nano
  Energy} \textbf{2021}, \emph{79}, 105433\relax
\mciteBstWouldAddEndPuncttrue
\mciteSetBstMidEndSepPunct{\mcitedefaultmidpunct}
{\mcitedefaultendpunct}{\mcitedefaultseppunct}\relax
\EndOfBibitem
\bibitem[Ruan \latin{et~al.}(2023)Ruan, Meng, Huang, Xu, Wen, Ba, Singh, Zhang,
  Zhang, Xie, \latin{et~al.} others]{ruan2023enhancing}
Ruan,~X.; Meng,~D.; Huang,~C.; Xu,~M.; Wen,~X.; Ba,~K.; Singh,~D.~J.;
  Zhang,~H.; Zhang,~L.; Xie,~T.; others Enhancing Photocatalytic Hydrogen
  Evolution by Synergistic Benefits of MXene Cocatalysis and Homo-Interface
  Engineering. \emph{Small Methods} \textbf{2023}, \emph{7}, 2300627\relax
\mciteBstWouldAddEndPuncttrue
\mciteSetBstMidEndSepPunct{\mcitedefaultmidpunct}
{\mcitedefaultendpunct}{\mcitedefaultseppunct}\relax
\EndOfBibitem
\bibitem[Ram{\'\i}rez \latin{et~al.}(2023)Ram{\'\i}rez, Melillo, Osella, Asiri,
  Garcia, and Primo]{ramirez2023green}
Ram{\'\i}rez,~R.; Melillo,~A.; Osella,~S.; Asiri,~A.~M.; Garcia,~H.; Primo,~A.
  Green, HF-Free Synthesis of MXene Quantum Dots and their Photocatalytic
  Activity for Hydrogen Evolution. \emph{Small Methods} \textbf{2023},
  \emph{7}, 2300063\relax
\mciteBstWouldAddEndPuncttrue
\mciteSetBstMidEndSepPunct{\mcitedefaultmidpunct}
{\mcitedefaultendpunct}{\mcitedefaultseppunct}\relax
\EndOfBibitem
\bibitem[Gu \latin{et~al.}(2023)Gu, Zhang, Wang, Li, Chang, Huang, Gao, Cui,
  Liu, and Dai]{gu2023robust}
Gu,~H.; Zhang,~H.; Wang,~X.; Li,~Q.; Chang,~S.; Huang,~Y.; Gao,~L.; Cui,~Y.;
  Liu,~R.; Dai,~W.-L. Robust construction of CdSe nanorods@ Ti3C2 MXene
  nanosheet for superior photocatalytic H2 evolution. \emph{Applied Catalysis
  B: Environmental} \textbf{2023}, \emph{328}, 122537\relax
\mciteBstWouldAddEndPuncttrue
\mciteSetBstMidEndSepPunct{\mcitedefaultmidpunct}
{\mcitedefaultendpunct}{\mcitedefaultseppunct}\relax
\EndOfBibitem
\bibitem[Wu \latin{et~al.}(2023)Wu, Huang, Liu, Lv, and Li]{wu2023insight}
Wu,~C.; Huang,~W.; Liu,~H.; Lv,~K.; Li,~Q. Insight into synergistic effect of
  Ti3C2 MXene and MoS2 on anti-photocorrosion and photocatalytic of CdS for
  hydrogen production. \emph{Applied Catalysis B: Environmental} \textbf{2023},
  \emph{330}, 122653\relax
\mciteBstWouldAddEndPuncttrue
\mciteSetBstMidEndSepPunct{\mcitedefaultmidpunct}
{\mcitedefaultendpunct}{\mcitedefaultseppunct}\relax
\EndOfBibitem
\bibitem[Wanasinghe \latin{et~al.}(2024)Wanasinghe, Gjoni, Burson, Majeski,
  Zaslona, and Rury]{wanasinghe2024motional}
Wanasinghe,~S.~T.; Gjoni,~A.; Burson,~W.; Majeski,~C.; Zaslona,~B.; Rury,~A.~S.
  Motional narrowing through photonic exchange: Rational suppression of
  excitonic disorder from molecular cavity polariton formation. \emph{The
  Journal of Physical Chemistry Letters} \textbf{2024}, \emph{15},
  2405--2418\relax
\mciteBstWouldAddEndPuncttrue
\mciteSetBstMidEndSepPunct{\mcitedefaultmidpunct}
{\mcitedefaultendpunct}{\mcitedefaultseppunct}\relax
\EndOfBibitem
\bibitem[Kaur \latin{et~al.}(2014)Kaur, Singh, Pathak, Wagner, and
  Nunzi]{kaur2014organic}
Kaur,~N.; Singh,~M.; Pathak,~D.; Wagner,~T.; Nunzi,~J. Organic materials for
  photovoltaic applications: Review and mechanism. \emph{Synthetic Metals}
  \textbf{2014}, \emph{190}, 20--26\relax
\mciteBstWouldAddEndPuncttrue
\mciteSetBstMidEndSepPunct{\mcitedefaultmidpunct}
{\mcitedefaultendpunct}{\mcitedefaultseppunct}\relax
\EndOfBibitem
\bibitem[Axelsson \latin{et~al.}(2024)Axelsson, Xia, Wang, Cheng, and
  Tian]{axelsson2024role}
Axelsson,~M.; Xia,~Z.; Wang,~S.; Cheng,~M.; Tian,~H. Role of the
  Benzothiadiazole Unit in Organic Polymers on Photocatalytic Hydrogen
  Production. \emph{JACS Au} \textbf{2024}, \emph{4}, 570--577\relax
\mciteBstWouldAddEndPuncttrue
\mciteSetBstMidEndSepPunct{\mcitedefaultmidpunct}
{\mcitedefaultendpunct}{\mcitedefaultseppunct}\relax
\EndOfBibitem
\bibitem[Moi \latin{et~al.}(2023)Moi, Chandra, Maity, Pradhan, and
  Biradha]{moi2023band}
Moi,~R.; Chandra,~M.; Maity,~K.; Pradhan,~D.; Biradha,~K. Band Gap Modulation
  in Fluorescein-Based Isostructural Coordination Polymers for Enhanced
  Photocatalytic Hydrogen Evolution under Visible Light. \emph{Crystal Growth
  \& Design} \textbf{2023}, \emph{23}, 8407--8414\relax
\mciteBstWouldAddEndPuncttrue
\mciteSetBstMidEndSepPunct{\mcitedefaultmidpunct}
{\mcitedefaultendpunct}{\mcitedefaultseppunct}\relax
\EndOfBibitem
\bibitem[Hai \latin{et~al.}(2023)Hai, Fang, Xiong, Zhou, Wang, Sun, Su, and
  Chen]{hai2023charge}
Hai,~X.; Fang,~L.; Xiong,~M.; Zhou,~X.; Wang,~S.; Sun,~H.; Su,~C.; Chen,~H.
  Charge Density Modulation of Pyrene-Related Small Molecules by Nitrogen
  Heteroatoms Precisely Regulates Photocatalytic Generation of Hydrogen.
  \emph{ACS nano} \textbf{2023}, \emph{17}, 20570--20579\relax
\mciteBstWouldAddEndPuncttrue
\mciteSetBstMidEndSepPunct{\mcitedefaultmidpunct}
{\mcitedefaultendpunct}{\mcitedefaultseppunct}\relax
\EndOfBibitem
\bibitem[Zhang \latin{et~al.}(2023)Zhang, Huang, Lin, Chen, Dai, and
  Lin]{zhang2023optical}
Zhang,~S.; Huang,~Y.; Lin,~F.; Chen,~Y.; Dai,~H.; Lin,~Y. Optical and Acoustic
  Synergetic Sensing Platform Enabled by a Pyrene-Based Conjugated Polymer
  Self-Circulating Amplified System for Lung Cancer Detection. \emph{Analytical
  Chemistry} \textbf{2023}, \emph{95}, 9967--9974\relax
\mciteBstWouldAddEndPuncttrue
\mciteSetBstMidEndSepPunct{\mcitedefaultmidpunct}
{\mcitedefaultendpunct}{\mcitedefaultseppunct}\relax
\EndOfBibitem
\bibitem[Zhang \latin{et~al.}(2020)Zhang, Wei, Wei, Pan, and
  Su]{zhang2020ultrathin}
Zhang,~J.-H.; Wei,~M.-J.; Wei,~Z.-W.; Pan,~M.; Su,~C.-Y. Ultrathin graphitic
  carbon nitride nanosheets for photocatalytic hydrogen evolution. \emph{ACS
  Applied Nano Materials} \textbf{2020}, \emph{3}, 1010--1018\relax
\mciteBstWouldAddEndPuncttrue
\mciteSetBstMidEndSepPunct{\mcitedefaultmidpunct}
{\mcitedefaultendpunct}{\mcitedefaultseppunct}\relax
\EndOfBibitem
\bibitem[Koutsouroubi \latin{et~al.}(2020)Koutsouroubi, Vamvasakis, Papadas,
  Drivas, Choulis, Kennou, and Armatas]{koutsouroubi2020interface}
Koutsouroubi,~E.~D.; Vamvasakis,~I.; Papadas,~I.~T.; Drivas,~C.;
  Choulis,~S.~A.; Kennou,~S.; Armatas,~G.~S. Interface Engineering of
  MoS2-Modified Graphitic Carbon Nitride Nano-photocatalysts for an Efficient
  Hydrogen Evolution Reaction. \emph{ChemPlusChem} \textbf{2020}, \emph{85},
  1379--1388\relax
\mciteBstWouldAddEndPuncttrue
\mciteSetBstMidEndSepPunct{\mcitedefaultmidpunct}
{\mcitedefaultendpunct}{\mcitedefaultseppunct}\relax
\EndOfBibitem
\bibitem[Li \latin{et~al.}(2020)Li, Lin, Wang, Wang, Zhang, Zhang, and
  Liu]{li2020rational}
Li,~K.; Lin,~Y.-Z.; Wang,~K.; Wang,~Y.; Zhang,~Y.; Zhang,~Y.; Liu,~F.-T.
  Rational design of cocatalyst system for improving the photocatalytic
  hydrogen evolution activity of graphite carbon nitride. \emph{Applied
  Catalysis B: Environmental} \textbf{2020}, \emph{268}, 118402\relax
\mciteBstWouldAddEndPuncttrue
\mciteSetBstMidEndSepPunct{\mcitedefaultmidpunct}
{\mcitedefaultendpunct}{\mcitedefaultseppunct}\relax
\EndOfBibitem
\bibitem[Bai \latin{et~al.}(2020)Bai, Wang, Zhang, Wen, Wang, and
  Yang]{bai2020carboxyl}
Bai,~J.~Y.; Wang,~L.~J.; Zhang,~Y.~J.; Wen,~C.~F.; Wang,~X.~L.; Yang,~H.~G.
  Carboxyl functionalized graphite carbon nitride for remarkably enhanced
  photocatalytic hydrogen evolution. \emph{Applied Catalysis B: Environmental}
  \textbf{2020}, \emph{266}, 118590\relax
\mciteBstWouldAddEndPuncttrue
\mciteSetBstMidEndSepPunct{\mcitedefaultmidpunct}
{\mcitedefaultendpunct}{\mcitedefaultseppunct}\relax
\EndOfBibitem
\bibitem[Samanta \latin{et~al.}(2021)Samanta, Battula, Sardana, and
  Kailasam]{samanta2021solar}
Samanta,~S.; Battula,~V.~R.; Sardana,~N.; Kailasam,~K. Solar driven
  photocatalytic hydrogen evolution using graphitic-carbon nitride/NSGQDs
  heterostructures. \emph{Applied Surface Science} \textbf{2021}, \emph{563},
  150409\relax
\mciteBstWouldAddEndPuncttrue
\mciteSetBstMidEndSepPunct{\mcitedefaultmidpunct}
{\mcitedefaultendpunct}{\mcitedefaultseppunct}\relax
\EndOfBibitem
\bibitem[Dong \latin{et~al.}(2022)Dong, Chen, Zhao, Zhang, Lu, Wang, Li, and
  Chen]{dong2022situ}
Dong,~Q.; Chen,~Z.; Zhao,~B.; Zhang,~Y.; Lu,~Z.; Wang,~X.; Li,~J.; Chen,~W. In
  situ fabrication of niobium pentoxide/graphitic carbon nitride type-II
  heterojunctions for enhanced photocatalytic hydrogen evolution reaction.
  \emph{Journal of Colloid and Interface Science} \textbf{2022}, \emph{608},
  1951--1959\relax
\mciteBstWouldAddEndPuncttrue
\mciteSetBstMidEndSepPunct{\mcitedefaultmidpunct}
{\mcitedefaultendpunct}{\mcitedefaultseppunct}\relax
\EndOfBibitem
\bibitem[Chang \latin{et~al.}(2023)Chang, Fan, Zhu, Lei, Wu, Feng, Wang, and
  Ma]{CHANG20236729}
Chang,~X.; Fan,~H.; Zhu,~S.; Lei,~L.; Wu,~X.; Feng,~C.; Wang,~W.; Ma,~L.
  Engineering doping and defect in graphitic carbon nitride by one-pot method
  for enhanced photocatalytic hydrogen evolution. \emph{Ceramics International}
  \textbf{2023}, \emph{49}, 6729--6738\relax
\mciteBstWouldAddEndPuncttrue
\mciteSetBstMidEndSepPunct{\mcitedefaultmidpunct}
{\mcitedefaultendpunct}{\mcitedefaultseppunct}\relax
\EndOfBibitem
\bibitem[Ullah \latin{et~al.}(2024)Ullah, Habib, Lu, Li, Chen, Habib, and
  Xu]{ullah2024bimetallic}
Ullah,~I.; Habib,~S.; Lu,~X.-J.; Li,~J.-H.; Chen,~S.; Habib,~A.; Xu,~A.-W.
  Bimetallic nitride NiMoN loaded on graphitic carbon nitride for
  plasmon-enhanced visible light-driven photocatalytic hydrogen evolution from
  water splitting. \emph{Catalysis Science \& Technology} \textbf{2024},
  \emph{14}, 912--918\relax
\mciteBstWouldAddEndPuncttrue
\mciteSetBstMidEndSepPunct{\mcitedefaultmidpunct}
{\mcitedefaultendpunct}{\mcitedefaultseppunct}\relax
\EndOfBibitem
\bibitem[Chen \latin{et~al.}(2017)Chen, Ma, Zhu, and Xia]{chen2017metal}
Chen,~B.; Ma,~G.; Zhu,~Y.; Xia,~Y. Metal-organic-frameworks derived cobalt
  embedded in various carbon structures as bifunctional electrocatalysts for
  oxygen reduction and evolution reactions. \emph{Scientific reports}
  \textbf{2017}, \emph{7}, 5266\relax
\mciteBstWouldAddEndPuncttrue
\mciteSetBstMidEndSepPunct{\mcitedefaultmidpunct}
{\mcitedefaultendpunct}{\mcitedefaultseppunct}\relax
\EndOfBibitem
\bibitem[Gong \latin{et~al.}(2022)Gong, Li, Li, and Jin]{gong20222d}
Gong,~H.; Li,~Y.; Li,~H.; Jin,~Z. 2D CeO2 and a partially phosphated 2D
  Ni-based metal--organic framework formed an S-scheme heterojunction for
  efficient photocatalytic hydrogen evolution. \emph{Langmuir} \textbf{2022},
  \emph{38}, 2117--2131\relax
\mciteBstWouldAddEndPuncttrue
\mciteSetBstMidEndSepPunct{\mcitedefaultmidpunct}
{\mcitedefaultendpunct}{\mcitedefaultseppunct}\relax
\EndOfBibitem
\bibitem[Han \latin{et~al.}(2022)Han, Liu, Yan, Jiang, Zhang, and
  Gu]{han2022integrating}
Han,~W.-K.; Liu,~Y.; Yan,~X.; Jiang,~Y.; Zhang,~J.; Gu,~Z.-G. Integrating
  Light-Harvesting Ruthenium (II)-based Units into Three-Dimensional Metal
  Covalent Organic Frameworks for Photocatalytic Hydrogen Evolution.
  \emph{Angewandte Chemie International Edition} \textbf{2022}, \emph{61},
  e202208791\relax
\mciteBstWouldAddEndPuncttrue
\mciteSetBstMidEndSepPunct{\mcitedefaultmidpunct}
{\mcitedefaultendpunct}{\mcitedefaultseppunct}\relax
\EndOfBibitem
\bibitem[Shen \latin{et~al.}(2023)Shen, Li, Qin, Zhang, and
  Li]{shen2023efficient}
Shen,~R.; Li,~X.; Qin,~C.; Zhang,~P.; Li,~X. Efficient Photocatalytic Hydrogen
  Evolution by Modulating Excitonic Effects in Ni-Intercalated Covalent Organic
  Frameworks. \emph{Advanced Energy Materials} \textbf{2023}, \emph{13},
  2203695\relax
\mciteBstWouldAddEndPuncttrue
\mciteSetBstMidEndSepPunct{\mcitedefaultmidpunct}
{\mcitedefaultendpunct}{\mcitedefaultseppunct}\relax
\EndOfBibitem
\bibitem[Li \latin{et~al.}(2024)Li, Liang, Zhou, Huang, Wang, Xiao, and
  Liu]{li2024hybridization}
Li,~X.-A.; Liang,~Z.-Z.; Zhou,~Y.-C.; Huang,~J.-F.; Wang,~X.-L.; Xiao,~L.-M.;
  Liu,~J.-M. Hybridization of covalent organic frameworks and photosensitive
  metal-organic rings: A new strategy for constructing supramolecular Z-scheme
  heterostructures for ultrahigh photocatalytic hydrogen evolution.
  \emph{Aggregate} \textbf{2024}, \emph{5}, e442\relax
\mciteBstWouldAddEndPuncttrue
\mciteSetBstMidEndSepPunct{\mcitedefaultmidpunct}
{\mcitedefaultendpunct}{\mcitedefaultseppunct}\relax
\EndOfBibitem
\bibitem[Porwal \latin{et~al.}(2022)Porwal, Paul, Dixit, Mishra, and
  Singh]{porwal2022investigation}
Porwal,~S.; Paul,~M.; Dixit,~H.; Mishra,~S.; Singh,~T. Investigation of defects
  in Cs2SnI6-based double perovskite solar cells via SCAPS-1D. \emph{Advanced
  Theory and Simulations} \textbf{2022}, \emph{5}, 2200207\relax
\mciteBstWouldAddEndPuncttrue
\mciteSetBstMidEndSepPunct{\mcitedefaultmidpunct}
{\mcitedefaultendpunct}{\mcitedefaultseppunct}\relax
\EndOfBibitem
\bibitem[Amano and Nakayama(2022)Amano, and Nakayama]{amano2022improvement}
Amano,~F.; Nakayama,~S. Improvement of water splitting activity of
  silver-excess AgTaO3 photocatalysts via nitric acid washing treatment.
  \emph{Journal of Environmental Chemical Engineering} \textbf{2022},
  \emph{10}, 108089\relax
\mciteBstWouldAddEndPuncttrue
\mciteSetBstMidEndSepPunct{\mcitedefaultmidpunct}
{\mcitedefaultendpunct}{\mcitedefaultseppunct}\relax
\EndOfBibitem
\bibitem[Dixit \latin{et~al.}(2022)Dixit, Porwal, Boro, Paul, Ghosh, Mishra,
  and Singh]{dixit2022theoretical}
Dixit,~H.; Porwal,~S.; Boro,~B.; Paul,~M.; Ghosh,~S.; Mishra,~S.; Singh,~T. A
  theoretical exploration of lead-free double perovskite La2NiMnO6 based solar
  cell via SCAPS-1D. \emph{Optical Materials} \textbf{2022}, \emph{131},
  112611\relax
\mciteBstWouldAddEndPuncttrue
\mciteSetBstMidEndSepPunct{\mcitedefaultmidpunct}
{\mcitedefaultendpunct}{\mcitedefaultseppunct}\relax
\EndOfBibitem
\bibitem[Peng \latin{et~al.}(2017)Peng, Ma, Hu, and Wu]{peng2017first}
Peng,~Y.; Ma,~Z.; Hu,~J.; Wu,~K. A first-principles study of anionic (S) and
  cationic (V/Nb) doped Sr 2 Ta 2 O 7 for visible light photocatalysis.
  \emph{RSC Advances} \textbf{2017}, \emph{7}, 40922--40928\relax
\mciteBstWouldAddEndPuncttrue
\mciteSetBstMidEndSepPunct{\mcitedefaultmidpunct}
{\mcitedefaultendpunct}{\mcitedefaultseppunct}\relax
\EndOfBibitem
\bibitem[Irani \latin{et~al.}(2020)Irani, Ahmet, Jang, Berglund, Plate,
  H{\"o}hn, B{\"o}ttger, Schmitt, Dubourdieu, Lardhi, \latin{et~al.}
  others]{irani2020nature}
Irani,~R.; Ahmet,~I.~Y.; Jang,~J.-W.; Berglund,~S.~P.; Plate,~P.; H{\"o}hn,~C.;
  B{\"o}ttger,~R.; Schmitt,~S.~W.; Dubourdieu,~C.; Lardhi,~S.; others Nature of
  nitrogen incorporation in BiVO4 photoanodes through chemical and physical
  methods. \emph{Solar RRL} \textbf{2020}, \emph{4}, 1900290\relax
\mciteBstWouldAddEndPuncttrue
\mciteSetBstMidEndSepPunct{\mcitedefaultmidpunct}
{\mcitedefaultendpunct}{\mcitedefaultseppunct}\relax
\EndOfBibitem
\bibitem[Idris \latin{et~al.}(2020)Idris, Liu, Shah, Zhang, Ma, Malik, Jin,
  Rasheed, Sun, Li, \latin{et~al.} others]{idris2020novel}
Idris,~A.~M.; Liu,~T.; Shah,~J.~H.; Zhang,~X.; Ma,~C.; Malik,~A.~S.; Jin,~A.;
  Rasheed,~S.; Sun,~Y.; Li,~C.; others A novel double perovskite oxide
  semiconductor Sr2CoWO6 as bifunctional photocatalyst for photocatalytic
  oxygen and hydrogen evolution reactions from water under visible light
  irradiation. \emph{Solar RRL} \textbf{2020}, \emph{4}, 1900456\relax
\mciteBstWouldAddEndPuncttrue
\mciteSetBstMidEndSepPunct{\mcitedefaultmidpunct}
{\mcitedefaultendpunct}{\mcitedefaultseppunct}\relax
\EndOfBibitem
\bibitem[Idris \latin{et~al.}(2020)Idris, Liu, Hussain~Shah, Han, and
  Li]{idris2020sr2cotao6}
Idris,~A.~M.; Liu,~T.; Hussain~Shah,~J.; Han,~H.; Li,~C. Sr2CoTaO6 double
  perovskite oxide as a novel visible-light-absorbing bifunctional
  photocatalyst for photocatalytic oxygen and hydrogen evolution reactions.
  \emph{ACS Sustainable Chemistry \& Engineering} \textbf{2020}, \emph{8},
  14190--14197\relax
\mciteBstWouldAddEndPuncttrue
\mciteSetBstMidEndSepPunct{\mcitedefaultmidpunct}
{\mcitedefaultendpunct}{\mcitedefaultseppunct}\relax
\EndOfBibitem
\bibitem[Wang \latin{et~al.}(2021)Wang, Zhang, Wang, Gao, Fan, Wu, Zong, and
  Li]{wang2021mechanistic}
Wang,~H.; Zhang,~H.; Wang,~J.; Gao,~Y.; Fan,~F.; Wu,~K.; Zong,~X.; Li,~C.
  Mechanistic Understanding of Efficient Photocatalytic H2 Evolution on
  Two-Dimensional Layered Lead Iodide Hybrid Perovskites. \emph{Angewandte
  Chemie} \textbf{2021}, \emph{133}, 7452--7457\relax
\mciteBstWouldAddEndPuncttrue
\mciteSetBstMidEndSepPunct{\mcitedefaultmidpunct}
{\mcitedefaultendpunct}{\mcitedefaultseppunct}\relax
\EndOfBibitem
\bibitem[Miody{\'n}ska \latin{et~al.}(2023)Miody{\'n}ska, Klimczuk, Lisowski,
  and Zaleska-Medynska]{miodynska2023bi}
Miody{\'n}ska,~M.; Klimczuk,~T.; Lisowski,~W.; Zaleska-Medynska,~A. Bi-based
  halide perovskites: Stability and opportunities in the photocatalytic
  approach for hydrogen evolution. \emph{Catalysis Communications}
  \textbf{2023}, \emph{177}, 106656\relax
\mciteBstWouldAddEndPuncttrue
\mciteSetBstMidEndSepPunct{\mcitedefaultmidpunct}
{\mcitedefaultendpunct}{\mcitedefaultseppunct}\relax
\EndOfBibitem
\bibitem[Gao \latin{et~al.}(2024)Gao, Wang, Chen, He, Zhang, Li, Li, Zhou,
  Feng, Mei, \latin{et~al.} others]{gao2024confinement}
Gao,~S.; Wang,~B.; Chen,~F.; He,~G.; Zhang,~T.; Li,~L.; Li,~J.; Zhou,~Y.;
  Feng,~B.; Mei,~D.; others Confinement of CsPbBr3 Perovskite Nanocrystals into
  Extra-large-pore Zeolite for Efficient and Stable Photocatalytic Hydrogen
  Evolution. \emph{Angewandte Chemie International Edition} \textbf{2024},
  \emph{63}, e202319996\relax
\mciteBstWouldAddEndPuncttrue
\mciteSetBstMidEndSepPunct{\mcitedefaultmidpunct}
{\mcitedefaultendpunct}{\mcitedefaultseppunct}\relax
\EndOfBibitem
\bibitem[Liao \latin{et~al.}(2012)Liao, Huang, and Wu]{liao2012hydrogen}
Liao,~C.-H.; Huang,~C.-W.; Wu,~J.~C. Hydrogen production from
  semiconductor-based photocatalysis via water splitting. \emph{Catalysts}
  \textbf{2012}, \emph{2}, 490--516\relax
\mciteBstWouldAddEndPuncttrue
\mciteSetBstMidEndSepPunct{\mcitedefaultmidpunct}
{\mcitedefaultendpunct}{\mcitedefaultseppunct}\relax
\EndOfBibitem
\bibitem[Song \latin{et~al.}(2022)Song, Bu, Fan, Wang, and
  Zhao]{song2022photocatalytic}
Song,~X.; Bu,~H.; Fan,~Y.; Wang,~J.; Zhao,~M. Photocatalytic hydrogen
  production and storage in carbon nanotubes: a first-principles study.
  \emph{RSC advances} \textbf{2022}, \emph{12}, 17029--17035\relax
\mciteBstWouldAddEndPuncttrue
\mciteSetBstMidEndSepPunct{\mcitedefaultmidpunct}
{\mcitedefaultendpunct}{\mcitedefaultseppunct}\relax
\EndOfBibitem
\bibitem[Yang \latin{et~al.}(2017)Yang, Li, Zhang, Cui, Wang, Jiang, Zhao, Luo,
  and Jiang]{yang2017combining}
Yang,~L.; Li,~X.; Zhang,~G.; Cui,~P.; Wang,~X.; Jiang,~X.; Zhao,~J.; Luo,~Y.;
  Jiang,~J. Combining photocatalytic hydrogen generation and capsule storage in
  graphene based sandwich structures. \emph{Nature communications}
  \textbf{2017}, \emph{8}, 16049\relax
\mciteBstWouldAddEndPuncttrue
\mciteSetBstMidEndSepPunct{\mcitedefaultmidpunct}
{\mcitedefaultendpunct}{\mcitedefaultseppunct}\relax
\EndOfBibitem
\bibitem[Zhang \latin{et~al.}(2015)Zhang, Tong, Zhang, Zhang, and
  Liu]{zhang2015porous}
Zhang,~H.; Tong,~C.-J.; Zhang,~Y.; Zhang,~Y.-N.; Liu,~L.-M. Porous BN for
  hydrogen generation and storage. \emph{Journal of Materials Chemistry A}
  \textbf{2015}, \emph{3}, 9632--9637\relax
\mciteBstWouldAddEndPuncttrue
\mciteSetBstMidEndSepPunct{\mcitedefaultmidpunct}
{\mcitedefaultendpunct}{\mcitedefaultseppunct}\relax
\EndOfBibitem
\end{mcitethebibliography}

\end{document}